\documentclass[titlepage,12pt]{article}
\usepackage{amssymb,amsmath,amsfonts}
\usepackage{graphicx}
\usepackage{cite}
\makeatletter
\def\@sect#1#2#3#4#5#6[#7]#8{\ifnum #2>\c@secnumdepth
  \def\@svsec{}\else
  \refstepcounter{#1}\edef\@svsec{\csname the#1\endcsname.\hskip0.5em}\fi
  \@tempskipa #5\relax
  \ifdim \@tempskipa>\z@
    \begingroup
      #6\relax
      \@hangfrom{\hskip #3\relax\@svsec}{\interlinepenalty \@M #8\par}%
    \endgroup
    \csname #1mark\endcsname{#7}\addcontentsline
      {toc}{#1}{\ifnum #2>\c@secnumdepth \else
        \protect\numberline{\csname the#1\endcsname}\fi #7}%
  \else
    \def\@svsechd{#6\hskip #3\@svsec #8\csname #1mark\endcsname
      {#7}\addcontentsline{toc}{#1}{\ifnum #2>\c@secnumdepth \else
        \protect\numberline{\csname the#1\endcsname}\fi #7}}%
  \fi \@xsect{#5}}
\@addtoreset{equation}{section}
\makeatother
\renewcommand\thesection{\Roman{section}}
\renewcommand\theequation{\ifnum \value{section}>0
 \thesection.\arabic{equation}%
\else
\arabic{equation}%
\fi}

\newcommand{\kh}{{{\bf \hat k}}}

\newcommand{\dhh}{{\bf \hat d}}
\newcommand{\one}{1\!\mbox{l}}
\newcommand{\als}{\alpha_s}
\newcommand{\al}{\alpha}
\newcommand{\MS}{\overline{\rm MS}}
\newcommand{\tbart}{t{\bar t}}
\newcommand{\mtt}{M_{t\bar t}}
\def\Sp{{\mathbf S}_t}
\newcommand{\Sm}{{\mathbf S_{\bar t}}}

\newcommand{\velp}{{\boldsymbol{\hat\ell^+}}}
\newcommand{\velm}{{\boldsymbol{\hat\ell^-}}}
\newcommand{\velpL}{{\boldsymbol{\hat\ell^+_L}}}
\newcommand{\velmL}{{\boldsymbol{\hat\ell^-_L}}}
\newcommand{\tatau}{{\boldsymbol{\tau}}}
\newcommand{\nn}{\nonumber}

\renewcommand{\thefootnote}{\small\fnsymbol{footnote}}

\begin{document}
\begin{titlepage}
  \begin{flushright}
    TTK-10-22
    \end{flushright}
\vspace{0.01cm}
\begin{center}
{\LARGE {\bf Distributions and
 correlations for top quark pair production and decay 
at the Tevatron and LHC}  \\
\vspace{1.5cm}
\large{\bf Werner Bernreuther\,$^{a}$\footnote{\tt breuther@physik.rwth-aachen.de} and 
 Zong-Guo Si\,$^{b}$}\footnote{\tt zgsi@sdu.edu.cn}
\par\vspace{1cm}
$^a$Institut f\"ur Theoretische Physik, RWTH Aachen University, 52056 Aachen, Germany\\
$^b$Department of Physics, Shandong University, Jinan, Shandong
250100, China
\par\vspace{1cm}
{\bf Abstract}\\
\parbox[t]{\textwidth}
{\small{We investigate a number of  observables  
that are and will be instrumental in the exploration of 
$\tbart$ production and decay at the Tevatron and the LHC. 
For this analysis we made a computer program that incorporates 
 besides the NLO QCD corrections to $\tbart$ production and decay also
   mixed weak-QCD corrections to the production amplitudes, and that allows for
 studies of correlated versus uncorrelated $\tbart$ events. In this
 paper we analyze and compute observables mostly for dileptonic $\tbart$ final
 states to next-to-leading order in the strong
   and weak gauge couplings (NLOW), with selection cuts.
  We calculate charge asymmetries of the top quark and of
 $\ell=e,\mu$  and we 
   compare, where possible, with 
   experimental results. We show that top-quark spin correlations
   affect a leptonic pair asymmetry (which has not been measured so
   far) by $\sim 7\%$. 
 We  determine several dileptonic angular correlations,
 which reflect $\tbart$ spin correlations, namely the beam, off-diagonal and helicity correlation,
  and the opening angle distribution (defined in the $t$, $\bar t$
  rest frames)  when selection cuts are applied.
Our NLOW predictions for the beam, off-diagonal, and helicity
correlation for the Tevatron agree with recent measurements 
  by the CDF and D0 experiments. In addition
 we make predictions for estimators   of these correlations  as functions
of $\mtt$. These estimators
    may prove useful for the Tevatron and 
 also  in the  early rounds of LHC data-analyses,
   where the  event numbers will  not be abundant.
Furthermore, we compute  to NLOW in the gauge couplings
  two dilepton angular correlations that are
defined in the laboratory frame, for correlated and
uncorrelated $\tbart$ events at the LHC.
Finally, based on our SM results, we identify 
several observables that allow to search  for non-standard parity-
and CP-violating interactions, especially with future LHC data.
}}
}
\end{center}
\vspace*{1.5cm}

PACS number(s): 12.38.Bx, 13.88.+e, 14.65.Ha 
\end{titlepage}
%
%
\setcounter{footnote}{0}
\renewcommand{\thefootnote}{\arabic{footnote}}
\setcounter{page}{1}

\section{Introduction} 
\label{introduction}
Exploration of the production and decay of  top-quark pairs $(\tbart)$
 is and will be among the central physics topics at the
 Tevatron and at the Large Hadron Collider (LHC). On the theoretical
 side, the investigation of these processes has reached quite a
 sophisticated 
 level. The next-to-leading order (NLO) QCD corrections to the
 $\tbart$ cross sections and several top-quark
  distributions 
 have been known for quite some time from
   the pioneering work of
   \cite{Nason:1987xz,Nason:1989zy,Beenakker:1988bq,Beenakker:1990ma}
   and \cite{Mangano:1991jk,Frixione:1995fj}, which was extended by threshold
   resummation calculations 
 \cite{Laenen:1993xr,Kidonakis:1997gm,Bonciani:1998vc,Kidonakis:2001nj}.
 The pair cross sections were recently
   updated by \cite{Moch:2008qy,Cacciari:2008zb,Kidonakis:2008mu}.
 The NLO QCD corrections including the full dependence of the matrix
 elements on the top-quark spin degrees of freedom were determined in
 \cite{Bernreuther:2000yn,Bernreuther:2001bx,Bernreuther:2001rq,Bernreuther:2004jv}. 
 The mixed weak-interaction QCD corrections
 \cite{Beenakker:1993yr,Bernreuther:2005is,Bernreuther:2006vg,Bernreuther:2008md,Kuhn:2005it,Kuhn:2006vh,Moretti:2006nf}and  the photonic
 corrections \cite{Hollik:2007sw} to $\tbart$ production and the
   order $\alpha_s^3$ nonfactorizable QCD corrections 
   \cite{Beenakker:1999ya} are also
 known. The computations of the NLO QCD corrections to $t {\bar t} \,+ \, {\rm jet}$,
  $t {\bar t} \,+ \, b {\bar b}$, and
     $t {\bar t} \,+ \, 2 \, {\rm jets}$,  which are important background
  processes in the search for (the) Higgs boson(s), were reported in
  \cite{Dittmaier:2007wz,Dittmaier:2008uj},  in
  \cite{Bredenstein:2009aj,Bredenstein:2010rs,Bevilacqua:2009zn}, and
  in \cite{Bevilacqua:2010ve}
  respectively.  Differential distributions 
 for polarized semileptonic and non-leptonic decays of the top quark
  were determined  order $\alpha_s$  by
    \cite{Czarnecki:1990pe} and   \cite{Brandenburg:2002xr},
  respectively, while its total width is known to order   $\alpha_s^2$
   \cite{Czarnecki:1998qc,Chetyrkin:1999ju}.

A number of observables   that are instrumental in exploring the
$\tbart$ production dynamics were investigated in detail\footnote{For more complete
    overviews, cf., e.g.,
    \cite{Bernreuther:2008ju,Incandela:2009pf,schwanen1}.}. These include the
  $W$-boson helicity fractions in $t\to bW$ at NLO in the
  gauge couplings \cite{Do:2002ky} and to  $\alpha_s^2$ QCD 
\cite{Piclum:2008zz}, and the 
  top-quark charge asymmetry at the Tevatron in the context of standard model
  (SM) interactions
  \cite{Halzen:1987xd,Nason:1989zy,Beenakker:1990ma,Kuhn:1998kw,Bowen:2005ap,Antunano:2007da,Almeida:2008ug,Dittmaier:2007wz}. This asymmetry was measured by the D0 and CDF experiments
  \cite{:2007qb,Aaltonen:2008hc,CDFpublic1}. Top-quark spin effects, in
 particular $\tbart$ spin correlations were analyzed in
 \cite{Kuhn:1983ix,Barger:1988jj,Kane:1991bg,Arens:1992fg,Mahlon:1995zn,Stelzer:1995gc,Brandenburg:1996df,Chang:1995ay,Bernreuther:1995cx,Dharmaratna:xd,Mahlon:1997uc,Uwer:2004vp,Nelson:2005jp,Godbole:2006tq,Mahlon:2010gw} at LO
 and in \cite{Bernreuther:2001rq,Bernreuther:2004jv}  at NLO QCD.
 Measurements of   $\tbart$ spin correlations at the Tevatron were 
 reported in \cite{D0public1,CDFpublic2,CDFpublic3}.
 Quite recently there has been an increased activity on
  theoretical and phenomenological issues, 
   including the first determination
   of the running mass ${\overline m}_t$ of the top quark
   \cite{Langenfeld:2009wd} from $\sigma_{\tbart}^{exp}$, analytic phase-space integration of the
   NLO QCD parton cross sections $gg,q {\bar q}, gq \to {\tbart}X$  \cite{Czakon:2008ii},  threshold
   expansion improvements of the partonic cross sections
   \cite{Czakon:2008cx,Czakon:2009zw,Beneke:2009ye,Beneke:2009rj}
     and of the pair-invariant mass distribution \cite{Ahrens:2009uz},
  an analysis of  quasi-bound-state effects
     in $\sigma_{\tbart}$ at the production threshold \cite{Hagiwara:2008df,Kiyo:2008bv},
 determination of various building blocks for the
   computation of $\sigma_{\tbart}$ at NNLO QCD 
\cite{Czakon:2007wk,Czakon:2008zk,Bonciani:2009nb,Kniehl:2008fd,Anastasiou:2008vd,Becher:2009kw,Ferroglia:2009ii},
  phenomenological analyses of boosted top-quark events 
\cite{Kaplan:2008ie,Almeida:2008tp} and of the 
   use of the LHC $\tbart$ cross  section for calibrating parton
   distribution functions \cite{Nadolsky:2008zw}. The NLO QCD
   Monte-Carlo generators \cite{Frixione:2008ym,Frixione:2007zp},
   \cite{MCFM}, and \cite{Frixione:2007nw} are important tools for the
    simulation of
   $\tbart$ production and decay including parton showering \cite{Frixione:2008ym,Frixione:2007nw}. Recently,
   results were presented based on a code which incorporates NLO QCD
   corrections to
    $\tbart$ production and semileptonic top-quark decay with $\tbart$ spin correlations 
    \cite{Melnikov:2009dn}. \\

\par
In this paper we analyze a number of distributions and correlations,
mostly for dileptonic  $\tbart$ partonic final states, at next-to-leading order in the strong and weak
 couplings. The spin-dependent terms are included to this order
   of the perturbation expansion in the production and decay matrix elements for the
  $\tbart$ intermediate state. For our analysis we made  a computer code   with
 input based on previous work 
\cite{Bernreuther:2001rq,Bernreuther:2004jv,Bernreuther:2005is,Bernreuther:2006vg,Bernreuther:2008md,Brandenburg:2002xr}
 where the authors of this paper were involved. As our approach is based on 
  spin density matrices being independent of  specific spin reference axes, 
    we can  switch off  the
   $t$- and/or $\bar t$-spin dependent terms in the matrix elements,
   in particular the $t{\bar t}$ spin correlations, also at NLO
   in the gauge couplings.  This seems to us quite useful for
   performing  comparative studies
   of final state  distributions  
     for  correlated and uncorrelated  ${t\bar t}$ events.
   This paper addresses the following
   issues: i) A study of the size of the 
  weak interaction corrections with
   respect to NLO QCD corrections for several distributions and
   spin correlations. So far, only comparisons with respect to LO QCD
   results were made in the above-mentioned literature. ii) An
   investigation  of the   top-quark charge  asymmetry and of two leptonic
   charge asymmetries \cite{Bowen:2005ap}
    for correlated and uncorrelated dileptonic $\tbart$ events at the Tevatron
    if acceptance cuts are applied. To our knowledge, this has not been
   analyzed so far to NLO in the gauge couplings. 
     iii) A detailed analysis of dileptonic angular 
    correlations induced by $\tbart$ spin correlations.

 In Sect.~\ref{sec:setup} we
 describe the ingredients of our analysis and list our input
 parameters. Sect.~\ref{sec.rett} contains results for several
 distributions and spin correlations at the level of the 
 $\tbart$ intermediate states. 
  The top-quark charge asymmetry at the Tevatron will be considered in
  some detail and a comparison with the results of
  \cite{Antunano:2007da} will be made. In Sect.~\ref{sec:diljet}
 we give our results for several distributions and angular
 correlations for dileptonic final states with phase-space cuts
   at NLO in the strong and
 weak couplings. Two distributions will be computed for semileptonic
 final states. We analyze the top-quark charge asymmetry and a related
  pair asymmetry at the
 Tevatron and two related
 leptonic charge asymmetries for correlated and uncorrelated dileptonic
 $\tbart$ events, and we compare the top-quark asymmetries with experimental 
 results \cite{:2007qb,Aaltonen:2008hc,CDFpublic1}. Our predictions of
 the leptonic asymmetries have to await confrontation with measurements.
 Then we investigate a number of dileptonic angular
 correlations,  for the Tevatron and the LHC at $\sqrt{s}=10$ and 14
 TeV, at NLO in the
 gauge couplings that are (in)sensitive to $t{\bar t}$ spin
 correlations. Moreover, we compare with the recent 
  Tevatron measurements \cite{D0public1,CDFpublic2,CDFpublic3}.
Based on our SM results we identify, in addition,
several observables that allow for searches for non-standard parity-
and CP-violating interactions, especially with future LHC data.
Sect.~\ref{sec.conc} contains a summary and outlook.

\section{Theoretical set-up}
\label{sec:setup}
We analyze hadronic top quark pair production and decay, at NLO in the
QCD and weak couplings, in the on-shell
approximation for the  $t, {\bar t}$ quarks, taking their
spin degrees  of freedom in the production and decay stage  into
account. To be specific we consider, at  the parton level, the following  reactions:
\begin{equation}
  gg, q{\bar q} \ {\buildrel
    t{\bar t}\over \longrightarrow} \  b {\bar b} + 4 f,
  \label{eq:ttrec1}
\end{equation}
\begin{equation}
  gg, q{\bar q} \  {\buildrel
    t{\bar t}\over \longrightarrow} \  b {\bar b} + 4f  + g,
  \label{eq:ttrec2}
\end{equation}
\begin{equation}
  g + q ({\bar q})\  {\buildrel
    t{\bar t}\over \longrightarrow}\   b {\bar b} + 4f  + q ({\bar q}),
  \label{eq:ttrec3}
\end{equation}
where  $f=q,\ell,\nu_{\ell}$. As the top 
 quark is a narrow resonance, we employ the narrow width
approximation $\Gamma_t/m_t\to 0$ and incorporate the factorizable QCD
 and mixed weak-QCD corrections, which form gauge-invariant sets. In this
approximation the squared matrix element $|{\cal M}|^2$ of
the respective reaction is of the form
\begin{equation}
  |{\cal M}|^2 = N_i\frac{1}{2\hat s} \frac{\pi^2}{m_t^2\Gamma_t^2} {\rm Tr}\;[\rho
  R{\bar{\rho}}]
  =  N_i \frac{1}{2\hat s} \frac{\pi^2}{m_t^2\Gamma_t^2} \rho_{\alpha'\alpha}
  R_{\alpha\alpha',\beta\beta'}{\bar{\rho}}_{\beta'\beta} .
  \label{eq:trace}
\end{equation}
 Here  $R, {\rho}, {\bar{\rho}}$  are the 
 production and $t$ and $\bar t$ decay density
 matrices  for (\ref{eq:ttrec1}) --  (\ref{eq:ttrec3}). The subscripts in  Eq.~(\ref{eq:trace}) denote the  top and antitop
spin indices, and the factor $N_i$ arises from averaging over the colors and
spins of the initial partons.

As already mentioned above,  in our spin density matrix approach 
  the $t, \bar t$ spin-dependent terms, in particular
$\tbart$ spin correlations can be switched off  also at
NLO in the gauge couplings. In addition, top-spin effects can be analyzed for
 arbitrary spin bases at NLO, in particular at the level of the
 $\tbart$  intermediate states. This is quite useful for studies of 
  non-standard interactions that change the SM-induced top-spin correlations.

As to the density matrices $R$ of the $t\bar t$  production reactions 
$q {\bar q} \to t {\bar t}(g)$, $gg\to t {\bar t}(g)$,
 $g q  \to t  {\bar t} q$, and $g {\bar q}  \to t  {\bar t} {\bar q}$:
 Besides taking into account the  ${\cal O}(\als^3)$  QCD corrections 
\cite{Bernreuther:2004jv}, we incorporate also
 the weak interaction corrections  of order $\als^2\al$,  $\als\al$,
 $\al^2$, and $\als\al^2$ to the spin density matrices of
  these processes as computed in
 \cite{Bernreuther:2005is,Bernreuther:2006vg,Bernreuther:2008md}.

 We use the decay density
matrices of the main semileptonic and non-leptonic
 SM decay modes of  polarized (anti)top quarks,
 $t\to b W^+  \to  b \ell^+ \nu_{\ell} (g), \,  b q {\bar q}' (g)$ 
 (where $q {\bar q}'= u {\bar d}, c {\bar s}$)
  to ${\cal O}(\als)$ in fully
 differential form. Specific distributions 
  were given in  \cite{Brandenburg:2002xr}.
 
In the numerical implementation of (\ref{eq:trace}) we use
 the expanded form of the right-hand side to NLO in the gauge
 couplings. We expand also the top width  $\Gamma_t$ 
 in the denominator of (\ref{eq:trace}) to  ${\cal O}(\als)$.  
 In Sections~\ref{sec.rett} and~\ref{sec:diljet}  below, the acronyms
   LO and NLO refer to the  ${\cal O}(\als^2)$
 and  ${\cal O}(\als^3)$ matrix elements in QCD, while NLOW refers to 
 the additional inclusion of the above-mentioned weak corrections.

The infrared (i.e. soft and collinear) singularities in the production
   density matrices  are taken care of by a phase-space slicing method, as worked
 out explicitly in \cite{Bernreuther:2004jv}. For semileptonic
   top-quark decay we use a slicing procedure, too, while for
   non-leptonic decay a hybrid scheme is employed
   (cf. \cite{Brandenburg:2002xr}).
     We checked that 
 the results of Sections~\ref{sec.rett} and~\ref{sec:diljet}
   are independent of the slicing parameter
  $x_{min}$ for $x_{min} \lesssim $ a few $\times 10^{-3}$.           
The results of  these sections were obtained with 
 $x_{min} = 10^{-3}$. 

With these building blocks we developed a numerical program for computing
 observables to fixed order in the gauge couplings as described above, at the
 level of the leptonic and partonic  final states that correspond to
 the dileptonic, lepton plus jets, and  all-jets final states. As we
 are primarily interested in modeling ``top as a signal'', we
 investigate in this paper only distributions mostly for  dileptonic and
  a few for semileptonic final states which have the highest
 signal-to-background ratios. 
 \begin{eqnarray}
p {\bar p}, p p &\rightarrow& t{\bar t}  + X 
\rightarrow \ell^+ \ell\,'^-  \, j_b \, j_{\bar b} \, + X, \label{eq:ttll} \\
p {\bar p}, p p &\rightarrow& t{\bar t}  + X 
\rightarrow \ell^+  \, j_b \, j_{\bar b} j_1 \, j_2  +
X, \qquad \ell^-  \, j_{\bar b} \, j_b \, j_1 \, j_2  +
X, 
\label{eq:ttlj1}
\end{eqnarray}
where
$\ell = e,\mu,\tau,$ and $j_b$ $(j_{1,2})$ denote (non) $b$ jets.

Our input parameters are as follows.
Throughout this paper the top-quark mass $m_t$ is defined in the
on-shell scheme. Adopting the common, albeit not rigorously
justifiable practice of identifying the experimentally determined
 top mass with the on-shell mass, we use  for $m_t$ the recent Tevatron average
 $m_t =  173.1  \pm  1.3$ GeV \cite{:2009ec}. Our main results,
 the charge asymmetries and the top-spin induced angular distributions 
 and correlations given in 
 Sect.~\ref{sec:diljet} are, in fact, not very sensitive to variations $\Delta m_t$
  of the top mass by a few GeV. The QCD coupling is defined, as usual,
  in the $\MS$ scheme.
We employ CTEQ6L1 and CTEQ6.6M parton distribution
  functions (PDF) for LO and NLO computations \cite{Nadolsky:2008zw}, which
  correspond to $\als(m_Z)=0.130$ and $\als(m_Z)=0.118$,
 respectively. The evolution of $\als = \alpha_{s,5}$ to some higher
 scale  $\mu_R$  is made within $5$-flavor QCD. In the parton matrix
  elements, which were computed in \cite{Bernreuther:2004jv}
 within 6-flavor QCD, 
the  $\MS$ coupling $\alpha_{s,6}(\mu_R)$ is replaced by
$\alpha_{s,5}(\mu_R)$ according to 
$\alpha_{s,6}/\pi = \alpha_{s,5}/\pi - \frac{1}{6}
    (\alpha_{s,5}/\pi)^2 \ln(m_t^2/\mu_R^2) + {\cal
      O}(\alpha_{s,5}^3)$
  \cite{Bernreuther:1981sg}.
 For the top width we use 
 $\Gamma_t =1.3882 - \als(m_t) \times 0.8076$ GeV.
  We use an on-shell
 mass $m_b=4.8$ GeV for the $b$ quark, while the masses of the lighter quarks and of the
 leptons are put equal to zero.
In the evaluation of the weak corrections
 we use  the QED coupling $\al(m_Z)$=0.008 and 
 for the  $W$- and $Z$-boson masses  and the $W$-boson width
 $\Gamma_W$ the measured values given in \cite{Amsler:2008zzb}. 
 We put the CKM matrix elements that appear
  in the CKM-unsuppressed semi- and non-leptonic top-quark decay amplitudes
 equal to one, $|V_{tb}|=|V_{qq'}|=1$. 
 The Higgs-boson mass is put equal to
 $m_H=120$ GeV. The dependence of the results of 
 Sections~\ref{sec.rett} and~\ref{sec:diljet}   on 
 $m_H \lesssim 200$ GeV is insignificant.

\section{Results for $t {\bar t}\, X$: }
\label{sec.rett}

In this section we give  results, at LO, NLO, and NLOW in the gauge
couplings,  for several
  distributions and correlations at the level of on-shell $t \bar t$
  intermediate states,
\begin{equation}
p {\bar p}, p p \rightarrow  t{\bar t}  + X \, ,
\label{eq:ttfins}
\end{equation}
both for the Tevatron ($\sqrt{s}=1.96$ TeV) and for the LHC at
  center-of-mass energies $\sqrt{s}=10$ and $14$ TeV. 
No phase space cuts are
  applied  in this section. The
  factorization and  renormalization scales are
  put equal, $\mu_F=\mu_R \equiv \mu$ and we choose $\mu=m_t/2, \,
  m_t,$ and $2m_t$.
   
 As was shown in 
 \cite{Bernreuther:2006vg}, the weak-interaction contributions to the
  total cross section $\sigma_{\tbart}$ at the Tevatron and at the LHC
  are marginal as compared with the NLO QCD corrections. On the other
   hand, for a number of distributions, for instance,
   the transverse momentum distribution of the
   (anti)top quark and the $\tbart$ invariant mass distribution,
      the weak-interaction corrections are of potential importance at
      large energies due to large Sudakov logarithms.
   For the $p_T$ $(\mtt)$  distribution at the 
   LHC  these  corrections amount to about $-10\%$ $(-6\%)$ of 
  the LO results for $p_T \sim 1$ TeV 
   ($\mtt \sim 2$ TeV),  and the respective
    ratios $d\sigma_{weak}/d\sigma_{LO}$  grow in 
    magnitude for
    larger $p_T$ and $\mtt$ \cite{Bernreuther:2005is,Bernreuther:2006vg,Bernreuther:2008md,Kuhn:2006vh,Kuhn:2005it}. Here we compare  the weak and  NLO  QCD 
  corrections for $\tbart$ production, to wit,  
   for the $p_T$ and $\mtt$ distributions and for several
   $\tbart$  spin correlation observables at the LHC, and 
   for the charge asymmetry at the Tevatron. 

\begin{figure}
\begin{center}
\includegraphics[width=8cm]{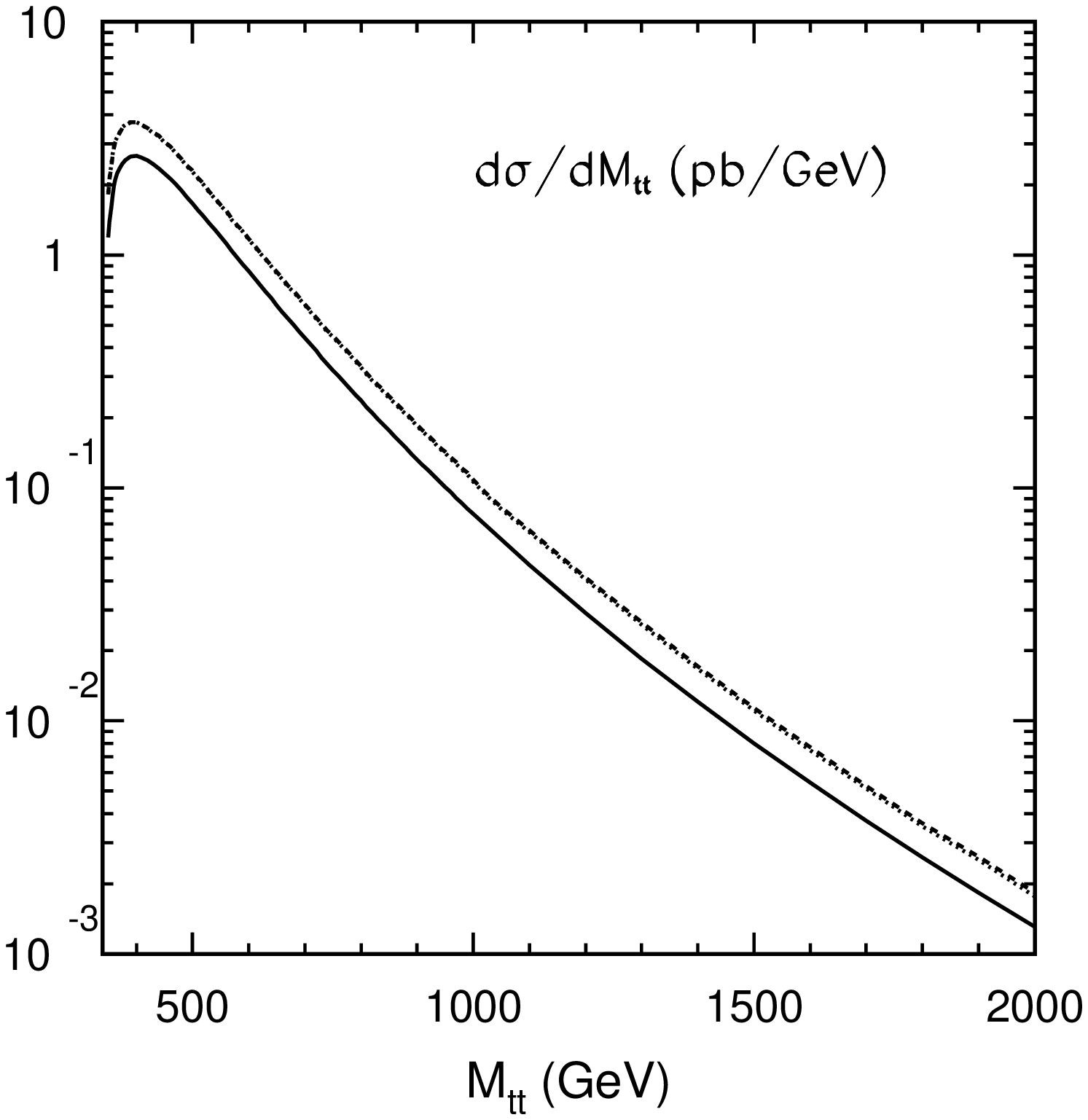}
\includegraphics[width=8cm]{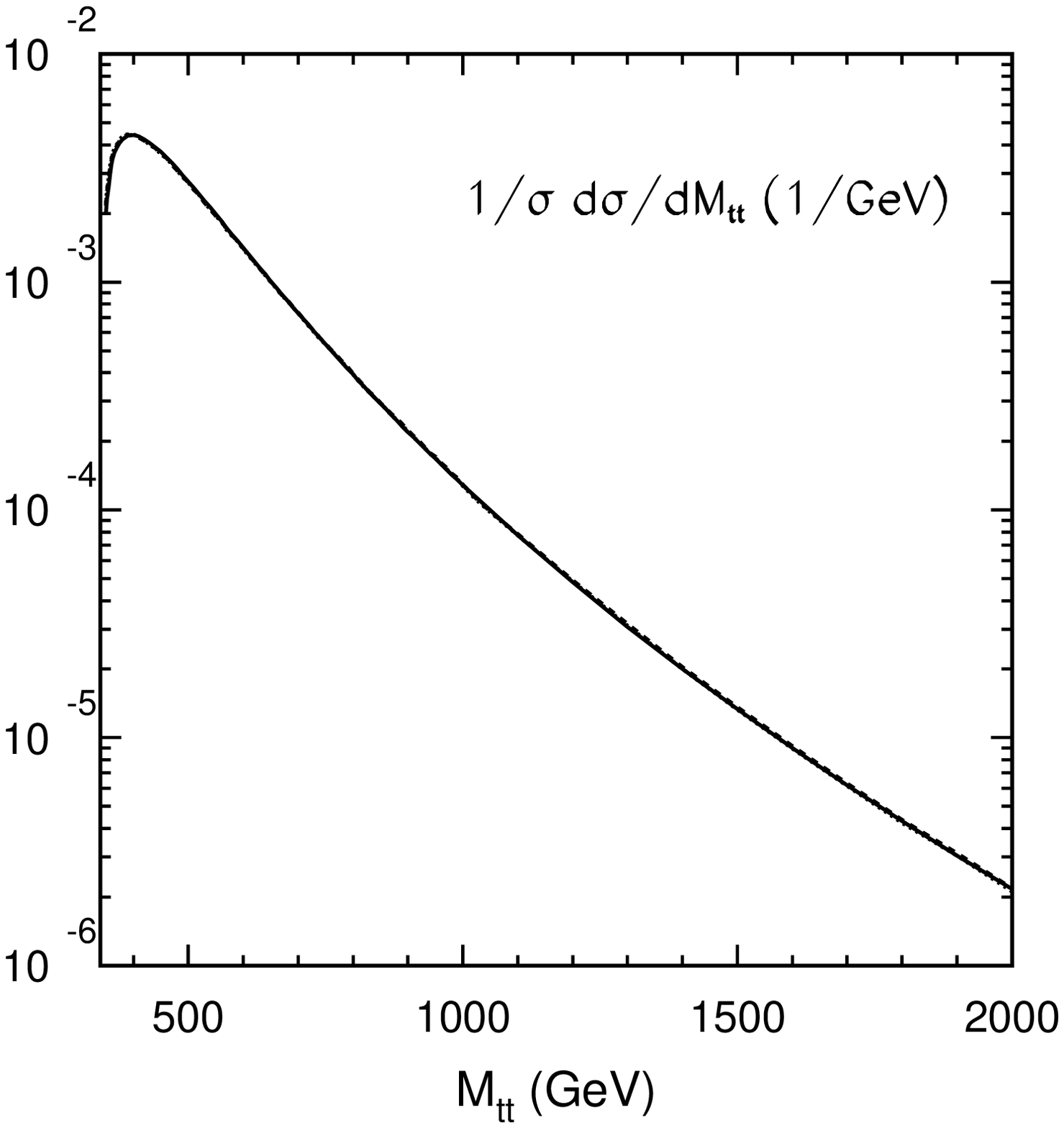}
\end{center}
\caption{The unnormalized (left panel) and normalized (right panel)
  $\tbart$ invariant-mass distribution for the LHC ($\sqrt{s}=14$ TeV)
   at LO (solid), NLO (dashed), and NLOW (dotted) for   $\mu =m_t$.}
 \label{fig:mttcomp}
\end{figure}

\begin{figure}
\begin{center}
\includegraphics[width=8cm]{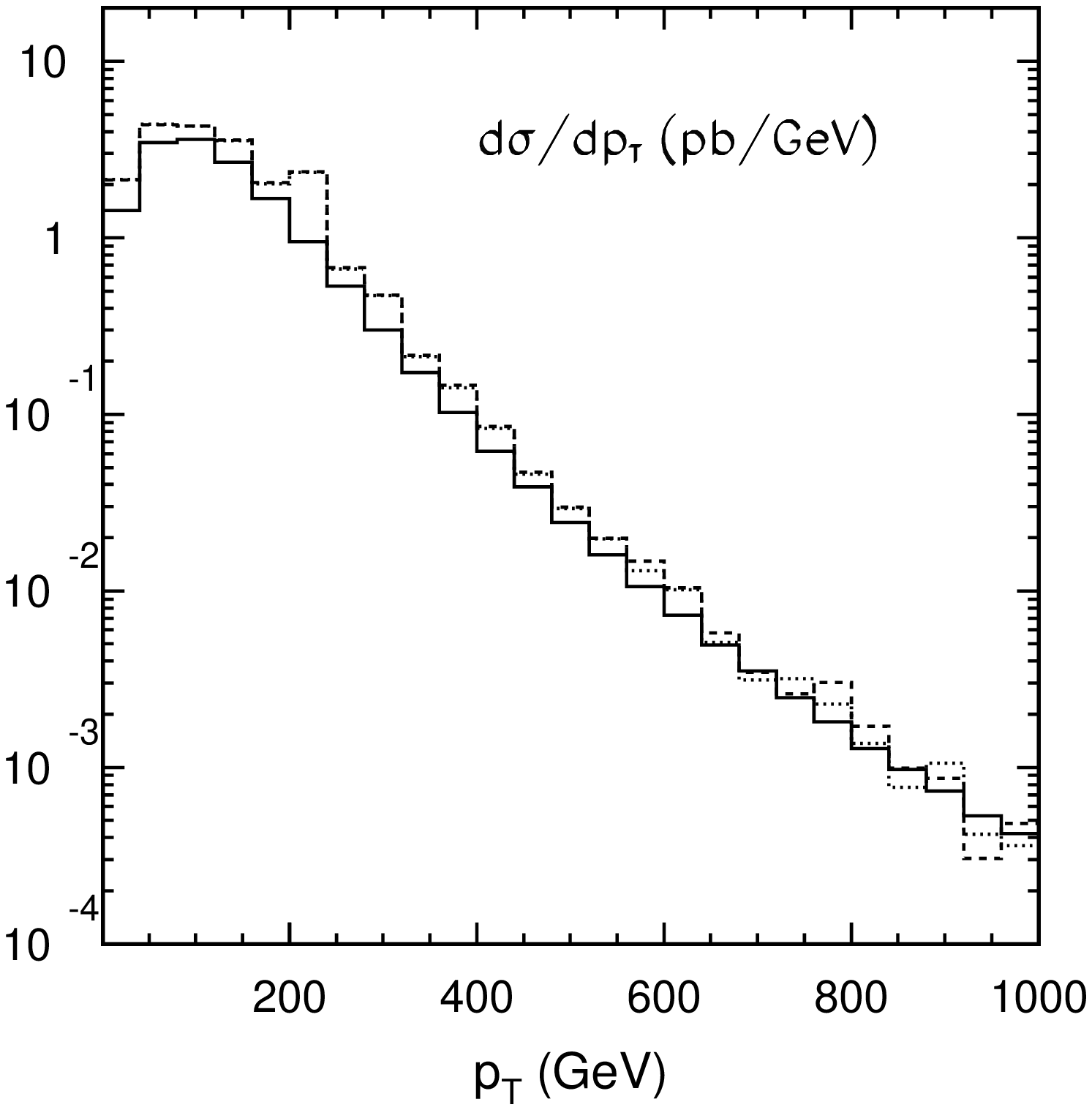}
\includegraphics[width=8cm]{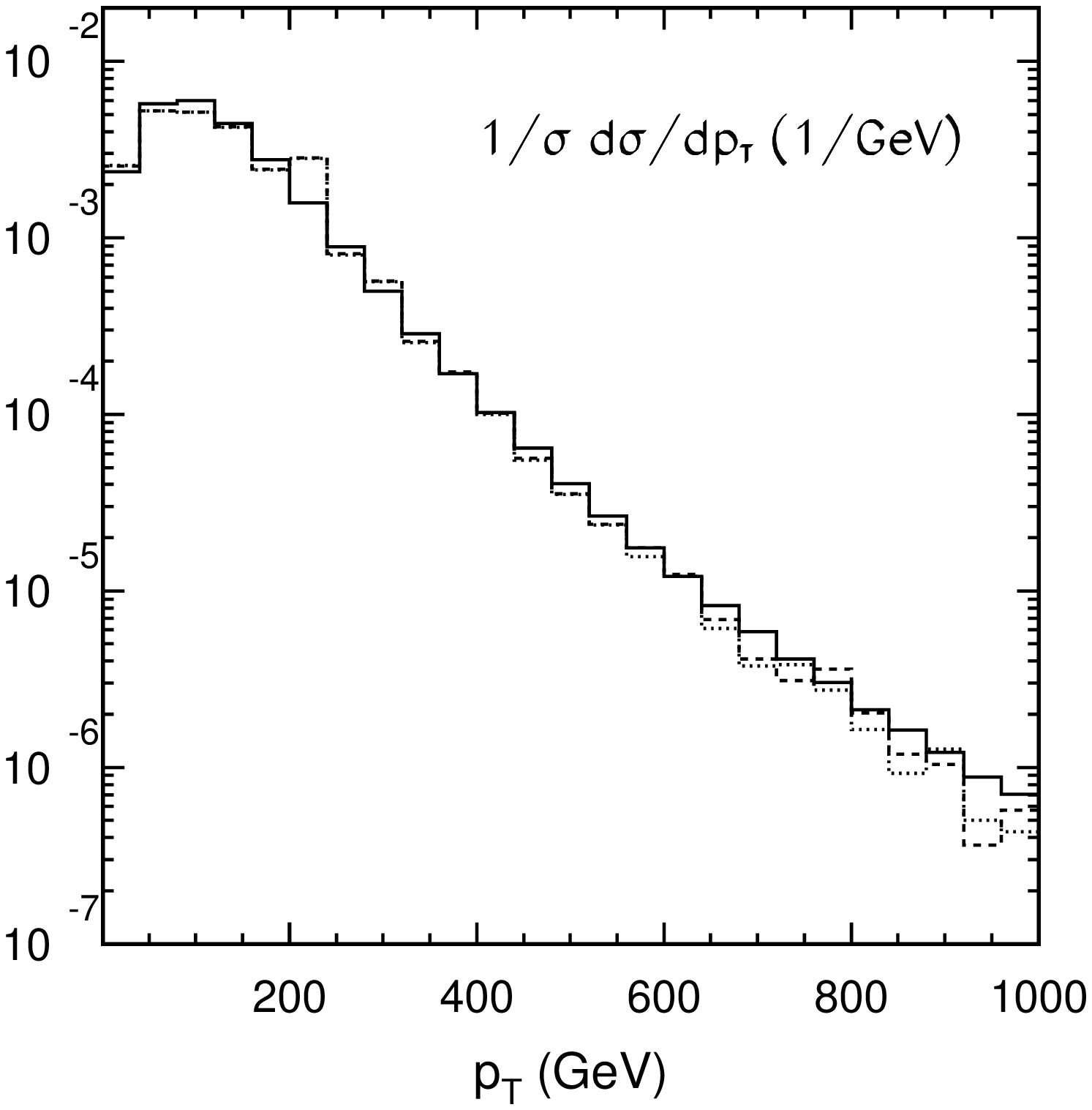}
\end{center}
\caption{The unnormalized (left) and normalized (right)
  $p_T$  distribution of the top quark for the LHC ($\sqrt{s}=14$ TeV)
   at LO (solid), NLO (dashed), and NLOW (dotted) for   $\mu =m_t$.}
 \label{fig:ptcomp}
\end{figure}

\begin{figure}
\begin{center}
\includegraphics[width=8cm]{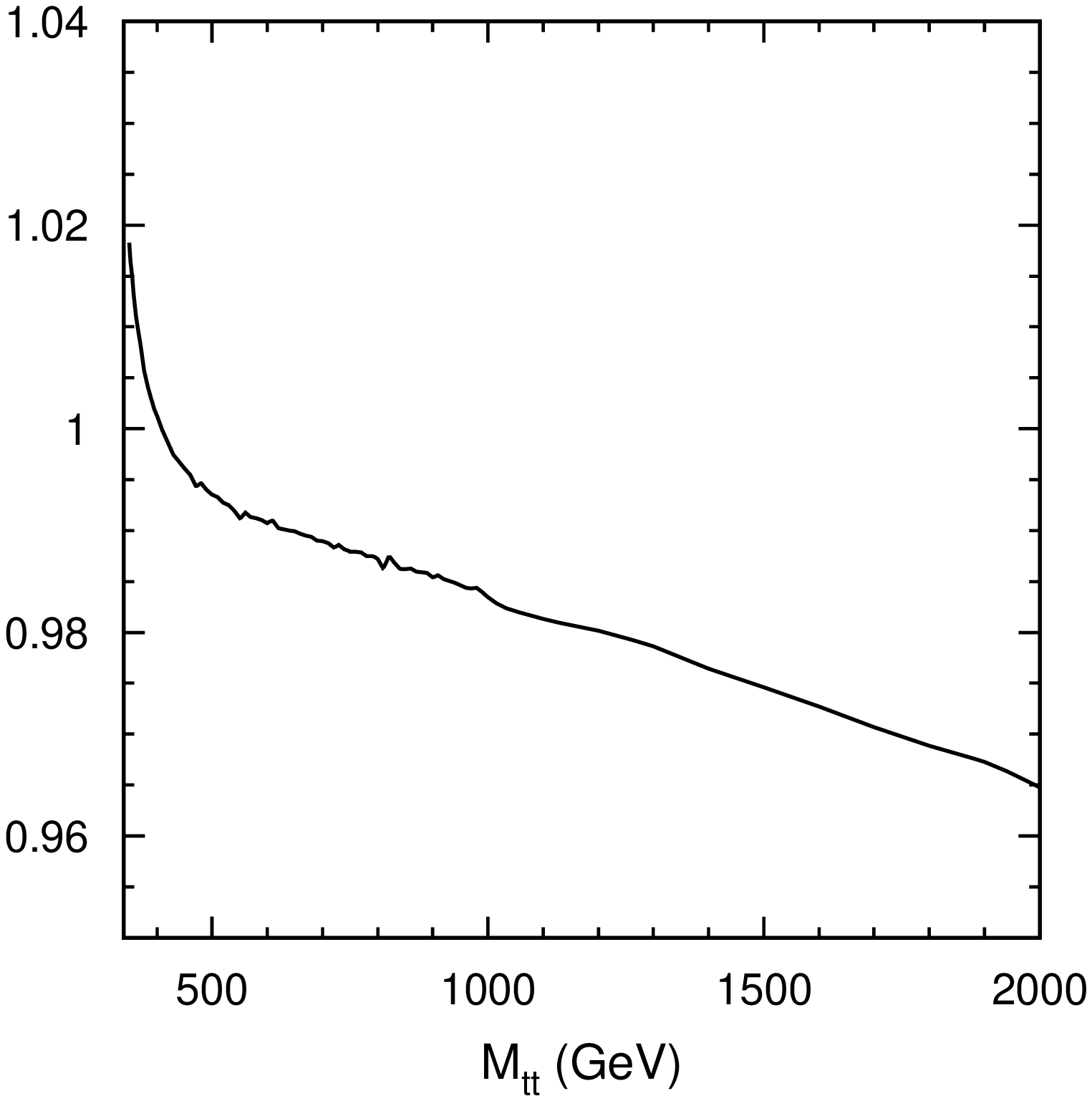}
\includegraphics[width=8cm]{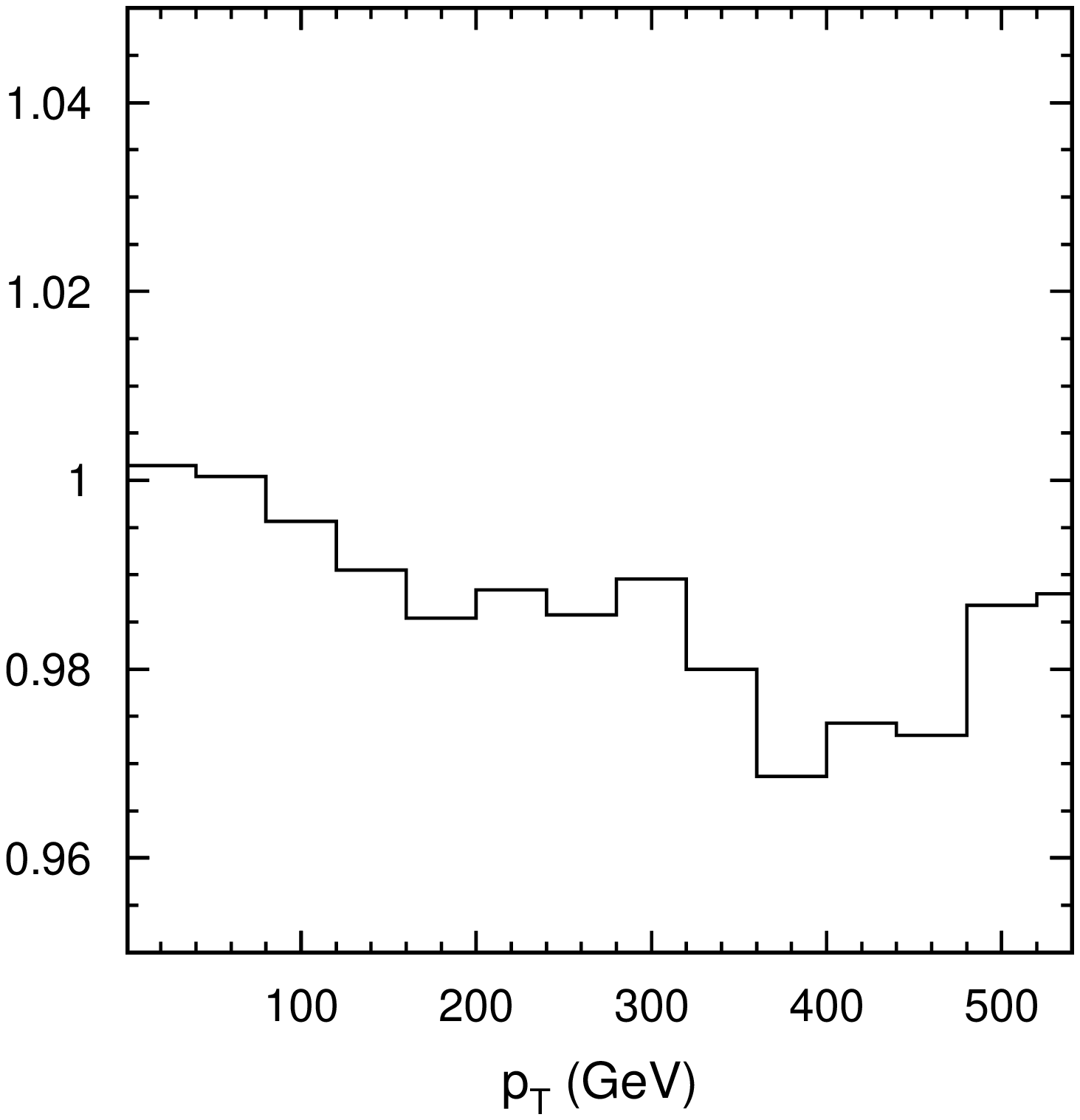}
\end{center}
\caption{ The ratio 
   of the $\mtt$ distribution evaluated at NLOW and NLO with $\mu=m_t$
    (left panel)
   and  the corresponding ratio of the $p_T$ distribution
   (right panel)  for the LHC (14 TeV). } 
\label{fig:ratioptmlhc}
\end{figure}

Fig.~\ref{fig:mttcomp} shows, for the LHC at $\sqrt{s}=14$ TeV, 
 the (un)normalized $\tbart$ invariant-mass distribution at 
 LO, NLO, and NLOW for $\mu =m_t$. 
  The analogous plots are displayed  in Fig.~\ref{fig:ptcomp}
for the (un)normalized $p_T$ distribution of the top
quark. 
In Fig.~\ref{fig:ratioptmlhc} the ratios of these distributions
   evaluated at  NLOW and NLO  are plotted for $\mu =m_t$. The weak-interaction
   corrections to the $\mtt$ distribution are negative except close to
    $2 m_t$ and become larger in magnitude than $\sim 2\%$ with
    respect to the NLO QCD corrections for $\mtt \gtrsim 1.2$ TeV.
     For the $p_T$ distribution the respective ratio is  also smaller 
     than one, except in the extreme forward- and backward region
     \cite{Bernreuther:2008md}, and the weak-interaction
      corrections grow to several percent in magnitude beyond $p_T
      \sim 500 $ GeV (not shown in Fig.~\ref{fig:ratioptmlhc}).
 A study of the scale dependence of the normalized $p_T$ and $\mtt$
  distributions in the range $p_T \lesssim 1$ TeV, $\mtt \lesssim 1.2$
  TeV  shows that for $m_t/2 \leq \mu \leq 2 m_t$ the scale variations
  are of the same order of magnitude as the weak-interaction
  contributions to the NLO QCD distributions.

%

%

%


Next we consider the asymmetry with respect to interchange of the
the $t$  and $\bar t$ charges, which is known to  be relatively large 
 at the Tevatron.  This  asymmetry is generated at NLO
QCD by the interference of  even and odd terms under $t\leftrightarrow
{\bar t}$ in the amplitudes
for $q {\bar q}$ annihilation and, likewise, for
$g q$ and $g {\bar q}$ fusion  
 \cite{Halzen:1987xd,Nason:1989zy,Beenakker:1990ma,Kuhn:1998kw,Bowen:2005ap,Antunano:2007da,Almeida:2008ug}.
(The contribution of the latter two processes to the charge asymmetry
 is one order of magnitude below the $q\bar q$ contribution.)
 The weak interaction corrections to the $q \bar q$ and flavor
 excitation processes contribute, too. 
The conventionally defined
    differential and integrated charge asymmetry, $A(y)$  and
$A$, are:
\begin{equation}
A(y) = \frac{N_t(y) - N_{\bar t}(y)}{N_t(y) + N_{\bar
    t}(y)}\, , \qquad 
A =\frac{\int\limits_{y>0} N_t(y) - \int\limits_{y>0}N_{\bar
    t}(y)}{\int\limits_{y>0} N_t(y) + \int\limits_{y>0} N_{\bar
    t}(y)}\, ,
\label{chasy}
\end{equation}
where $y$ denotes  the  rapidities  $y_t$, $y_{\bar t}$ of 
the $t$ and $\bar t$ quark  defined in the laboratory frame, 
 and $N(y)=d\sigma_{\tbart} /d y$. 

As the  $p{\bar p}$ state at the Tevatron is a CP eigenstate in the laboratory
frame,  CP invariance of the SM corrections\footnote{The
  CP-violating terms in the matrix elements for $p {\bar p} \to t {\bar t} \to$ final state
  induced by the non-zero SM KM phase $\delta_{KM}$ in higher orders of
  perturbation theory are of course numerically irrelevant.}
implies that $N_{\bar t}(y_{\bar t}) = N_t(-y_t)$, which in turn  
 implies that  $A$
is equal to the forward-backward asymmetry of the top quark:
\begin{equation}
A_{FB}^t = \frac{\int\limits_{y>0}N_t(y) -\int\limits_{y<0} N_{t}(y)}{\int\limits_{y>0}N_t(y) +
  \int\limits_{y<0}N_{t}(y)}\, .
\label{fobasy}
\end{equation}
In the case of the Tevatron, $CP$ invariance tells us also that
$A_{FB}^{\bar t}= - A_{FB}^{t}$. Non-standard $CP$-violating
interactions, if existent, may invalidate 
this relation by a small amount. As to $\tbart$ produced by $gg$
fusion: the Bose symmetry of the $gg$ state precludes a  contribution
from this sub-process  to the {\it integrated} asymmetries $A$, $A_{FB}^t$ --
 irrespective of whether or not the production density matrix $R_{gg}$
  contains P- and/or CP-violating pieces.

The D0 and CDF  experiments \cite{:2007qb,Aaltonen:2008hc} and 
  \cite{Antunano:2007da} considered  
  the following observable  which  also 
 reflects the charge asymmetry in $\tbart$ production
 at the  Tevatron:
\begin{equation} 
 A^{\tbart} = \frac{\int N(\Delta y > 0) - \int N(\Delta y <
       0)}{\int N (\Delta y > 0 ) +
  \int N(\Delta y  < 0)}\, ,
\label{fobattasy}
\end{equation}
where $\Delta y = y_{t}- y_{\bar t}$. This asymmetry is, for kinematical reasons, 
 larger than $A$.

It should be recalled that the 
  numerators of the above asymmetries start at order $\alpha_s^3$ in
  the QCD perturbation expansion. Thus, the above inclusive 
  asymmetries are of order $\alpha_s$. They were computed to this order, 
including weak corrections, by \cite{Kuhn:1998kw,Antunano:2007da}.  The
order  $\alpha_s^4$ QCD corrections are not known.
Also within our framework, the determination of these asymmetries is 
 a LO calculation. However, we  compute here the numerators of 
   $A$  and $A^{\tbart}$
  by taking into account the  NLO respectively NLOW
   $\tbart$ parton matrix elements\footnote{In the calculation
  of \cite{Kuhn:1998kw,Antunano:2007da} just the relevant 
${\cal O}(\alpha_s^3)$  terms  that generate the asymmetries
 were taken  into account in the numerator. That is why they used LO
 PDF also in the numerator.} both from 
  $q {\bar q}, \, gq \, ({\bar q}),$ and $gg$ initial states.
(In this way, the computation of 
  these asymmetries provides also a check of our code.) We adopt
  the following procedure. We evaluate the numerators   at NLO and NLOW, respectively,
  with NLO PDF and the denominators at LO with LO PDF, using however the
   same value of $\alpha_s$ in both numerator and
   denominator. (Evaluating the denominator including the NLO
   corrections would be in conflict with the perturabtion expansion.)
 We label our results with the acronyms NLO' and NLOW', respectively. 

\begin{figure}
\begin{center}
\includegraphics[width=8cm]{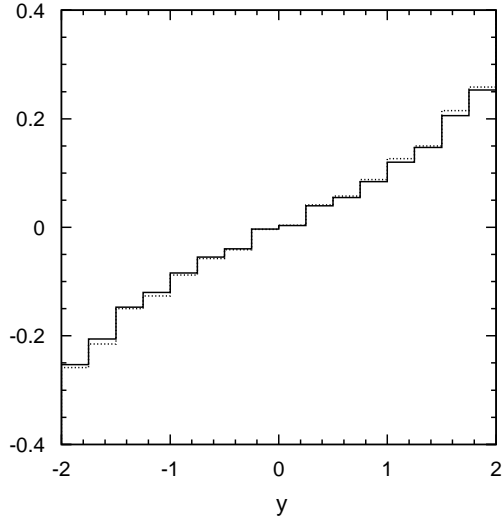}
\end{center}
\caption{
 The differential charge asymmetry $A(y)$, eq. (\ref{chasy}),
    for the Tevatron 
   as a function of the 
   (anti)top-quark rapidity $y$, for  $\mu =m_t.$  The solid
   and dotted lines correspond to NLO' and NLOW', respectively. 
  }
 \label{fig:chasytev}
\end{figure}

We have plotted in Fig.~\ref{fig:chasytev} the differential charge
  asymmetry (\ref{chasy}) for the Tevatron 
 at NLO' and NLOW'  for a fixed scale, $\mu=m_t$. 
 Table~\ref{chasyTEVLHC} contains our results 
 for the integrated charge asymmetry $A$ 
  and  the  pair asymmetry $A^{\tbart}$  for
 the for three
 different scales. 

\begin{table}[hp]
\begin{center}
\caption{
 Results 
 for the  charge asymmetry $A$ 
 and the  pair asymmetry $A^{\tbart}$  at NLO' and NLOW' for 
      $\mu =m_t/2, m_t$, and $2m_t$. The  acronyms NLO' and
    NLOW' are explained in the text. 
 }
\vspace*{0.5cm}

\label{chasyTEVLHC} 
\begin{tabular}{||c|c|c|c||}\hline \hline
  &\multicolumn{3}{c|}{Tevatron} \\ \hline \hline
$\mu$ & $m_t/2$ & $m_t$ & $2m_t$  \\ \hline \hline
$A$ (NLO') &0.054 &0.049 &0.045  \\ \hline 
$A$ (NLOW')&0.056 &0.051 &0.048  \\ \hline \hline 
$A^{\tbart}$ (NLO')&0.084&0.076 &0.071 \\ \hline 
$A^{\tbart}$ (NLOW')&0.087 &0.080 &0.075 \\ \hline\hline 
\end{tabular}
\end{center}
\end{table}


%

Our results for $A$ and  $A^{\tbart}$ given in Table~\ref{chasyTEVLHC}
agree  with the results $A=0.051(6)$ and $A^{\tbart} = 0.078(9)$ of 
 \cite{Antunano:2007da}. The following remarks are, however, in
 order. 
  We use  a different set of PDF and a slightly
  different value of $m_t$. The contribution of the
 mixed electroweak-QCD corrections to the charge asymmetries 
  were determined in \cite{Kuhn:1998kw} 
  for the $q\bar q$ initiated production process
   which, as found by  these authors, increases the QCD asymmetry $A$ by a
  factor 1.09.
   We have taken into account in addition the weak-interaction contributions
   to $g q\ ({\bar q}) \to t {\bar t} q\ ({\bar q})$ and a specific contribution
    to  $b {\bar b} \to  t {\bar t}$  \cite{Bernreuther:2008md}.
  In  Born approximation, the amplitude of the latter reaction
  consists of 
 $t$-channel $W$ boson exchange
$b {\bar b} {\stackrel{ W}{\longrightarrow}} t {\bar t}$
 and  $s$-channel photon,  $Z$ boson, and gluon exchanges $b {\bar b}
{\stackrel{\gamma, Z, g}{\longrightarrow}} t {\bar t}$.
     The  terms in $|{\cal M}(b {\bar b} \to t \bar t)|^2_B$
that involve the electroweak interactions are of order $\alpha^2$ and
$\alpha_s\alpha$, respectively. They contain also
charge-asymmetric pieces which contribute to $A$ and $A^{\tbart}$
 with opposite sign as compared to the contributions from the order
    $\alpha_s^2 \alpha$ terms in the $q \bar q\to {\tbart}$ 
 matrix element. At present the size of these $b{\bar b}$ induced 
 contributions can not be determined precisely in view of the rather
 poorly known $b$-quark PDF.  In any case, these additional
  terms diminish the overall size of the  weak-interaction
  contributions, as the numbers of Table~\ref{chasyTEVLHC} show.
  
The spread of the asymmetries  in Table~\ref{chasyTEVLHC} 
  due to scale variations certainly underestimates
   the true uncertainties of these predictions of the
   inclusive asymmetries\footnote{For  ${\tbart}$ + jet final states
   the  charge asymmetry was  calculated to NLO QCD
 in \cite{Dittmaier:2007wz}.}. A more realistic
   estimate of the theory uncertainty, namely  $\sim 30 \%$, is  provided by the calculation of $A$
 in \cite{Almeida:2008ug} where NLL QCD threshold resummation corrections were
   included. Comparisons with the experimental results from the
   Tevatron will be made in the next section.

At the LHC, the
 initial $pp$ state  is an eigenstate of parity.
 Thus, $A_{FB}^t=A_{FB}^{\bar t}=0$ in the laboratory frame, 
 as long as only parity-invariant interactions -- more general, only
 parity-even terms in the scattering operator -- are taken into account.
 The parity-violating terms of the weak corrections
  appear in fact only in the $t$- and/or $\bar t$-spin dependent
 terms of the partonic production density matrices and do, therefore,
 not contribute when making predictions for top quarks summed over
 their spins.  As a consequence, the 
  differential charge asymmetry  $A(y)$ at the LHC induced  
  by the SM interactions must be symmetric with respect to $y=0$. 
Thus, in the absence of cuts, the integrated charge asymmetry $A=0$.
  We have checked that our code reproduces this conclusion.
  Nevertheless, there are non-zero effects  in suitably defined
     distributions \cite{Kuhn:1998kw,Antunano:2007da}. \\


Next we consider top-spin induced effects.
  The polarization of the $t$ or $\bar t$ quarks
   in the hadronically  produced
  $\tbart$   sample   due to weak interactions or
  QCD absorptive parts is small 
\cite{Bernreuther:1995cx,Bernreuther:2006vg}. On the other hand, the
 QCD dynamics leads to correlations of the $t$, $\bar t$ spins already
 at tree level which
 are characteristic of the production mode. One may
  investigate   $\tbart$ spin correlations  by the following 
 expectation values of spin operators ${\cal{\bf O}}$:  
\begin{equation}
{\cal O}_{ab} = \langle 4(\Sp\cdot {\bf{\hat a}})
(\Sm\cdot {\bf{\hat b}})\rangle  \quad  {\rm and} \quad
   {\cal O}_{\rm spin}= \langle \frac{4}{3}\, \Sp \cdot \Sm  \rangle \, .
\label{genob}
\end{equation}
Here $\Sp$, $\Sm$ are the top and antitop spin operators, and 
the unit vectors ${\bf{\hat a}}$ and ${\bf{\hat b}}$ are 
arbitrary reference directions, which may be interpreted as 
  spin axes.  As usual, the normalization of the averages is such
that $\langle 1 \rangle = 1.$ We use the following spin
  bases \cite{Bernreuther:2004jv} in the zero-momentum frame (ZMF)  of the $\tbart$ pair:
\begin{eqnarray}
     {\bf\hat a} = -  {\bf\hat b} = \kh,&&
    \mbox{(helicity basis)}
\label{helbasis}, \\
     {\bf\hat a} = {\bf\hat b} = {\bf\hat p}, &&\mbox{(beam\
      basis)}
\label{beambasis},\\
     {\bf\hat a} =  {\bf\hat b} = \dhh,&&
    \mbox{(off-diagonal\ basis)}.
\label{offbasis}
\end{eqnarray}   
Here $\kh $ denotes the direction of
flight of the top quark in the  $t\bar{t}$-ZMF
and ${\bf\hat p}$  is the direction of one of the colliding
hadrons in that frame.
 The unit vector $\dhh$, which we use in the
convention of \cite{Bernreuther:2004jv}, defines the so-called
off-diagonal basis \cite{Mahlon:1997uc}. 
 The correlations ${\cal O}_{ab}$ (${\cal O}_{\rm
  spin}$) are equivalent to (a sum of) double spin asymmetries.
 At the Tevatron,
 the strength of the
  spin correlations induced by the SM interactions 
 is largest with respect to the beam and off-diagonal
 basis.
 For the LHC the helicity correlation and
   ${\cal O}_{\rm spin}$ are good choices\footnote{For $gg\to t {\bar t}$
   an optimal axis was constructed in 
   \cite{Uwer:2004vp}. It was reconsidered
   recently in \cite{Mahlon:2010gw}. Cf. the next section.}.

We consider  the variation 
  of these correlations  with $\mtt$,  which we define by
\begin{equation}
\frac{d{\cal O}}{d{\mtt}} = \frac{1}{\sigma} \int 
  d\sigma  \, {\cal{\mathbf O}} \; \delta (\sqrt{(p_t+p_{\bar t})^2} -\mtt
      )\, .
\label{diffgenob}
\end{equation}
 We normalize  to the integrated
 cross section. 

\begin{figure}
\begin{center}
\includegraphics[width=8cm]{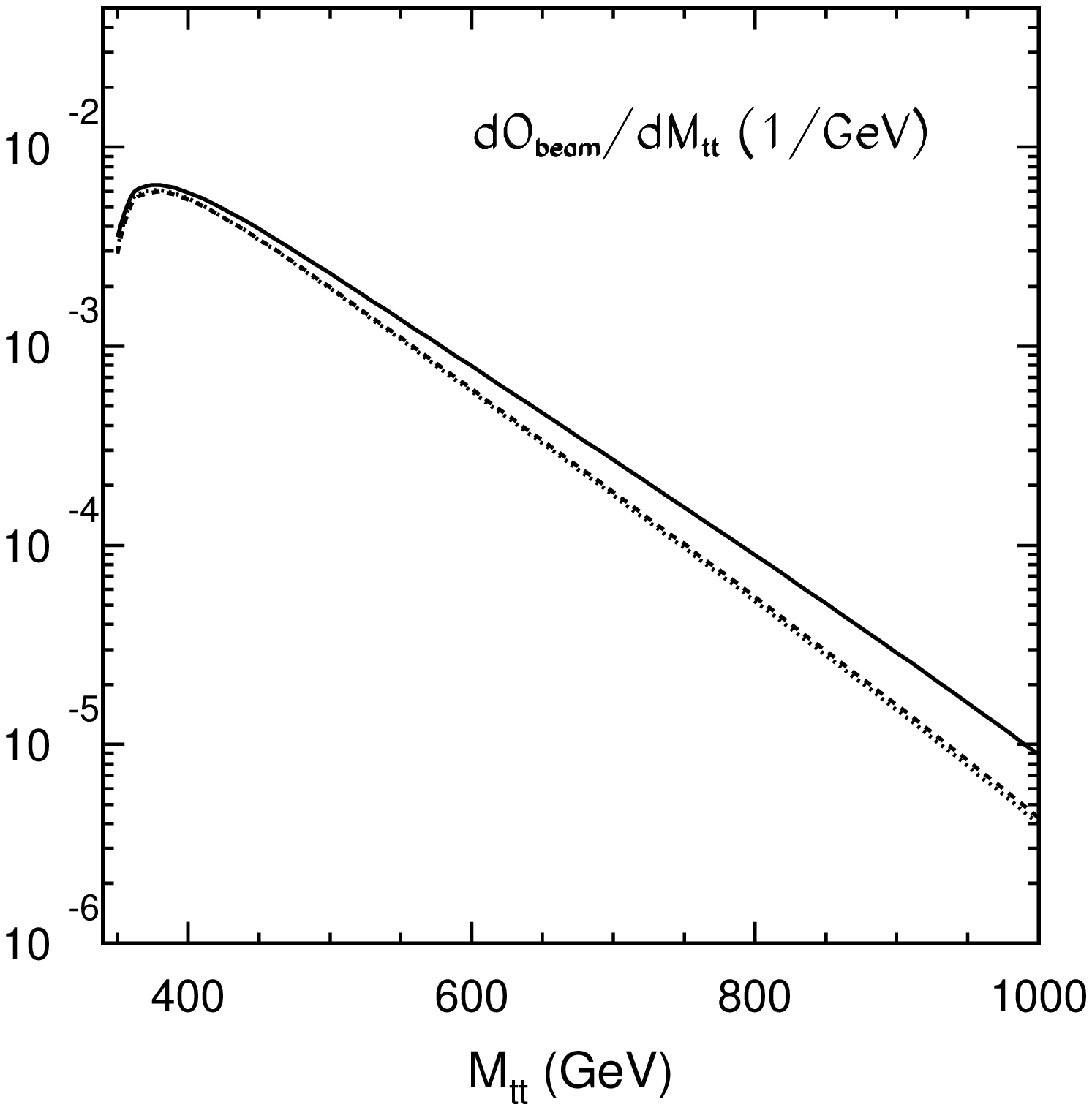}
\includegraphics[width=8cm]{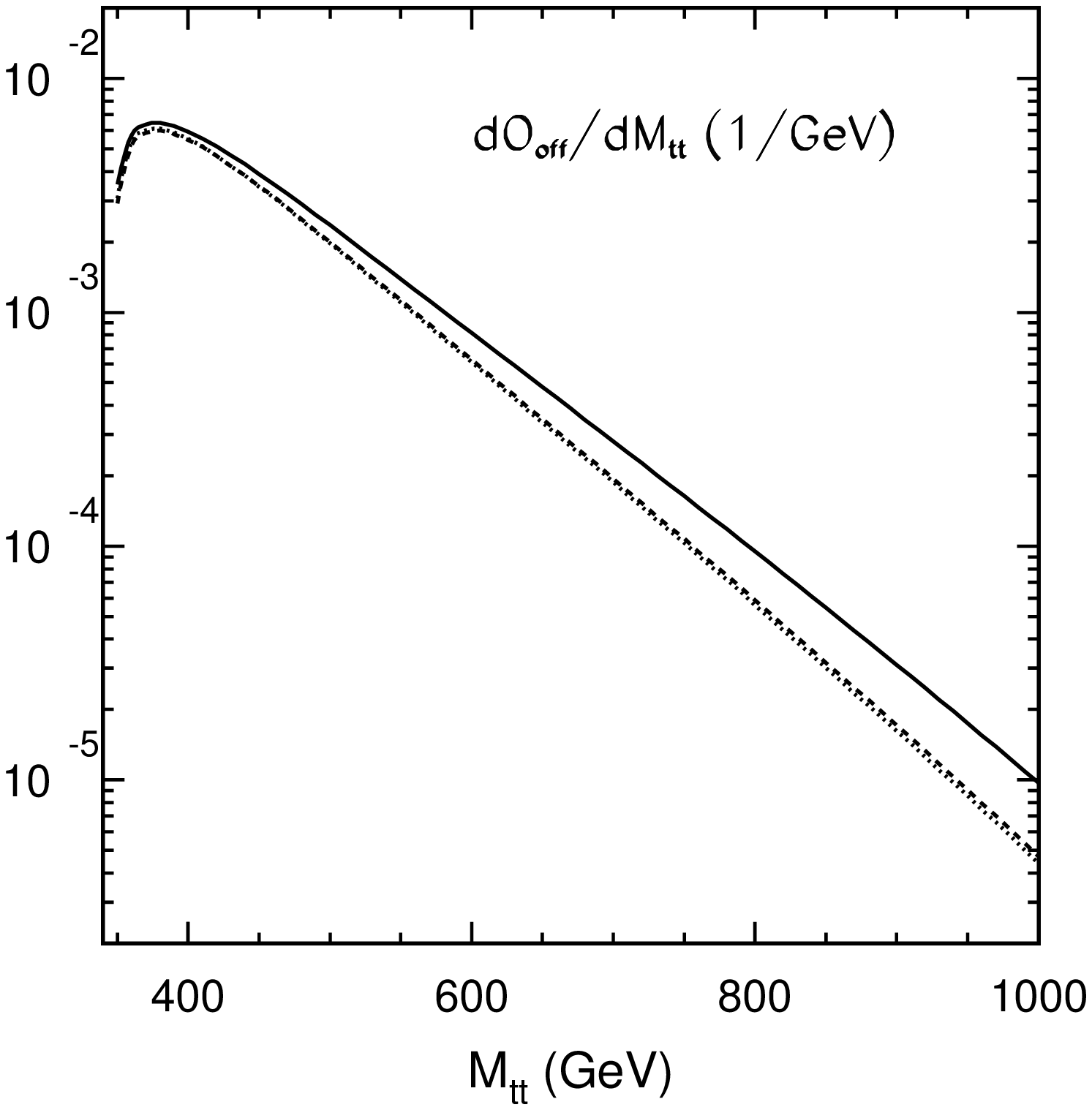}
\end{center}
\caption{The distributions $d{\cal
   O}_{\rm beam}/d\mtt$ (left) and $d{\cal
   O}_{\rm off}/d\mtt$ (right)  for the Tevatron 
   at LO (solid), NLO (dashed), and NLOW (dotted) for   $\mu =m_t$.}
 \label{fig:ttspintevcom}
\end{figure}

\begin{figure}
\begin{center}
\includegraphics[width=8cm]{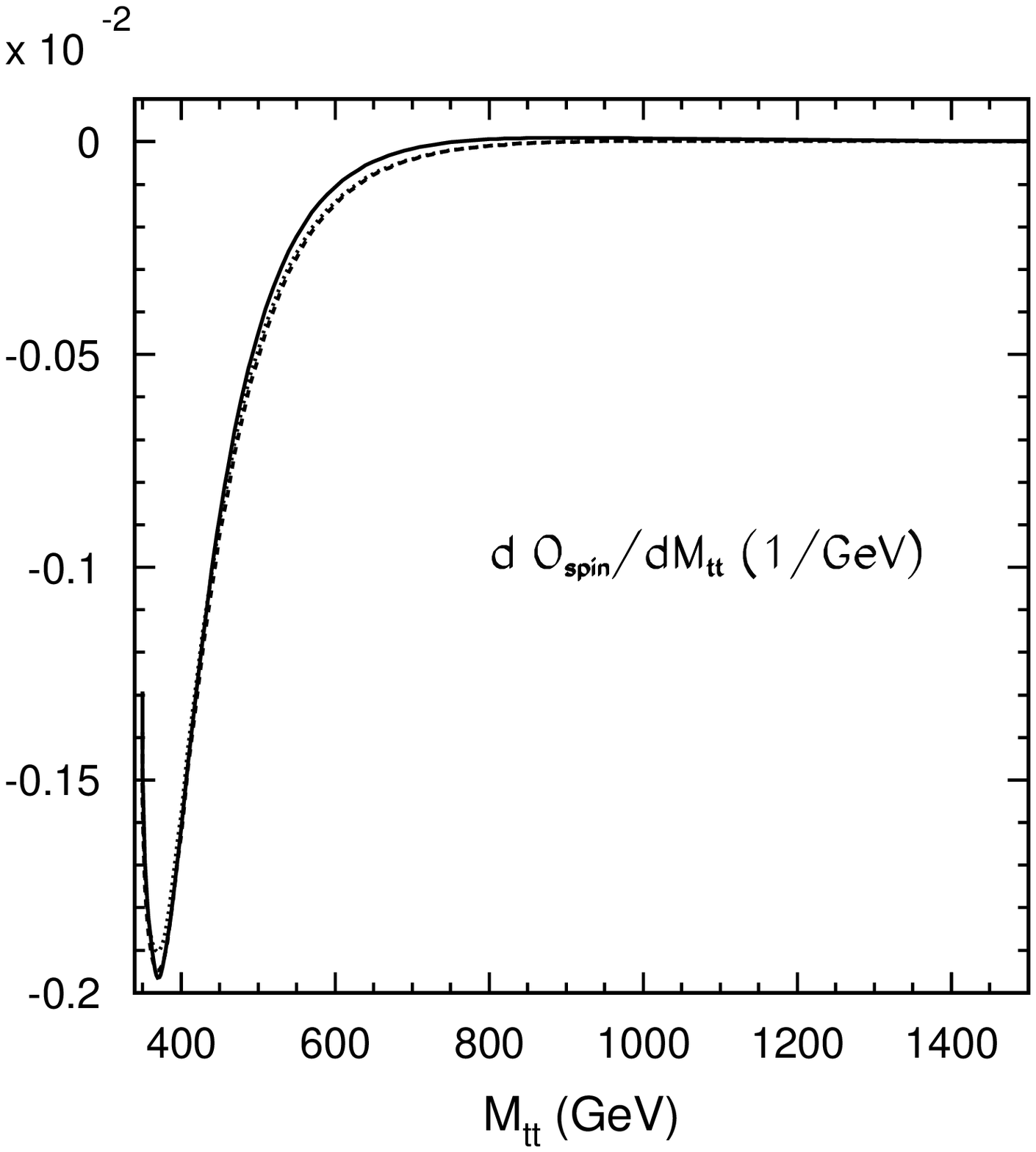}
\includegraphics[width=8cm]{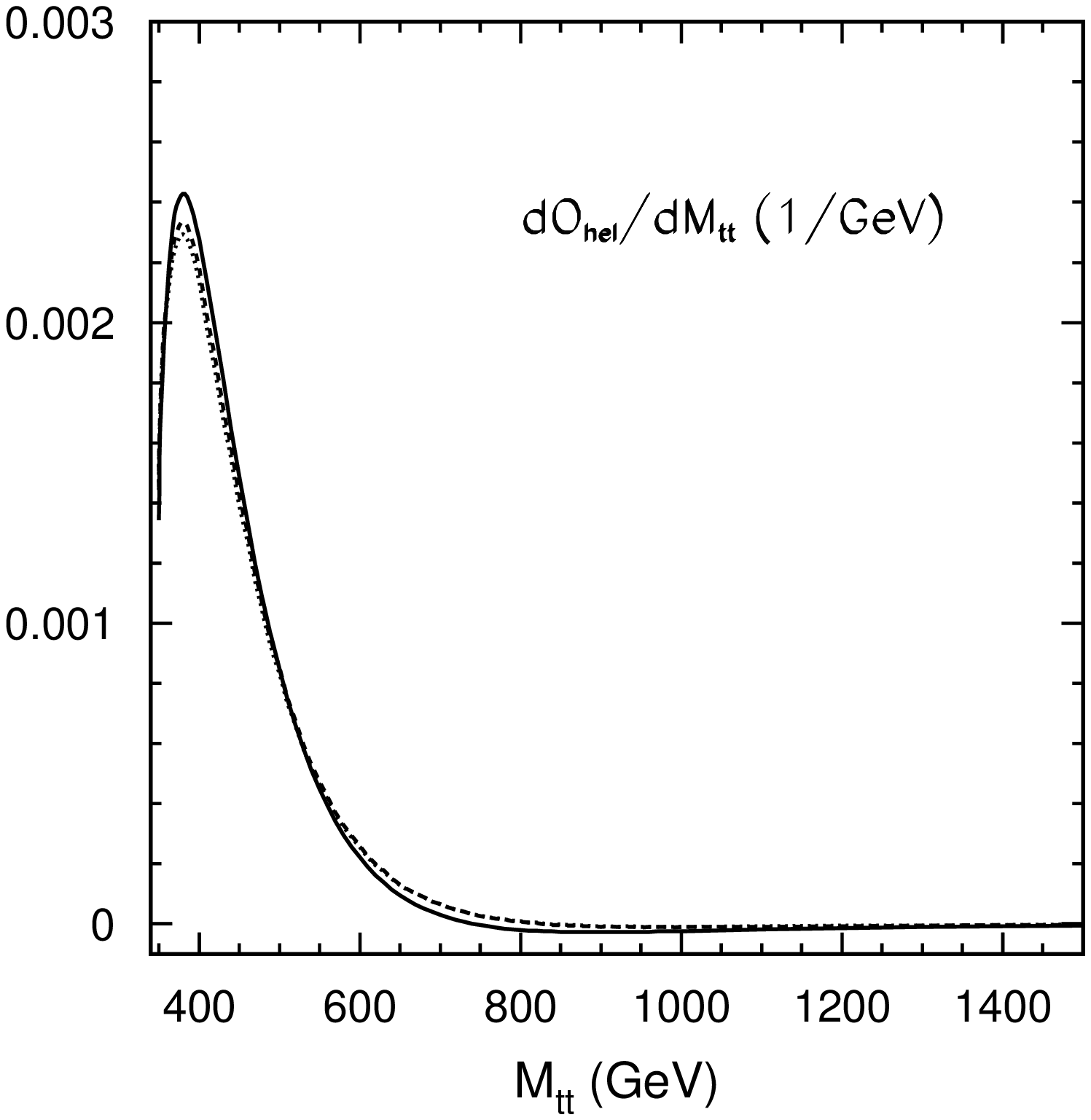}
\end{center}
\caption{The distributions $d{\cal
   O}_{\rm spin}/d\mtt$ (left) and $d{\cal
   O}_{\rm hel}/d\mtt$ (right)  for the LHC ($\sqrt{s}=14$ TeV)
   at LO (solid), NLO (dashed), and NLOW (dotted) for   $\mu =m_t$.}
 \label{fig:ttspinlhccom}
\end{figure}

\begin{figure}
\begin{center}
\includegraphics[width=8cm]{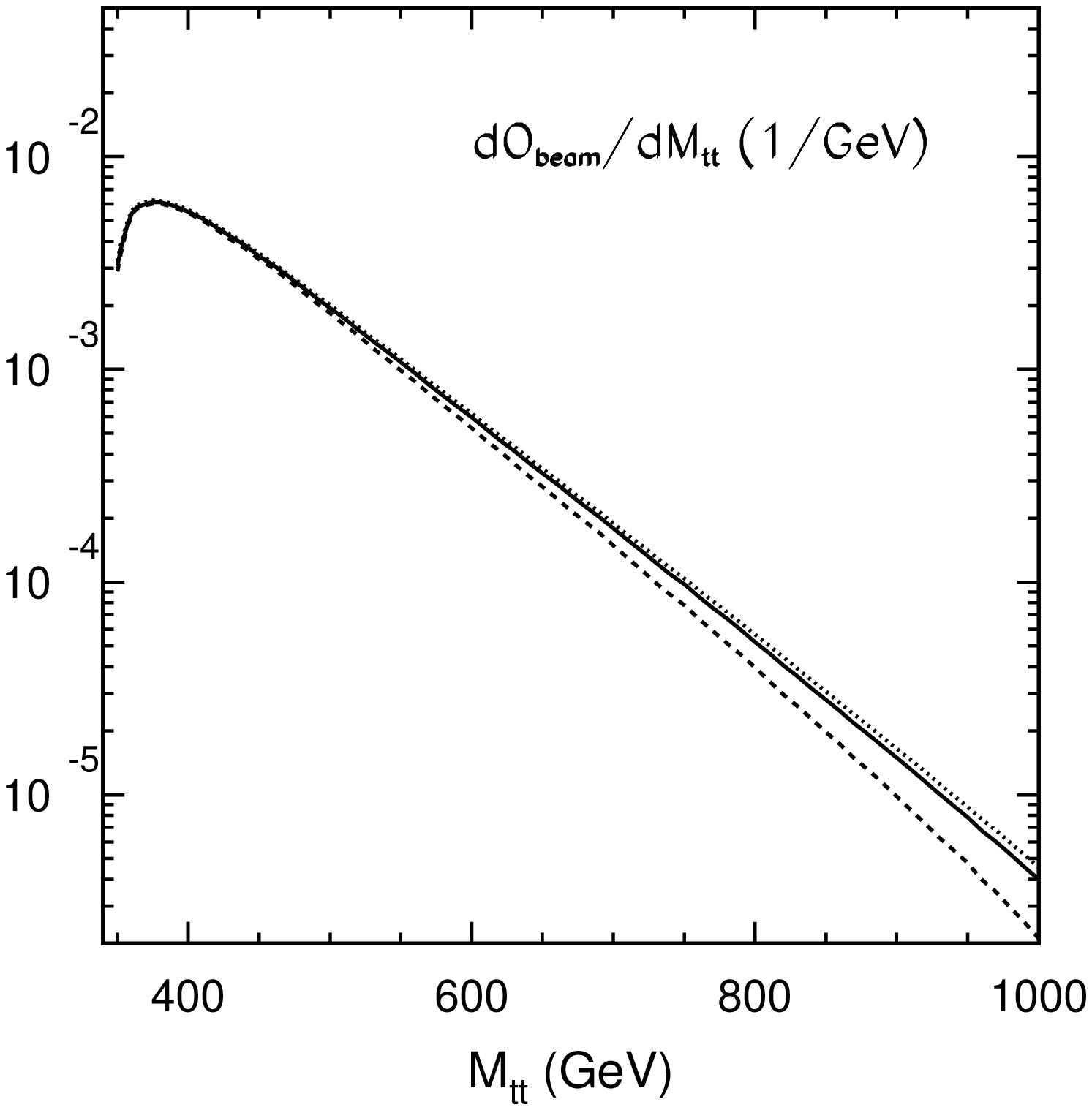}
\includegraphics[width=8cm]{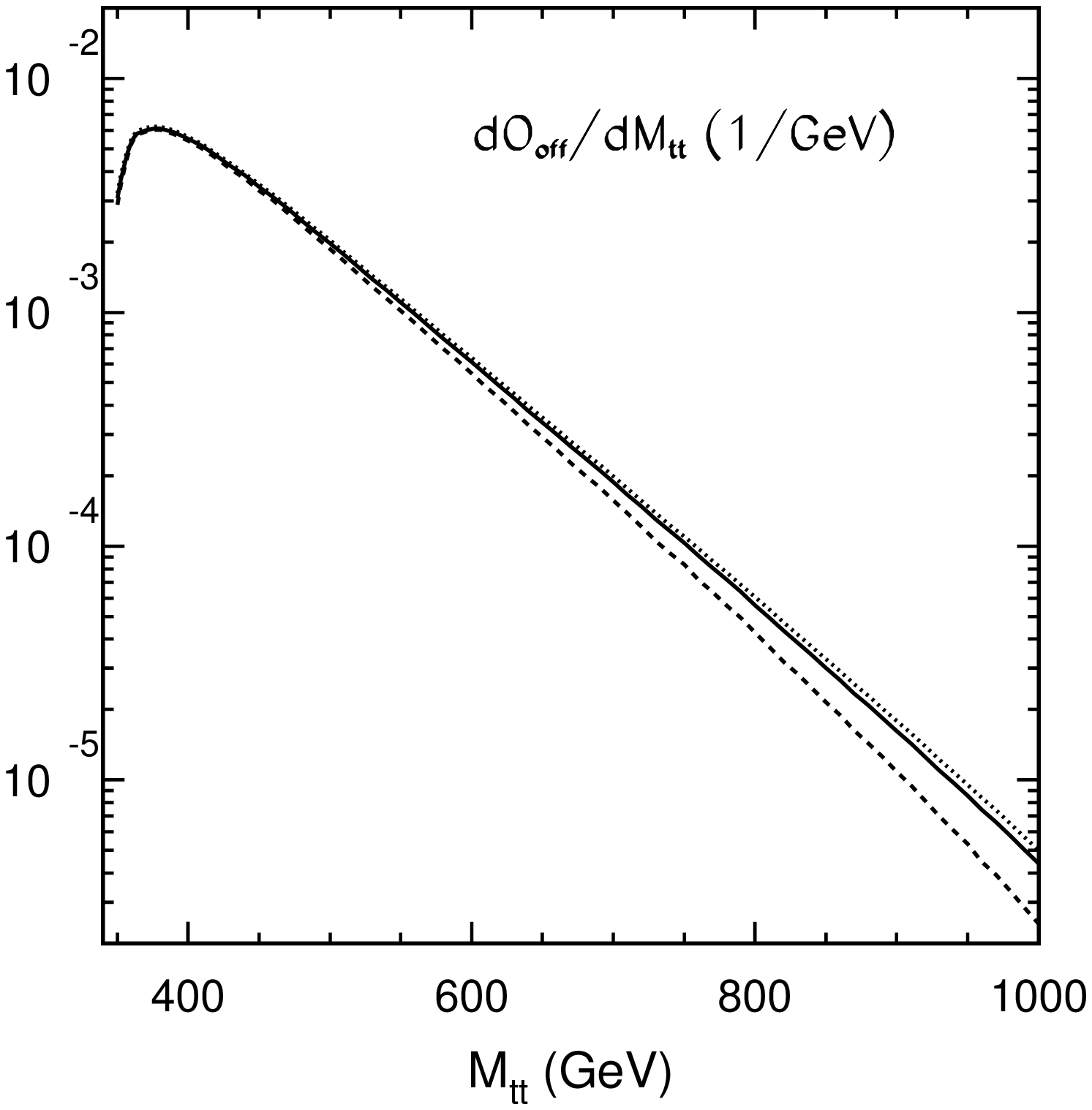}
\end{center}
\caption{The distributions $d{\cal
   O}_{\rm beam}/d\mtt$    (left) and  $d{\cal
   O}_{\rm off}/d\mtt$
 (right)  for the Tevatron
   at NLOW for $\mu =m_t/2$ (dashed), $\mu =m_t$ (solid),
   and $\mu=2 m_t$ (dotted).}
 \label{fig:ttspintevfull}
\end{figure}

\begin{figure}
\begin{center}
\includegraphics[width=8cm]{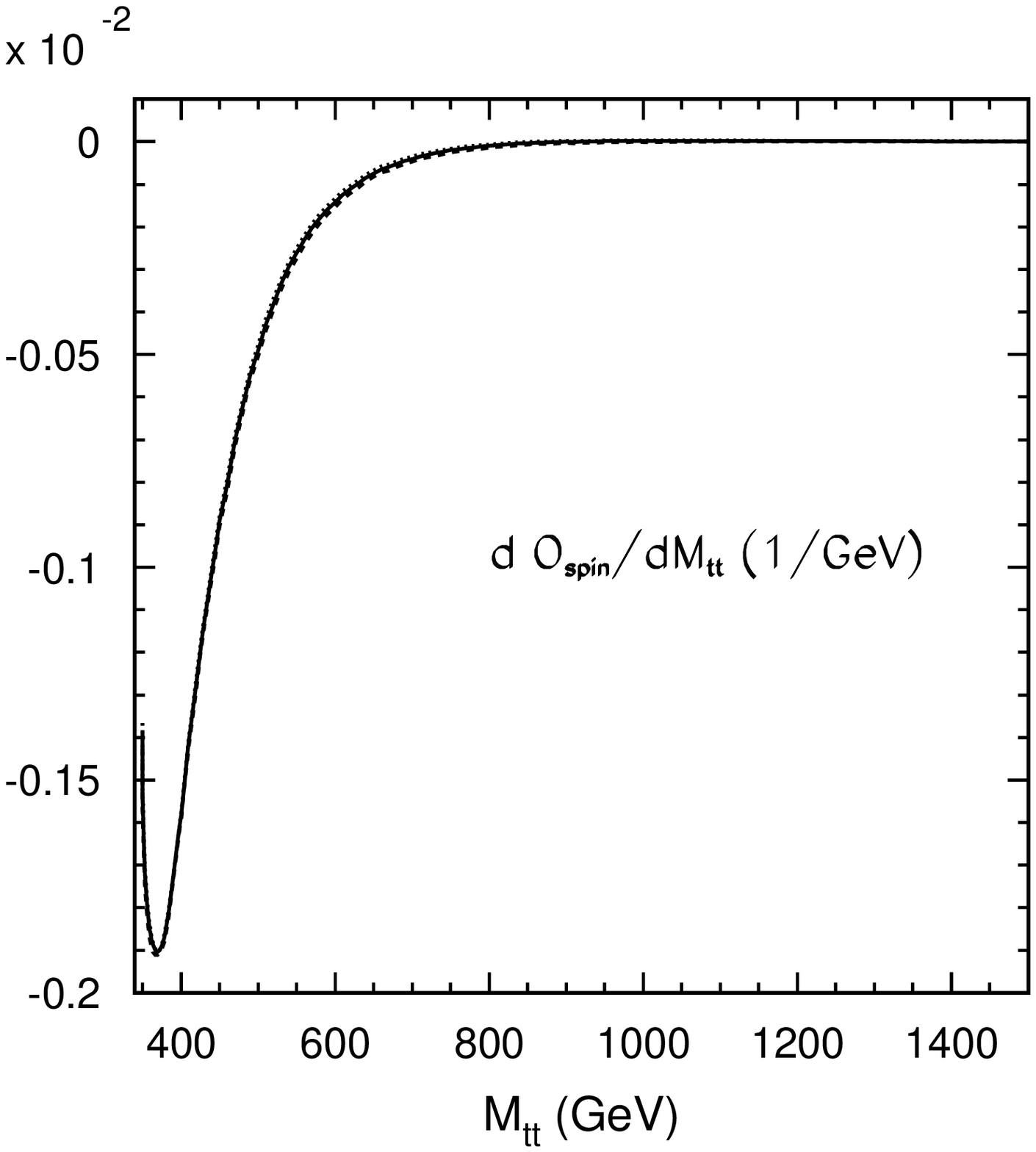}
\includegraphics[width=8cm]{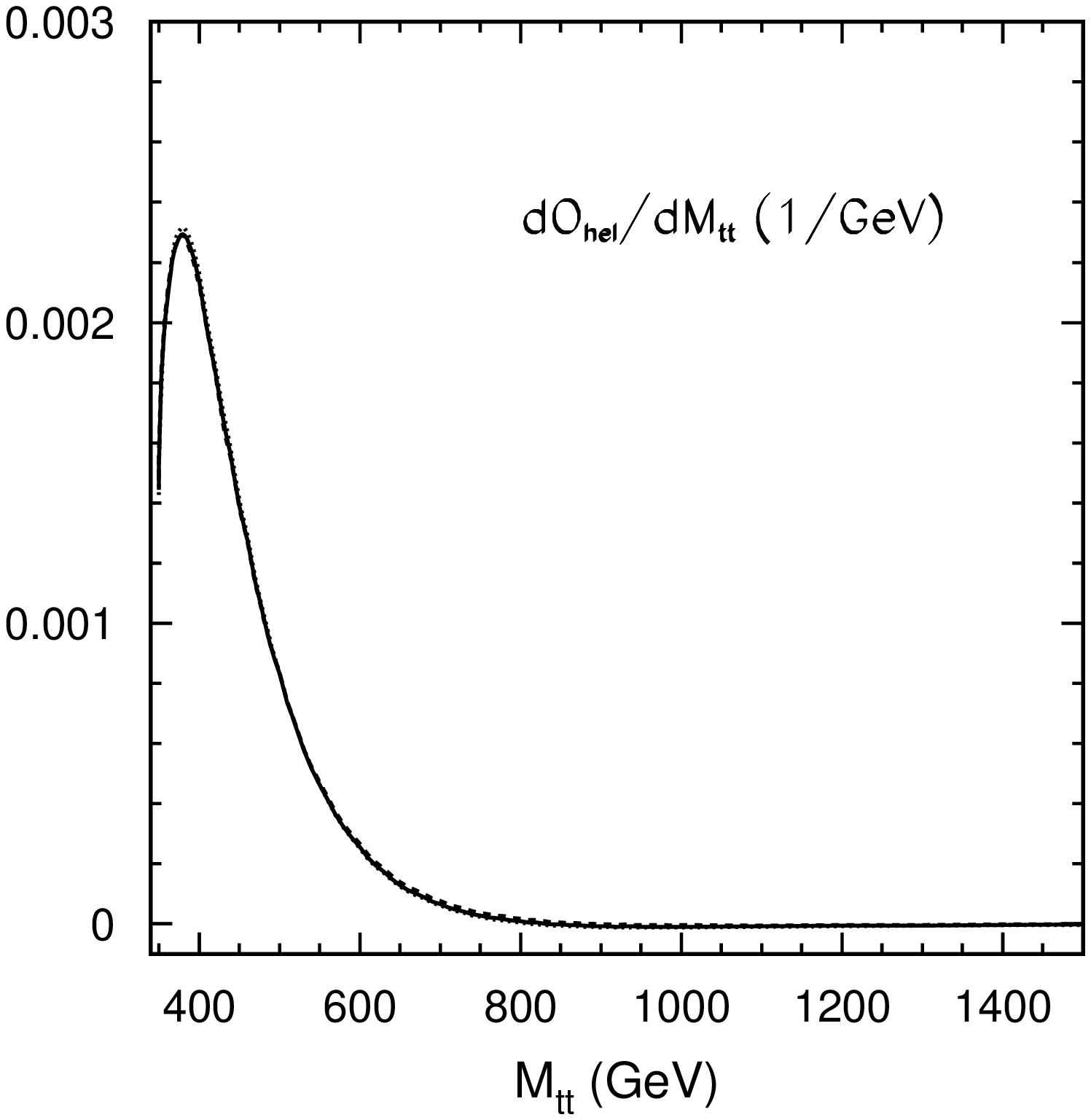}
\end{center}
\caption{The distributions $d{\cal
   O}_{\rm spin}/d\mtt$    (left) and  $d{\cal
   O}_{\rm hel}/d\mtt$
 (right)  for the LHC ($\sqrt{s}=14$ TeV)
   at NLOW for $\mu =m_t/2$ (dashed), $\mu =m_t$ (solid),
   and $\mu=2 m_t$ (dotted).}
 \label{fig:ttspinlhcfull}
\end{figure}

The distributions are displayed
   in Fig.~\ref{fig:ttspintevcom} to LO, NLO, and NLOW for the beam and
 off-diagonal basis  at the Tevatron.
   For the helicity correlation and for  ${\cal O}_{\rm spin}$
  the analogous plots  are given in Fig.~\ref{fig:ttspinlhccom}
  for the  LHC (14  TeV). The latter
  distributions pass through zero\footnote{Their detailed behaviour can
    be read off from plots of scaling functions given
    in\cite{Bernreuther:2004jv}.}  near $\mtt \sim 800$ GeV. 
   These figures  show  the known feature
     that  the $\tbart$ 
  spin correlations (\ref{genob}) at the Tevatron and LHC 
  receive  most of their contributions from $\tbart$ pairs with rather
  low $\mtt$. In Figs.~\ref{fig:ttspintevfull} and
  \ref{fig:ttspinlhcfull} the respective distributions are given
   for  $\mu = m_t/2,\ m_t,$ and $2 m_t$. 

\begin{figure}
\begin{center}
\includegraphics[width=8cm]{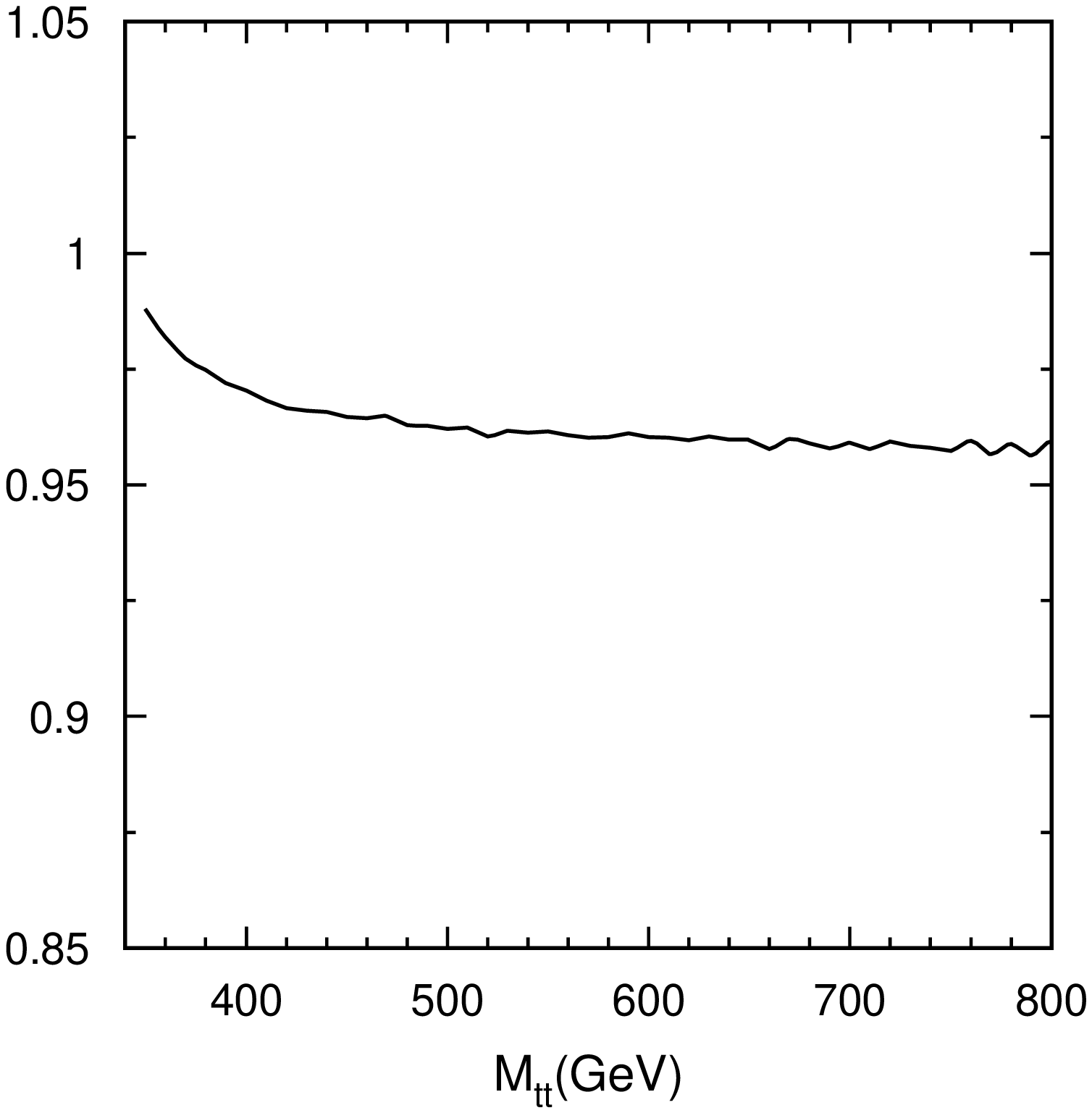}
\includegraphics[width=8cm]{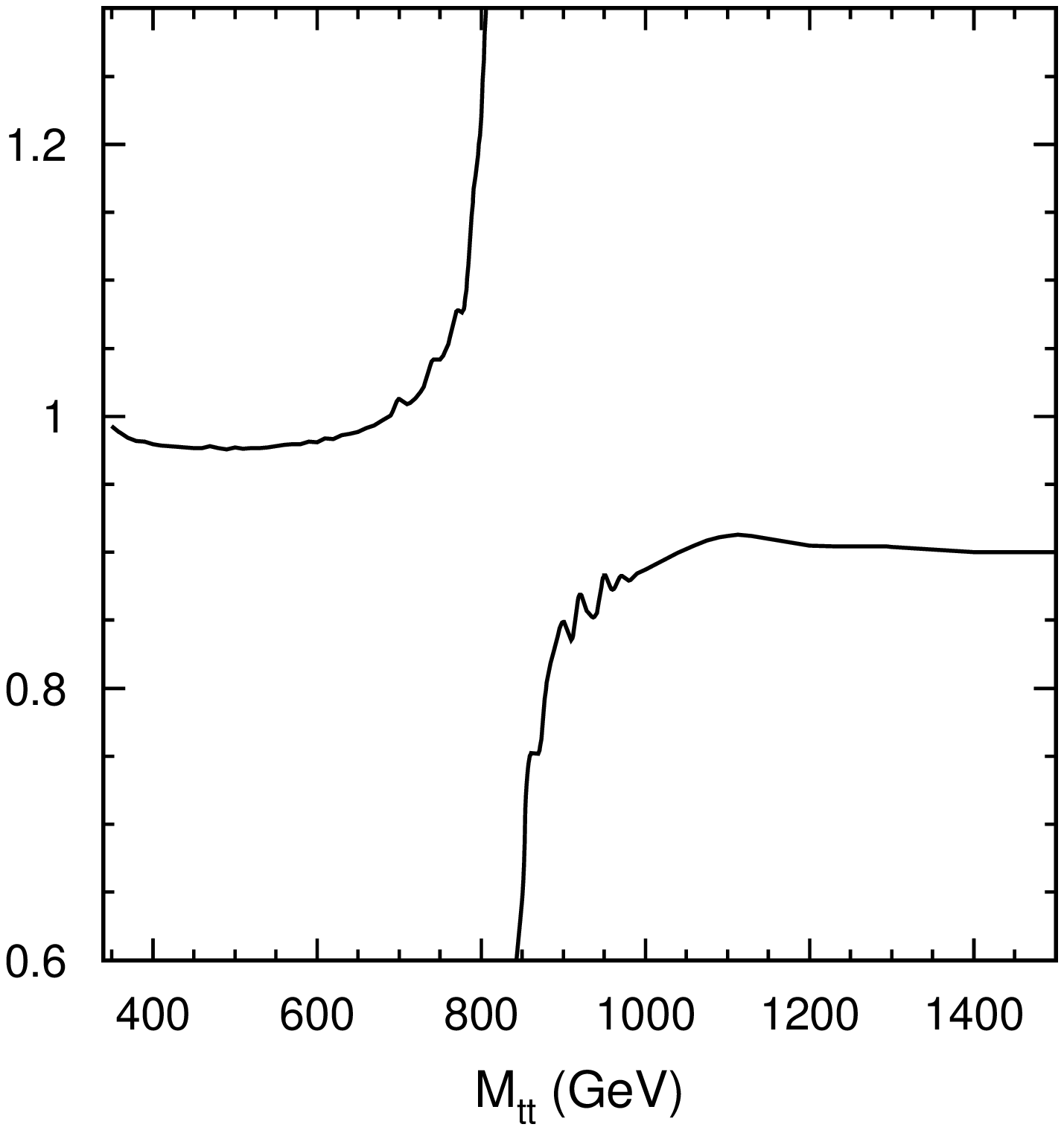}
\end{center}
\caption{The ratios 
   of $d{\cal O}/d{\mtt}$ evaluated at NLOW and NLO with $\mu=m_t$
   for ${\cal O}_{\rm spin}$  (left panel) 
 and   ${\cal O}_{\rm hel}$ (right panel) for the LHC (14 TeV).}
\label{fig:ratiiosplhcfull}
\end{figure}

In order to quantify the size of the weak corrections
   we have plotted in Fig.~\ref{fig:ratiiosplhcfull} the ratios 
   of $d{\cal O}/d{\mtt}$ evaluated at NLOW and NLO
   for ${\cal O}_{\rm spin}$ and   ${\cal O}_{\rm hel}$ at the 
 LHC (14 TeV). This figure shows that the weak-interaction corrections
   to ${\cal O}_{\rm spin}$ are negative and $\lesssim 4\%$ with
   respect to NLO QCD for  $\mtt < 800$ GeV,  while these corrections
   are smaller for   ${\cal O}_{\rm hel}$. In the latter case, the increase of the
   magnitude of the ratio in the vicinity of 
   $\mtt \sim 800$ GeV is just a consequence of  the NLO correlation
   passing through zero  and changing sign afterwards.
  
In Table~\ref{TabTevLHCsptt} the values of these variables, integrated
   over the $\mtt$ spectrum, are given at NLOW for three scales.
   The spin correlations will be
    analyzed further  for dileptonic final states in the next section.
  There we will also compare the numbers of Table~\ref{TabTevLHCsptt}
  with the results of \cite{Bernreuther:2004jv}.


\begin{table}[hp]
\begin{center}
\caption{Results for several spin correlation
  variables  at NLOW at the level of $\tbart$
  final  states (no cuts) for  the Tevatron and LHC. At the LHC the
  SM-induced $\tbart$ spin correlations are very small with respect to the beam
  and off-diagonal bases and are therefore not given. Their SM values
   were given in \cite{Bernreuther:2001rq,Bernreuther:2004jv}.}
\label{TabTevLHCsptt} 
\vspace*{0.5cm}
\begin{tabular}{||c|c|c|c||c|c|c||c|c|c||}\hline \hline
  &\multicolumn{3}{c|}{Tevatron} 
&\multicolumn{3}{c|}{LHC (10 TeV)} 
&\multicolumn{3}{c|}{LHC (14 TeV)} \\ \hline \hline
$\mu$ & $m_t/2$ & $m_t$ & $2m_t$ & $m_t/2$ & $m_t$ & $2m_t$ &$m_t/2$ & $m$ & $2m$ \\ \hline \hline
${\cal O}_{\rm beam}$  &0.777 &0.791 &0.804 &  &  &  &  &  & \\ \hline
 ${\cal O}_{\rm off}$  &0.783 &0.798 &0.810 &  &  &  &  &  &  \\ \hline \hline
${\cal O}_{\rm spin}$  &0.216 &0.218 &0.220 &-0.236&-0.233&-0.229 &-0.239 &-0.236 &-0.234  \\ \hline
 ${\cal O}_{\rm hel}$  &-0.358 &-0.368 &-0.376 &0.326 &0.326 &0.325 &0.327 &0.328 &0.331  \\ \hline \hline
\end{tabular}
\end{center}
\end{table}

%

%

%

%

%


\newpage

\section{Results for dilepton and semileptonic
   final states}
\label{sec:diljet}

In this section we give results for  several distributions,  correlations, and
 asymmetries  for 
 dileptonic  final states (\ref{eq:ttlj1}), and for a few 
 observables for  semileptonic final states (\ref{eq:ttll}),
   at NLO in the strong and weak couplings (NLOW). The
 radiative corrections were implemented in our computer code as
 described in Section~\ref{sec:setup}. 

We use the following cuts for  the dileptonic final states ($\ell=e,\mu$,
$\not{E}_T$ denotes the missing transverse energy, and $\eta$ is the
 pseudorapidity):
\begin{eqnarray}
{\rm Tevatron:} \qquad p_T^{\ell}\geq 20 \, {\rm GeV},  \quad |\eta_{\ell}| \leq 2.0, \quad
p_T^{j}\geq 20  \,  {\rm GeV},  \quad |\eta_{j}| \leq 2.0, \quad
\not{E}_T\geq 25  \, {\rm GeV}, \nonumber \\
{\rm LHC:} \qquad p_T^{\ell }\geq 20\, {\rm GeV}, \quad
|\eta_{\ell}|\leq 2.5, \quad
p_T^{j}\geq 20\, {\rm GeV},  \quad   |\eta_{j}|\leq 2.4, \quad
 \not{E}_T\geq 40 \, {\rm GeV} . \label{cut:dilep}
\end{eqnarray}
For the semileptonic $\tbart$  final states  we use
\begin{eqnarray}
{\rm Tevatron:} \qquad p_T^{\ell}\geq 20 \, {\rm GeV},  \quad |\eta_{\ell}| \leq 2.0, \quad
p_T^{j}\geq 20  \,  {\rm GeV},  \quad |\eta_{j}| \leq 2.0, \quad
\not{E}_T\geq 20  \, {\rm GeV}, \nonumber \\
{\rm LHC:} \qquad p_T^{\ell }\geq 20\, {\rm GeV}, \quad
|\eta_{\ell}|\leq 2.5, \quad
p_T^{j}\geq 20\, {\rm GeV},  \quad   |\eta_{j}|\leq 2.4, \quad
 \not{E}_T\geq 20 \, {\rm GeV} . \label{cutseml}
\end{eqnarray}
We put again $\mu_R=\mu_F=\mu$ and evaluate the distributions for 
 $\mu = m_t/2, \, m_t,$ and $2 m_t$. We calculated the  observables
 below inclusively.  We
 checked for some dilepton final-state distributions that the results do not change when using 
 instead the $k_\perp$  jet algorithm \cite{Catani:1992zp}. This 
 is to say that we checked an inclusive calculation
  against $\ell^+ \nu_\ell  \ell^- {\bar \nu}_\ell \, j_b j_{\bar b}$ (NLO)
   and  $\ell^+ \nu_\ell \ell^- {\bar \nu}_\ell \, j_b j_{\bar b} j$
   (LO), where $j$ denotes a gluon  or light quark jet.

\begin{figure}
\begin{center}
\includegraphics[width=8cm]{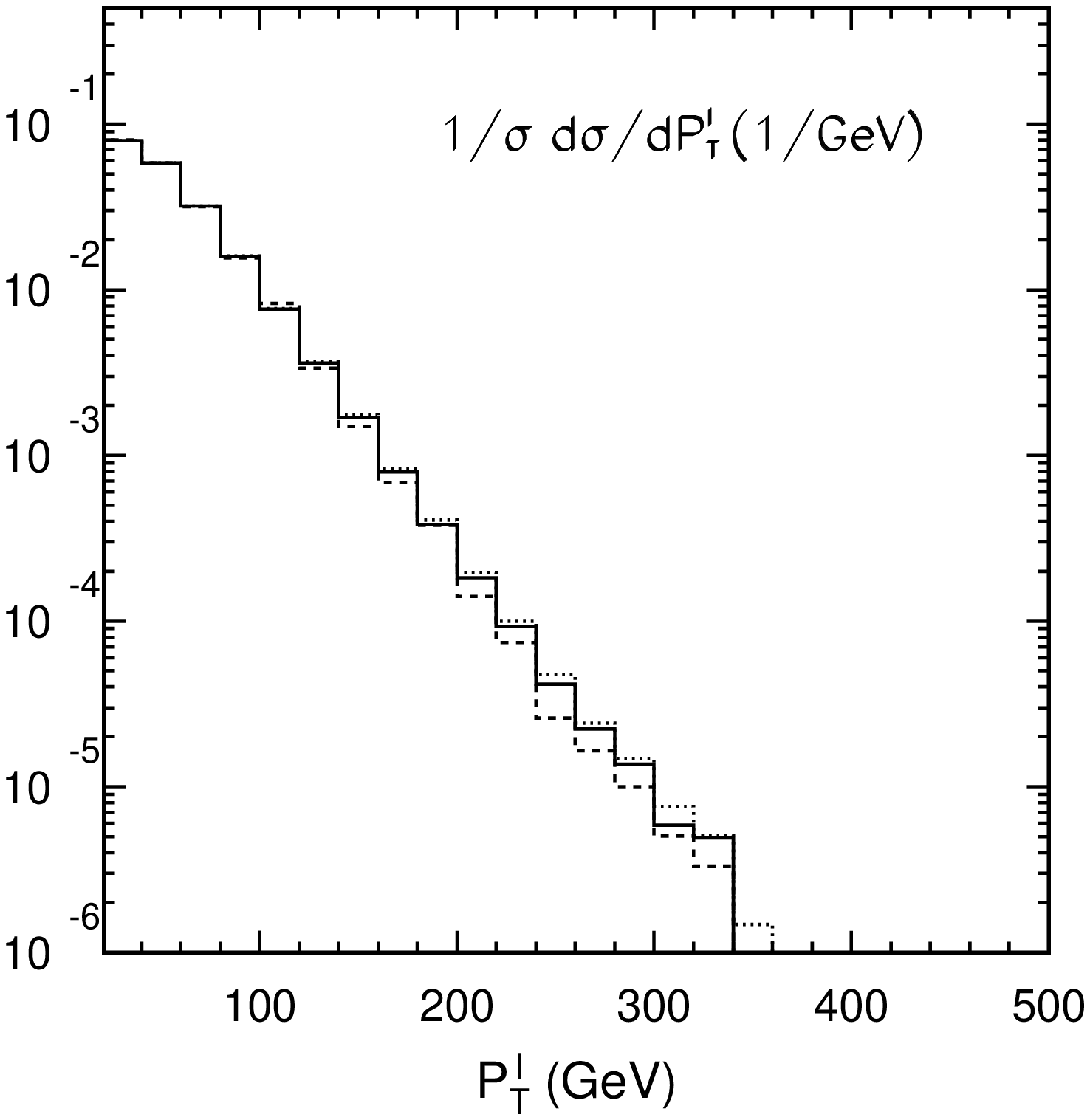}
\includegraphics[width=8cm]{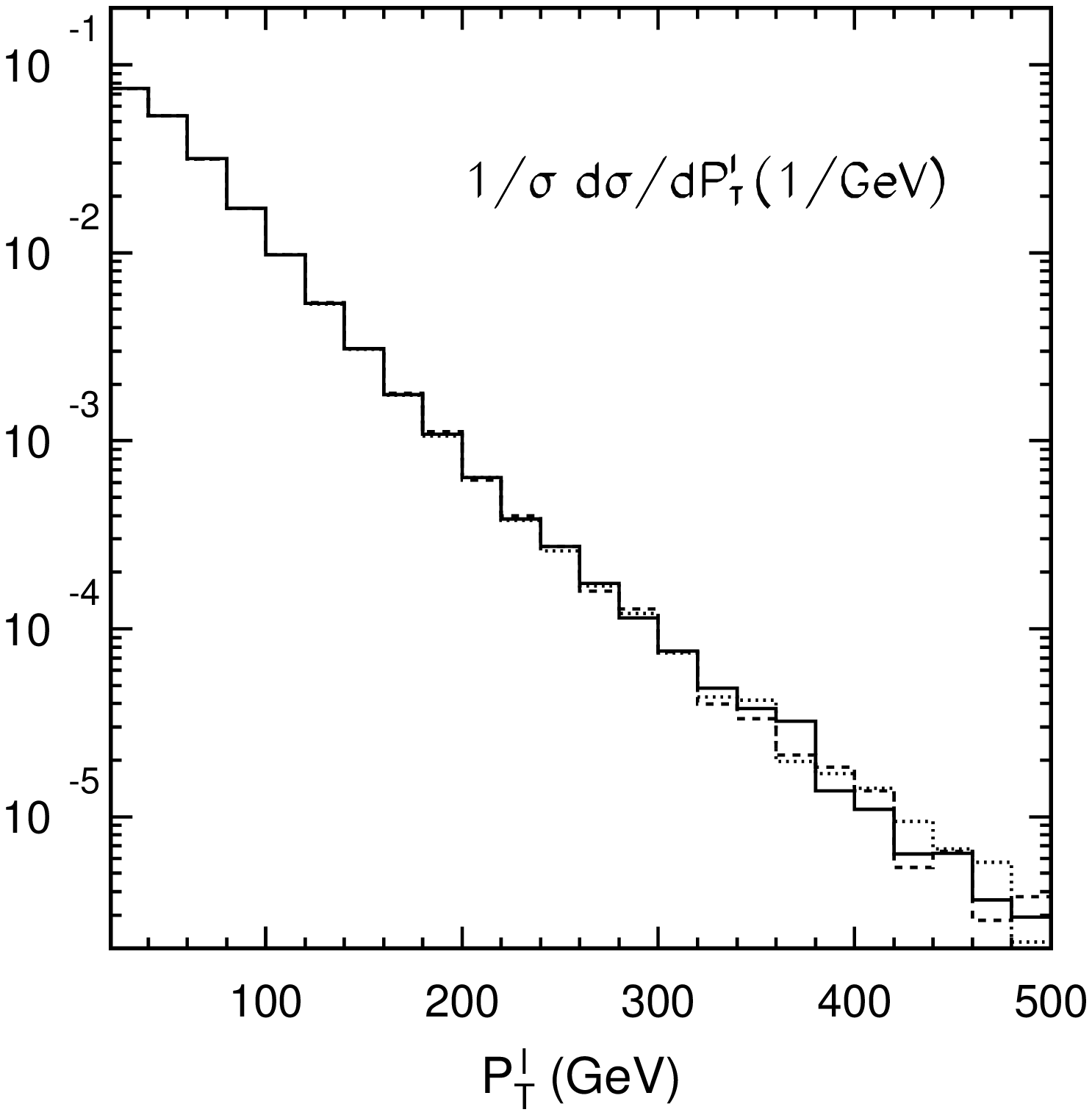}
\end{center}
\caption{The normalized transverse momentum distribution of $\ell^+$ in the 
 $\ell \ell$ channels at NLOW, for the Tevatron (left) and the
  LHC at $\sqrt{s}=10$ TeV (right). The solid, dashed, and
  dotted lines correspond to $\mu =m_t, m_t/2, 2m_t$, respectively.}
 \label{fig:ptlept} 
\end{figure}

\begin{figure}
\begin{center}
\includegraphics[width=8cm]{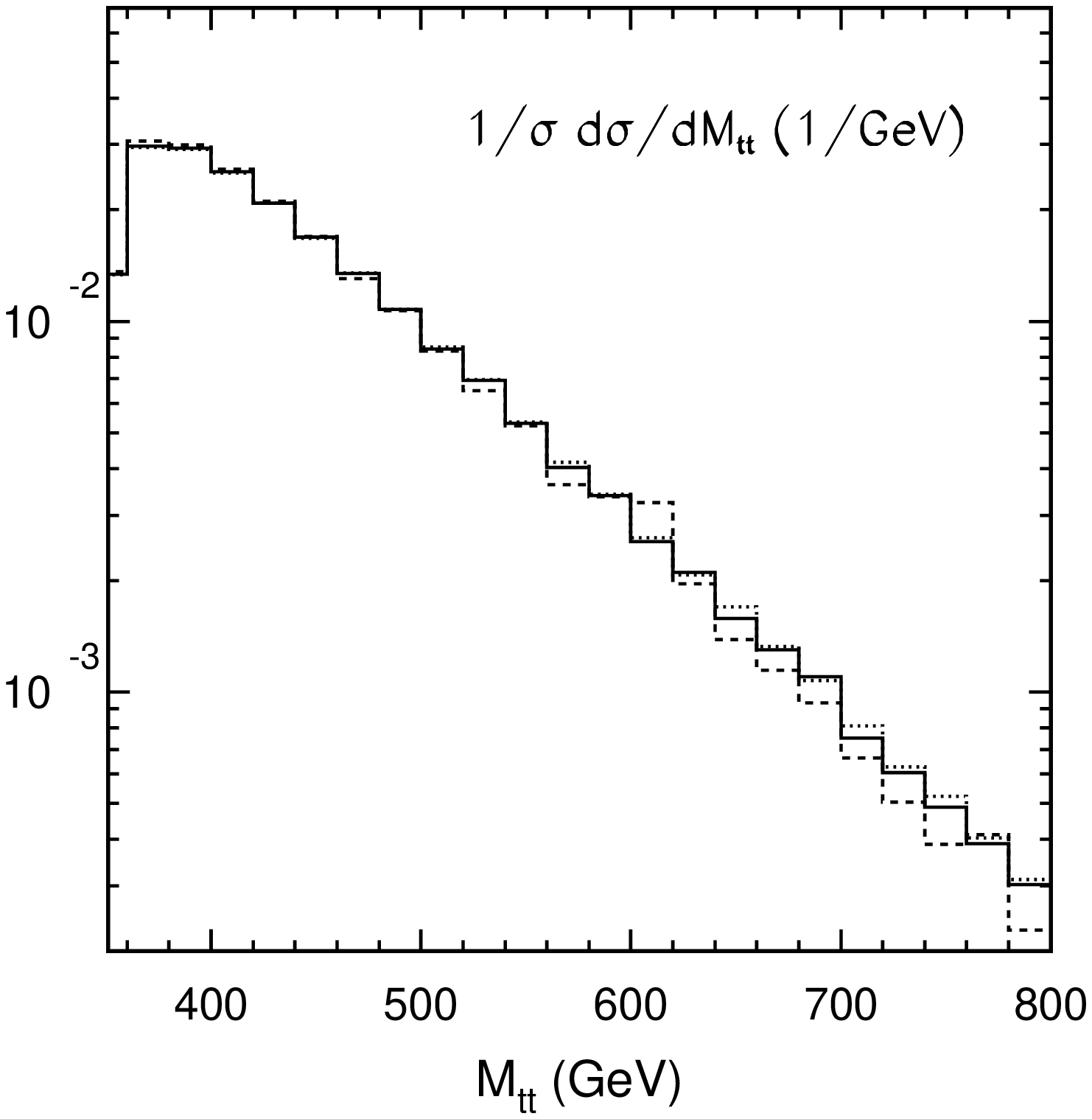}
\includegraphics[width=8cm]{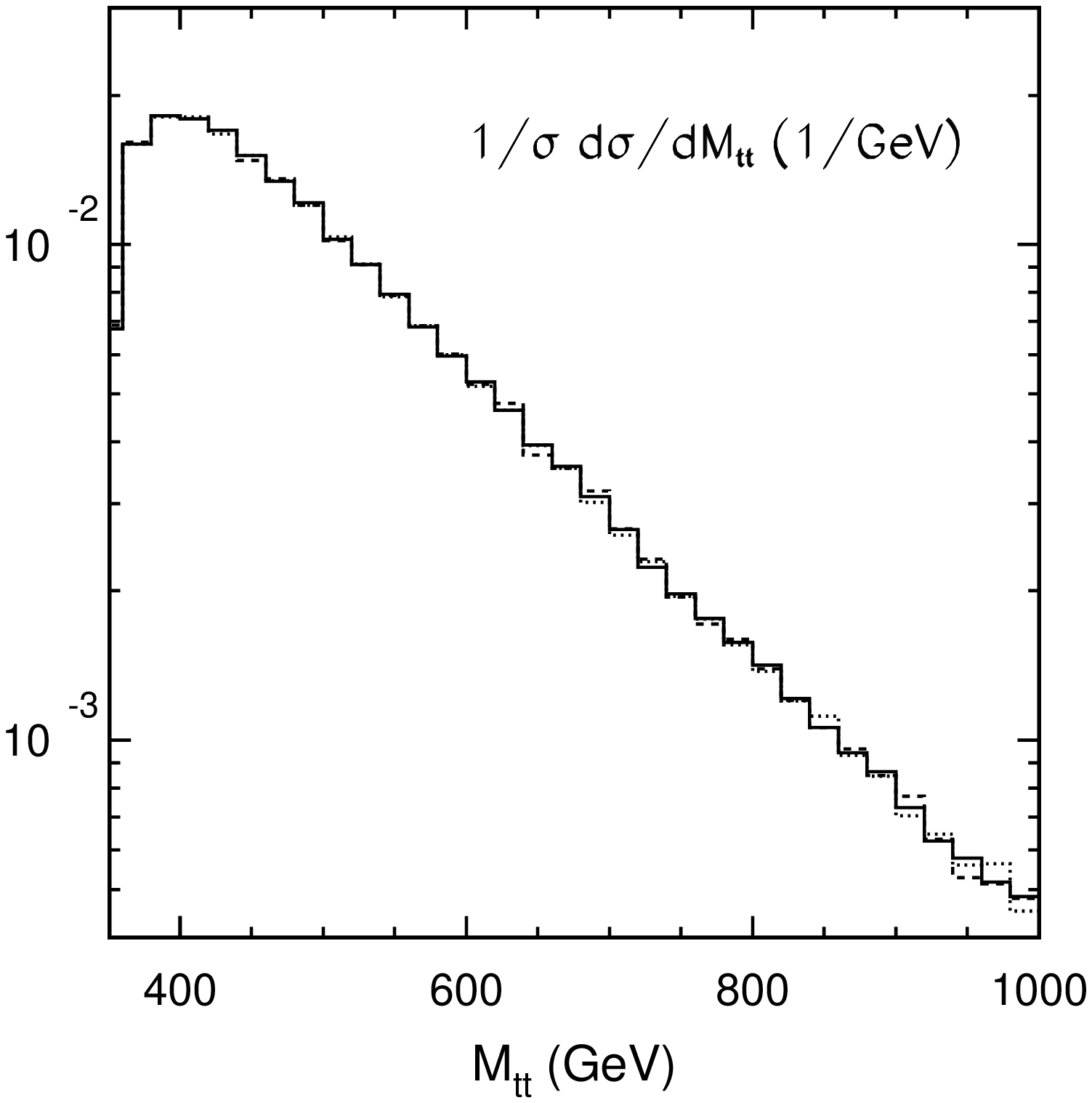}
\end{center}
\caption{The normalized $\tbart$ invariant mass distribution 
  for semileptonic final states  at NLOW, for the Tevatron (left) and the
  LHC at $\sqrt{s}=10$ TeV (right). The solid, dashed, and
  dotted lines correspond to $\mu =m_t, m_t/2, 2m_t$, respectively.}
 \label{fig:mttljet}
\end{figure}

Fig.~\ref{fig:ptlept} shows, for the dilepton channels
   at the Tevatron and LHC (10 TeV),
 the $p^\ell_T$ distribution of the charged
lepton $\ell^+$ for three scales $\mu$. We have normalized to the
  respective cross section $\sigma(\tbart \to \ell^+\ell'^- \, X)$ in order 
   to reduce the scale uncertainties. Nevertheless, as is
   well-known, the scale
   variations become significant at large $p^\ell_T$.
 Fig.~\ref{fig:mttljet} shows the analogous plots for the normalized
$\tbart$ invariant mass distribution for the semileptonic
 final states. The only purpose of showing these plots is to
 demonstrate that our code produces also established fixed order NLO QCD results. 
 The $\mtt$ distribution  is, especially for the semileptonic channels,
   an important observable in the  search for
  resonances that (strongly) couple to $\tbart$ -- for investigations
   within various models, cf., for 
instance, \cite{Bernreuther:1997gs,Dicus:1994bm,Barger:2006hm,Agashe:2007ki,Baur:2008uv,Frederix:2007gi}.

%

%

\subsection{Charge  asymmetries}
\label{sub:chaasy}


\begin{table}[]
\caption{Results 
 for the charge asymmetry $A$ and the  pair asymmetry $A^{\tbart}$ of top quarks,
 and the  leptonic  charge asymmetry  $A^\ell$ and the pair asymmetry  $A^{\ell\ell}$,
 for  dileptonic final states at the Tevatron, with the
   cuts (\ref{cutseml}),  at NLO' and NLOW' for 
      $\mu =m_t/2, \, m_t$, and $2m_t$. The  acronyms NLO' and
    NLOW' are explained in Section~\ref{sec.rett}.
    The acronym {\it $\tbart$ uncorrelated} refers
   to the case when all $t$- and $\bar t$-spin dependent 
 terms in the spin density matrices  are switched off at NLOW. }
\label{chasTEVdilep}
\begin{center}
\begin{tabular}{||c|c|c|c||c|c|c|}\hline \hline
  &\multicolumn{3}{c|}{Tevatron (($\tbart$ correlated))} &\multicolumn{3}{c|}{Tevatron ($\tbart$ uncorrelated))}\\ \hline \hline
$\mu$ & $m_t/2$ & $m_t$ & $2m_t$ & $m_t/2$ & $m_t$  &$2m_t$  \\ \hline \hline
$A$ (NLO')            &0.053 &0.048 &0.044 &0.053 &0.047 &0.043  \\ \hline 
$A$ (NLOW')           &0.054 &0.049 &0.046 &0.054 &0.049 &0.046  \\ \hline \hline
$A^{\tbart}$ (NLO')   &0.074 &0.068 &0.062 &0.075 &0.067 &0.061 \\ \hline 
$A^{\tbart}$ (NLOW')  &0.078 &0.071 &0.066 &0.077 &0.070 &0.065  \\ \hline \hline 
$A^{\ell}$ (NLO')     &0.038 &0.033 &0.031 &0.037 &0.033 &0.030 \\ \hline
$A^{\ell}$ (NLOW')    &0.039 &0.034 &0.032 &0.038 &0.035 &0.032 \\ \hline  
$A^{\ell\ell}$ (NLO') &0.047 &0.042 &0.038 &0.050 &0.045 &0.041  \\ \hline 
$A^{\ell\ell}$ (NLOW')&0.048 &0.044 &0.040 &0.052 &0.047 &0.043  \\ \hline \hline
\end{tabular}
\end{center}
\end{table}

\begin{figure}
\begin{center}
\includegraphics[width=8cm]{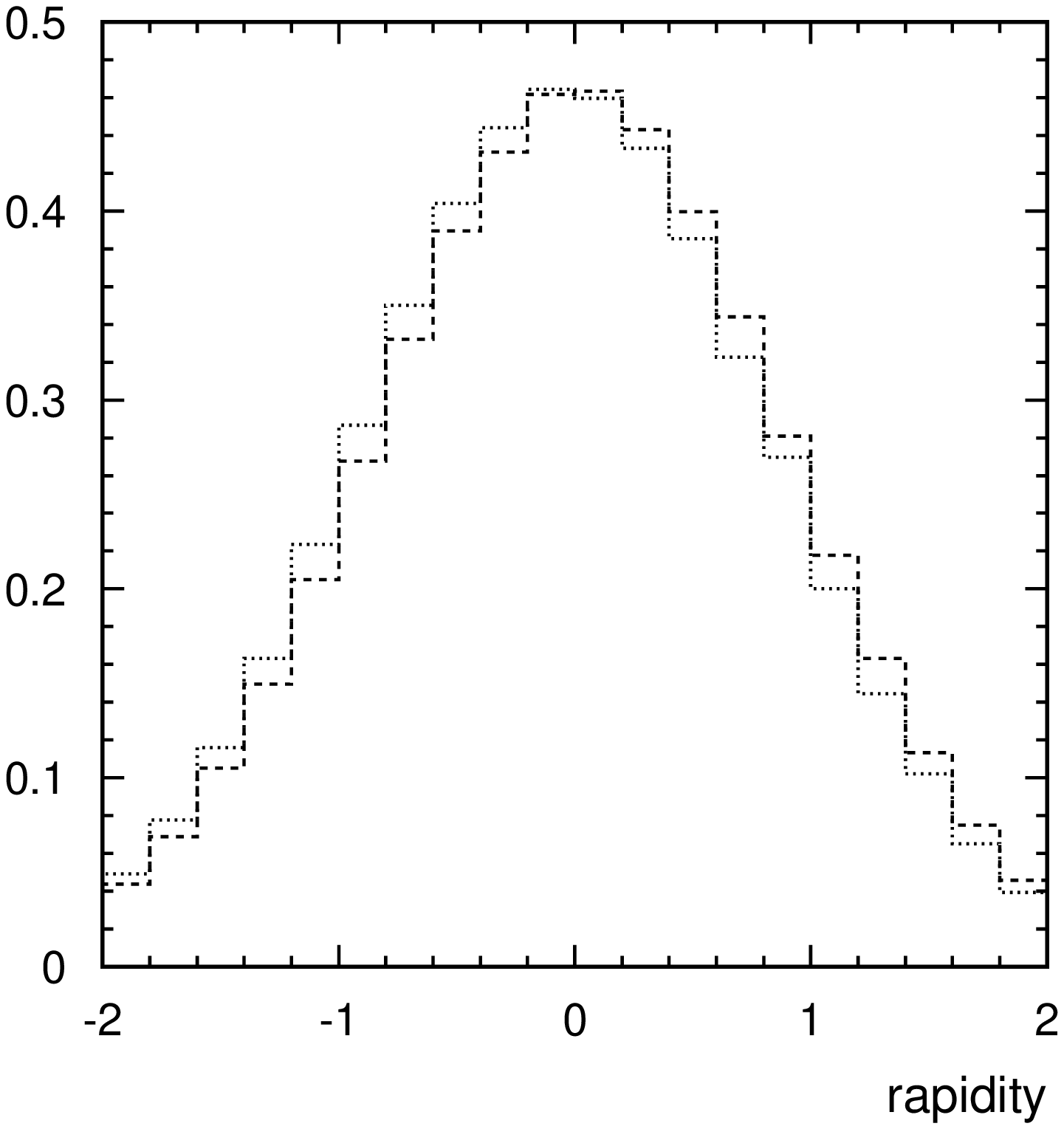}
\includegraphics[width=8cm]{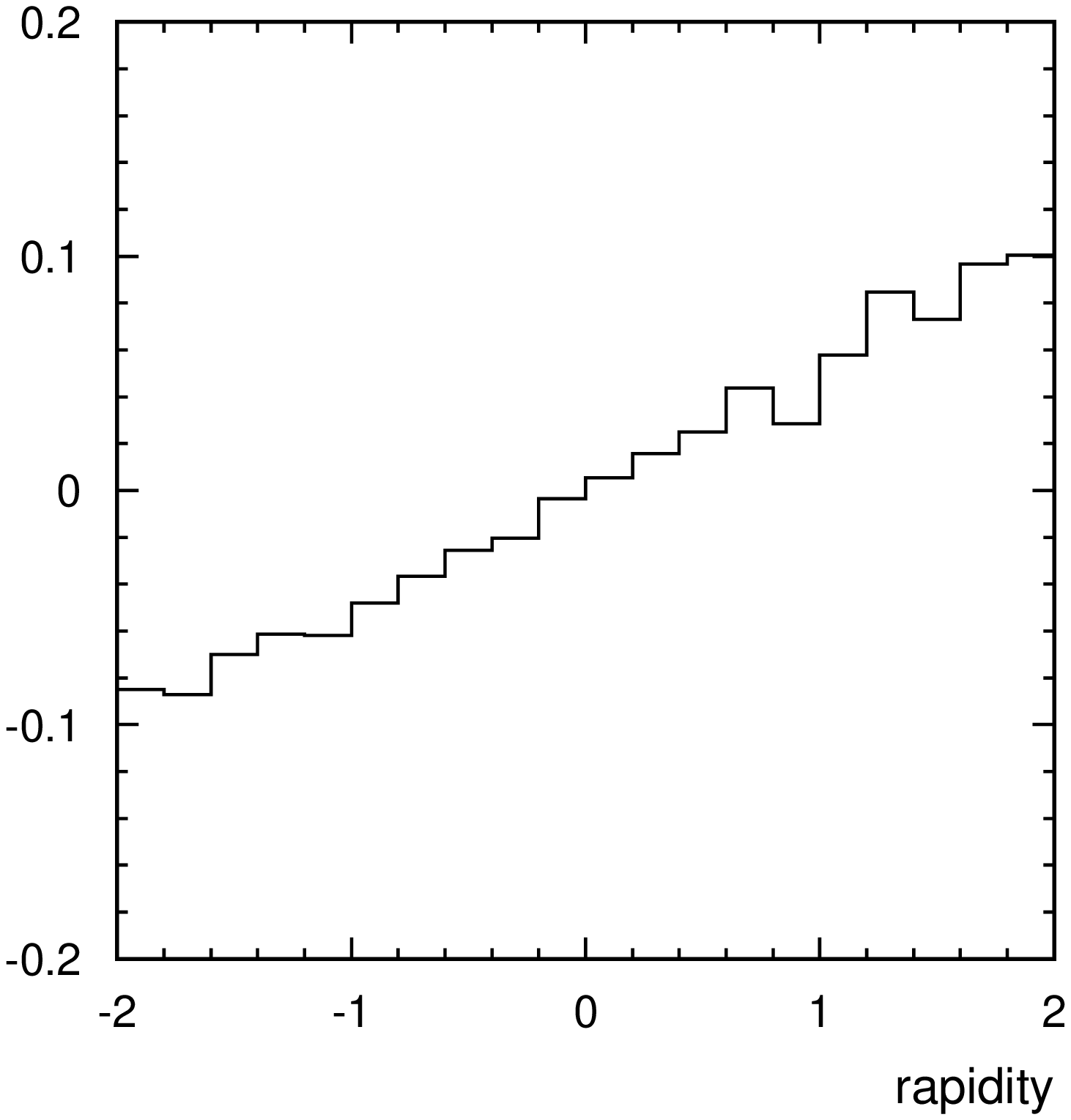}
\end{center}
\caption{
 Left:
 The rapidity distributions  (\ref{rapdischl}), normalized to $\sigma_{\ell\ell}$,
  for $\ell^+$ (dashed) and $\ell^-$ (dotted) for the dileptonic
   final states  at NLOW for the Tevatron  for $\mu=m_t$.
     Right: The differential leptonic charge asymmetry defined in   
    (\ref{rapdiasle}) 
  at NLOW' for the Tevatron  for $\mu =m_t.$ 
}
\label{fig:rapdischll}
\end{figure}

Next we compute, for the dilepton final states, 
 the  charge
 asymmetry $A$  (\ref{chasy}), and the pair asymmetry $A^{\tbart}$
   (\ref{fobattasy}) with the above cuts.
 The  rapidity distributions of the charged
 leptons,
\begin{equation} \label{rapdischl}
 N_{\ell^+}(y_{\ell^+}) = \frac{d\sigma}{dy_{\ell^+}}
\, , \qquad  N_{\ell^-}(y_{\ell^-}) = \frac{d\sigma}{dy_{\ell^-}} \,,
\end{equation}
 are defined with
  polar angles of $\ell^\pm$ and 
  the rapidities $y_{\ell^\pm}$  in the laboratory frame. 
 They are shown in Fig.~\ref{fig:rapdischll} (left) for the Tevatron, computed at NLOW 
  with the cuts (\ref{cut:dilep}). In this figure we have actually
  divided  $N_{\ell^\pm}$ by the dileptonic cross section at NLOW.         \\
One may define a differential leptonic
 charge asymmetry by  
 \begin{equation} \label{rapdiasle}
   A^\ell (y) = \frac{N_{\ell^+}(y)-
     N_{\ell^-}(y)}{N_{\ell^+}(y)+ 
    N_{\ell^-}(y)} \, .
\end{equation}
where $y$ denotes  the  rapidities  $y_{\ell^+}$ and $y_{\ell^-}$.
 This distribution is of course not identical to the  
 differential  top-quark charge asymmetry $A(y)$.
It is shown in Fig.~\ref{fig:rapdischll} (right) 
   for the Tevatron, computed at NLOW' (i.e., by
  the procedure described in Sect.~\ref{sec.rett})
  with the cuts (\ref{cut:dilep}). \\
 The associated leptonic charge asymmetry is
 \begin{equation} \label{rapdiaslin}
   A^\ell  = \frac{\int\limits_{y>0} N_{\ell^+}(y) \, - \, 
     \int\limits_{y>0} N_{\ell^-}(y)} 
{\int\limits_{y>0} N_{\ell^+}(y) \, +  \, 
    \int\limits_{y>0} N_{\ell^-}(y)}   \; .
\end{equation}

If CP invariance holds, then $N_{\ell^+}(y_{\ell^+}) = N_{\ell^-}(-y_{\ell^-})$ and  
 $A^\ell$ is equal to the leptonic forward-backward asymmetry,
 $A^\ell = A^{\ell^+}_{FB}  = -A^{\ell^-}_{FB}$, where
 \begin{equation} \label{rapdiaslfb}
   A^{\ell^\pm}_{FB}  = \frac{\int\limits_{y>0} N_{\ell^\pm}(y) \, - \, 
     \int\limits_{y<0} N_{\ell^\pm}(y)} 
{\int\limits_{y>0} N_{\ell^\pm}(y) \, +  \, 
     \int\limits_{y<0} N_{\ell^\pm}(y)}   \; .
\end{equation}
 In analogy to the $\tbart$ pair asymmetry $A^{\tbart}$ one may
 consider a leptonic pair asymmetry
\begin{equation} 
 A^{\ell\ell} = \frac{\int N(\Delta y_\ell > 0) - \int  N(\Delta y_\ell < 0)}{\int N(\Delta y_\ell > 0 ) +
  \int N(\Delta y_\ell < 0)}\, ,
\label{fobaelelasy}
\end{equation}
where $\Delta y_\ell = y_{\ell^+}- y_{\ell^-}$.
The observables (\ref{rapdiaslfb}),  (\ref{fobaelelasy}) and related, more exclusive asymmetries were
  analyzed in \cite{Bowen:2005ap}  for dileptonic and 
   semileptonic final states at the Tevatron with an approximate NLO
   QCD procedure (for details,see \cite{Bowen:2005ap}). 
    Here we compute (\ref{rapdiaslin}) and (\ref{fobaelelasy}) 
   at NLOW' with $t$, $\bar t$ spin effects included and, for
   comparison, also for uncorrelated $\tbart$ events.
 For the evaluation of $A$, $A^{\tbart}$, $A^\ell(y)$,
  $A^\ell$, and  $A^{\ell\ell}$  we use the
  procedure described in the previous section. 
Results are given in
 Table~\ref{chasTEVdilep}, at NLO'
   and NLOW',  for the dilepton final states at the Tevatron. \\
The distributions (\ref{rapdischl}) on which $A^\ell$ is based are displayed 
    in Fig.~\ref{fig:rapdischll} (left). The figure shows that the CP-symmetry
    relation   $N_{\ell^+}(y) = N_{\ell^-}(-y)$ is satisfied. The figure
    displays also the consequence of the  $q\bar q$ initiated SM top-quark charge asymmetry
 which tells us that the $t$ $(\bar t)$ is preferentially
  emitted into the (anti)proton direction.  
 The numbers in Table~\ref{chasTEVdilep} show that  the charge
 asymmetries $A$ and  $A^{\tbart}$ become smaller by several percent
 as compared to the values given in Table~\ref{chasyTEVLHC}  due to the acceptance cuts.
 Kinematic arguments suggest 
 that  $A^\ell$ has the same sign than $A$ but is smaller in
 magnitude, while  the SM value of $A^{\ell\ell}$  is larger than
 $A^\ell$, in analogy to  $A^{\tbart}$ versus $A$.
  This is is corrobated by the the numbers in
 Table~\ref{chasTEVdilep}. We give in this table also the values of
 these asymmetries when all the spin-dependent terms, in particular
 the $t\bar t$ spin correlations  in the matrix elements
 (\ref{eq:trace})  are switched off at next-to-leading order
  in the gauge couplings. This corresponds to the
 production of uncorrelated $\tbart$ pairs, followed by their decays
 which are spherical in the $t$ and $\bar t$ rest frames. One sees that 
 the values of the inclusive asymmetries $A$ and $A^\ell$ remain
 essentially unchanged, which is not surprising.  
As to $A^{\tbart}$, one may expect an effect of
  spin correlations on this ordered pair
 asymmetry in the presence of acceptance cuts.  
 Yet the
 numbers for the correlated case are larger than those for 
 uncorrelated events by only about $1\%$. An inspection of the top-charge
 asymmetric terms  in the partonic matrix elements reveals that the
 spin-dependent terms make indeed only a small contribution to this
  rather inclusive observable. 
 However, as  the numbers for the leptonic pair asymmetry
 $A^{\ell\ell}$ show, there  is a moderate  effect of $\tbart$ spin correlations
   on this more exclusive observable. The difference between our 
 predictions for
  correlated and uncorrelated  $\tbart$ events amounts to about $7\%$.

In view of theoretical uncertainty of $\delta A \sim 30\%$ 
that is implied by
the  analysis of \cite{Almeida:2008ug} at the level of $\tbart$
intermediate states, 
 one may assign an uncertainty of this order of
magnitude also to the numbers of Table~\ref{chasTEVdilep}.
The investigation of more exclusive asymmetries at NLOW', e.g. those
proposed in  \cite{Bowen:2005ap}, including semileptonic $\tbart$ events and the
corresponding asymmetries at the LHC is beyond the scope of this paper
 and has to be left for a future investigation.

The  D0 experiment at the Tevatron obtained, for lepton plus jets
final states,  a pair
 asymmetry $A^{\tbart} =0.12 \pm 0.08 \pm 0.01 $ \cite{:2007qb}.
This result must be unfolded for detector efficiencies and migration
effects \cite{:2007qb}. The pair asymmetry
$A^{\tbart}=0.071(7)$ given in Table~\ref{chasTEVdilep}  was
 computed for dileptonic final states with the above cuts, but we expect that it
 does not change much  when computed for semileptonic events with the
 cuts (\ref{cutseml}). 
With this proviso,  it can be compared  with the D0 result and the
   numbers agree within experimental and theoretical uncertainties. 

The CDF collaboration measured  $A^{\tbart}=0.24 \pm 0.14$
\cite{Aaltonen:2008hc}, and the recent result of this experiment
 on the forward-backward asymmetry is
 $A_{FB}^t=0.193\pm 0.065 \pm  0.024$ \cite{CDFpublic1}. The CDF
 results are actually  unfolded by using SM hypotheses, i.e. they
 can be compared with the predictions of Table~\ref{chasyTEVLHC}.
 The central values of the CDF results are higher than but still consistent
with the above SM expectations. The leptonic asymmetries   $A^{\ell}$ and $A^{\ell\ell}$, which were
predicted above at NLOW', have not yet been measured to our
knowledge.  

Although there is no statistically significant discrepancy between the
   SM expectations and
the above cited experimental results, the  numbers leave room for speculations
about new physics contributions.  One class of  possible new interactions
  involve axial vector  couplings to quarks. At the level of
  uncorrelated $\tbart$ final states, investigations were made 
 for instance by \cite{Sehgal:1987wi,Antunano:2007da}.

%

\subsection{Top-spin induced distributions and correlations}
\label{sub:tspidico}

There are a number of angular distributions and correlations
 with which $t$, $\bar t$ spin effects, in particular
  $\tbart$ spin correlations can  
 be analyzed  in the $\ell\ell'$ and $\ell + j$ final states.
It is well-known that the charged lepton from top-decay is, due to
 the $V-A$ law (which in the meantime is known 
 to hold for $t\to b$ to an accuracy\footnote{In 
  many SM  extensions new-physics
  contributions to the $tbW$ vertex are quite small, see e.g. \cite{Bernreuther:2008us}.}
of $\sim 10\%$ \cite{schwanen1}),  the most efficient top-spin analyzer. 
 Thus $\tbart$ spin-correlation effects are most pronounced in   $\ell^+
 \ell^{'-}$ angular correlations in the dilepton
 final states.  A correlation of the $t$ or $\bar t$ spin with some axis is
 best analyzed by a corresponding lepton angular distribution in  the
 semileptonic $\tbart$ modes. \\
For qualitative reasoning below we recall here the form of the
 one-particle
inclusive decay density matrices integrated over the energies:
\begin{eqnarray}
  \rho^{t\to a_1}
  & = \frac{\Gamma^{(1)}}{2}(\one +{\kappa}_1\, {\tatau} \cdot
  {{\bf{\hat q}}_1}),  \label{rhoa1} \\
 {\bar \rho}^{{\bar t}\to {a_2}}
  & = \frac{\Gamma^{(2)}}{2}(\one - {\kappa}_2 \, {\tatau} \cdot
  {{\bf{\hat q}}}_2),  \label{rhoa2}
\end{eqnarray}
where ${\bf{\hat q}}_{1,2}$
are the directions of flight of $a_1$ and $a_2$ in the rest frame of the top
and antitop quarks, respectively, and the ${\tau}^i$ are the Pauli matrices.  
As we incorporate top decay
   to lowest order in the electroweak couplings,
$\Gamma^{(2)} = \Gamma^{(1)}$ and 
${\kappa}_2 = {\kappa}_1$ holds to all orders in $\alpha_s$, if 
 $a_2$ is the
charge-conjugate of $a_1$. The spin-analyzing power of leptons 
 is essentially one, ${\kappa}_{\ell^+}= {\kappa}_{\ell^-}=0.999$,
  to order $\alpha_s$ \cite{Czarnecki:1990pe}. The spin-analyzing 
 powers of other final state
  particles/jets are given to order $\alpha_s$ in \cite{Brandenburg:2002xr}. \\
We analyze the following  correlations:
\begin{enumerate}
\item The double distributions: 
  \begin{eqnarray}
    {1\over \sigma}{d\sigma\over d\cos\theta_1 d\cos\theta_2}=
    {1\over 4} (1 +
    {B}_1 \cos\theta_1
    + {B}_2 \cos\theta_2 
    - {C} \cos\theta_1 \cos\theta_2)\,\, ,
    \label{eq:ddist1}
  \end{eqnarray}
  where $\sigma$ denotes the cross section of the respective reaction.
  The functional form of the right-hand side holds only
  if cuts are not applied. Here  $\theta_1$ ($\theta_2$) is the angle 
 between the
  direction of flight of the lepton $\ell^+$ or jet $j_1$ ($\ell\,'^-$ or
  $j_2$) from $t$ $({\bar t})$ decay in the $t$ ($\bar{t}$)
  rest frame  and a reference 
  direction ${\bf \hat a}$ (${\bf \hat b}$). 
 Below we will choose 
   the axes defined in the last section. The coefficient $C$ is
   a measure of $\tbart$ spin correlations -- it is induced by the
   correlation (\ref{genob}). In the absence of
   cuts we have the formula \cite{Bernreuther:2001rq}:
\begin{equation} \label{allofoab}
 C=  \kappa_1 \kappa_2 \langle 4 \Sp\cdot {\bf{\hat a}}) (\Sm\cdot
 {\bf{\hat b}})\rangle \, ,
\end{equation} 
  which holds
 for factorizable corrections to all orders\footnote{In fact, the
  nonfactorizable ${\cal O}(\alpha_s^3)$ QCD corrections
    \cite{Beenakker:1999ya}  do not change this result; cf. the
    comment in   \cite{Bernreuther:2004jv}.}.
  The expectation values on the right-hand side of this equation
   are given in Table~\ref{TabTevLHCsptt}.
   The coefficients $B_{1,2}$ will be discussed in items 4 and 5.
\item  The opening angle distribution  \cite{Bernreuther:2004jv}
  \begin{eqnarray}
    {1\over \sigma}{d\sigma \over d\cos\varphi}=
    {1\over 2} (1 - {D} \cos\varphi)\,\, ,
    \label{eq:ddist2}
  \end{eqnarray}
  where $\varphi$ denotes the angle 
  between the
  direction of flight of the lepton $\ell^+$ (or jet $j_1$)
  and  of $\ell\,'^-$ (or $j_2$), defined  in the $t$ and $\bar{t}$
  rest frame, respectively.  The functional form of the right-hand side applies only
  in the absence of cuts. The coefficient $D$ is induced by a term
  proportional to 
 $\Sp \cdot \Sm$ in the  $\tbart$ production density matrix. If no cuts are applied the formula
 \begin{equation} \label{allofoD} 
D= \kappa_1 \kappa_2 \langle  \frac{4}{3} \Sp \cdot \Sm \rangle
\end{equation} 
 holds.  The expectation values on the right-hand side of this equation
   are given in Table~\ref{TabTevLHCsptt}. Below we comment also on the leptonic opening angle
  distribution in the laboratory frame. 
\item The dileptonic azimuthal angle correlation, 
   $\sigma^{-1}{d\sigma/ d\Delta\phi},$
   \cite{Barger:1988jj,Arens:1992fg}.  Here
   $\Delta\phi$ is the difference of the 
      azimuthal angles of the $\ell^+$ and $\ell\,'^-$ directions
    of flight, which can be measured in  the laboratory
    frame. This distribution was recently analyzed by \cite{Mahlon:2010gw} 
       for the case of dileptonic $\tbart$ events at the LHC.
    It was shown by these authors that the  $\Delta\phi$ distribution 
      is sensitive  to SM spin correlations if  only
   events with $\mtt < 400$ GeV are taken into account. 
   The price one pays for excluding  with this cut 
  a large fraction of the dileptonic events 
   is balanced by the fact that
 the  $\Delta\phi$  distribution 
 is easier to measure in the dilepton channels than (\ref{eq:ddist1})
  or  (\ref{eq:ddist2}). 

  \item Nonzero  coefficients
   $B_1$ and $B_2$   (\ref{eq:ddist1}) would 
  be due to a correlation $\Sp\cdot {\bf{\hat a}}$ and $\Sm\cdot
  {\bf{\hat b}}$ of the $t$ and
   $\bar t$ spin with the respective axis. 
  If polar vectors ${\bf{\hat a}}$, ${\bf{\hat b}}$
 are chosen then $B_i\neq 0$ would signal  parity violation if no cuts
  were made. 
   For the
  axes defined in Section~\ref{sec.rett} 
  the weak-interaction corrections 
  induce non-zero, albeit
 very small $B_i$ \cite{Bernreuther:2006vg,Bernreuther:2008md}. 
 If cuts are not applied  one obtains from (\ref{eq:ddist1}):
\begin{equation} \label{eq:ddistcosl}
 {1\over \sigma}{d\sigma\over d\cos\theta_{1,2}} =   {1\over 2} (1 +
    {B}_{1 ,2}\cos\theta_{1,2}) \, .
\end{equation}
 In view of the very small parity-violating effects just mentioned,
 the SM predicts this distribution to be essentially flat
  \cite{Bernreuther:2008md}  if cuts could be
  applied that are  parity-symmetric with respect to the $t$,
  respectively $\bar t$ rest frame. However, realistic cuts severely
  distort  (\ref{eq:ddistcosl}). We compute for the 
  semileptonic $\tbart$  modes at the LHC:
\begin{equation} \label{eq:ddistcosl2}
 G_{\ell^+}(z_+) = \frac{1}{\sigma_{\ell^+}}{d\sigma\over d z_+}
 \, , \qquad  G_{\ell^-}(z_-) = \frac{1}{\sigma_{\ell^-}}{d\sigma\over d z_-} \, ,
\end{equation}
 where $z_\pm =\cos\theta_{\ell^\pm}$ and
  $\cos\theta_{\ell^+}={\bf{\hat a}}\cdot{\velp}$,
     $\cos\theta_{\ell^-}={\bf{\hat b}}\cdot{\velm}$.     
             Below we use for definiteness
  the helicity basis. 
  One can nevertheless use these distributions for parity
   and also for CP tests, cf. the
   comments at the end of this section.  
\item The functional forms of the r.h.s. of Eqs. 
(\ref{eq:ddist1}),  (\ref{eq:ddist2}), and  (\ref{eq:ddistcosl}) do not hold
in the presence of kinematic cuts. Kinematic considerations make it clear 
that the forms (\ref{eq:ddist1}) and (\ref{eq:ddistcosl}) will be stronger
  distorted than (\ref{eq:ddist2}).
   Nevertheless, the right-hand side of (\ref{eq:ddist1}) -- or of 
    (\ref{eq:ddist2}) --  
 may be used in fits  to data
   for determining the correlation coefficients $C$ or $D$. One may 
    also divide the events  in an experimental analysis 
   into spin-enhanced (low $\mtt$) and spin-depleted (high $\mtt$)
   samples and form ratios in order to reduce systematic uncertainities
    \cite{DHondt}. \\
 Here we use    the estimators \cite{Bernreuther:2004jv}:
\begin{eqnarray}\label{expval}
{\hat C} = -9\langle \cos\theta_1\cos\theta_2 \rangle \, , \qquad
& {\hat D} = -3  \langle \cos\varphi\rangle \, ,\nn \\ 
{\hat B}_1 = 3 \langle \cos\theta_1\rangle \, , \qquad 
 & {\hat B}_2 = 3 \langle \cos\theta_2 \rangle  \, .
\end{eqnarray}
 In the absence of cuts, ${\hat C}=C$, ${\hat D}=D$, 
${\hat B}_i =B_i$. Below we compute  the expectation values
  (\ref{expval}) with the above cuts. We drop
  the hats from now on for ease of notation. In fact, we
  compute the  variation of the correlations with $\mtt$. 
   We define it in the case of $D$ by
\begin{equation}
\frac{d D}{d{\mtt}} = \frac{ -3}{\sigma} \int 
  d\sigma  \,  \cos\varphi \, \delta\left(\sqrt{(\sum{p_f})^2} -\mtt \,
      \right)\, \Theta_{\rm cut} ,
\label{diffgenobfin}
\end{equation}
Here $\Theta_{\rm cut}$ denotes the product of theta functions
implementing the cuts and, in case of an exclusive 
  computation, of jet functions.
 For the other angular variables in (\ref{expval})
  the variations are defined in analogous fashion.
We normalize  to the respective integrated
 cross section. The computations  of (\ref{diffgenobfin}) and 
 of the other observables concerning their variation with
  respect to $\mtt$ made in this paper 
    are theory predictions; i.e. the complications
  in the actual measurement of the invariant mass of the $\tbart$ pair
   are not taken into account; cf. the comments below.

\end{enumerate}

\begin{figure}
\begin{center}
\includegraphics[width=8cm]{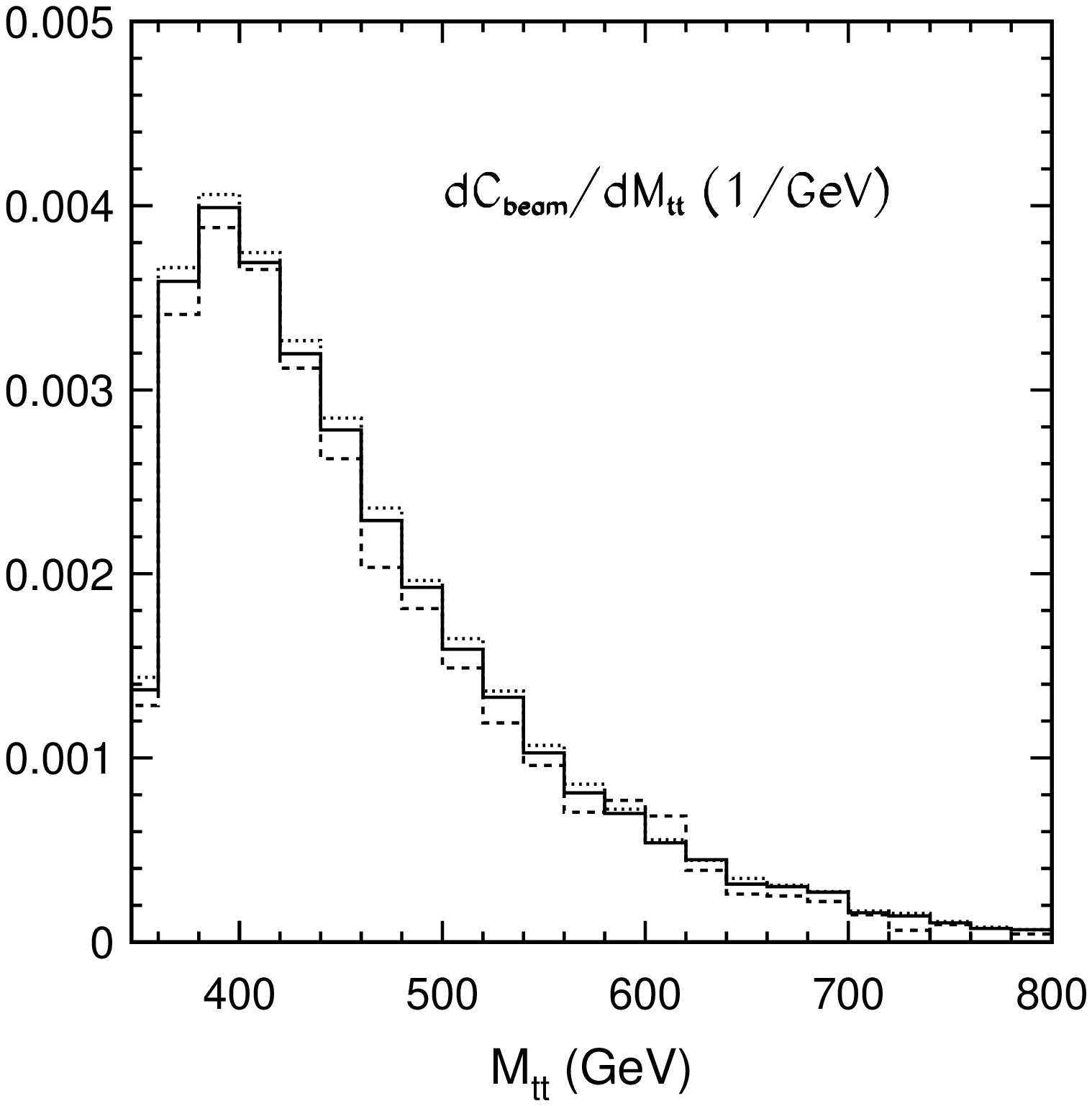}
\includegraphics[width=8cm]{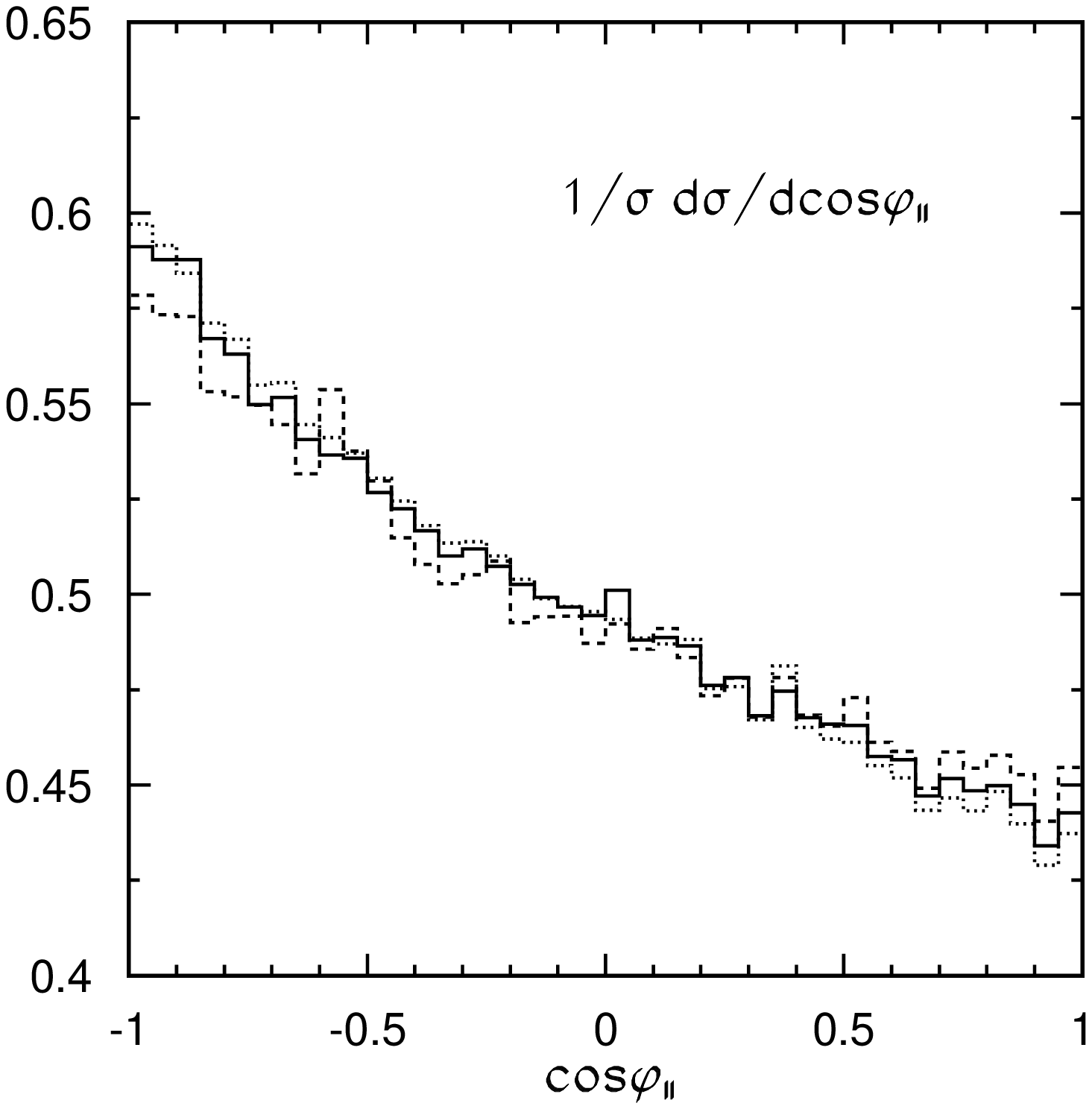}
\end{center}
\caption{Distributions  at NLOW for  $\ell \ell$ final states  at  the Tevatron:
 $dC_{\rm beam}/ d\mtt$
   (left panel) and the opening angle distribution  (\ref{eq:ddist2}) (right panel). The solid, dashed, and
  dotted lines correspond to $\mu =m_t, \, m_t/2$, and $2m_t$, respectively.}
 \label{fig:spin1tev}
\end{figure}

Fig.~\ref{fig:spin1tev} shows
  $dC_{\rm beam}/ d\mtt$ and the opening angle 
  distribution  (\ref{eq:ddist2})
  for the Tevatron for three scales. The resulting 
  integrated correlation coefficients
   are given in Table~\ref{TabTEVLHC}. Comparing
    $C_{\rm beam}=0.614(10)$  with the ``no cut'' value $0.791(14)$ 
    shows the significant effects of the
    cuts on this correlation. 
   The distribution $dC_{\rm off}/ d\mtt$, which is not shown here,
   has essentially the same form as $dC_{\rm beam}/ d\mtt$. 
   The correlation coefficient
    $C_{\rm off}$ in the off-diagonal basis  is 
   also  reduced  by the cuts to  $C_{\rm off}=0.621(10)$, 
    cf. Table~\ref{TabTEVLHC}. 
    Without cuts $C_{\rm off} = 0.798(15)$.

  The correlations can be understood qualitatively in terms
   of simple spin arguments. We recall this here only for the opening angle
  distribution,  using
    (\ref{allofoD}) and  
   Born level considerations. The $\tbart$ pair produced by the
   dominant LO reaction
    $q {\bar q}\to t {\bar t}$ is in a $^3S_1$ state near threshold.  
  Thus, $\langle \Sp \cdot \Sm \rangle_{q\bar q} = 1/4$ and 
$D_{q\bar q}= 1/3$. The ``true''  $\tbart$ intermediate state is
  of course a mixed state. The component from
    $gg$ fusion, the radiative corrections, and the cuts dilute this 
  correlation to $D=0.145(6)$, cf. 
  Table~\ref{TabTEVLHC}. Comparing with the ``no cut'' value
  $0.218(2)$ 
   shows again the significant  effect of the cuts.

  On may attempt to optimize these correlation coefficients by applying
  an additional cut on $\mtt$, but in view of the not very large  number
  of dileptonic $\tbart$ events at the Tevatron this seems not
   very  useful. In Table~\ref{TabTEVLHC} we give the values of several
   correlation coefficients for dileptonic events at the Tevatron with 
  ${M}_{\rm max} = 550$ GeV. This value has been chosen here rather
  arbitrarily; in fact, the correlations do not increase when this cut
  is tightened to 400 GeV. With even tighter cuts one would discard
  too much dileptonic events.

\begin{figure}
\begin{center}
\includegraphics[width=8cm]{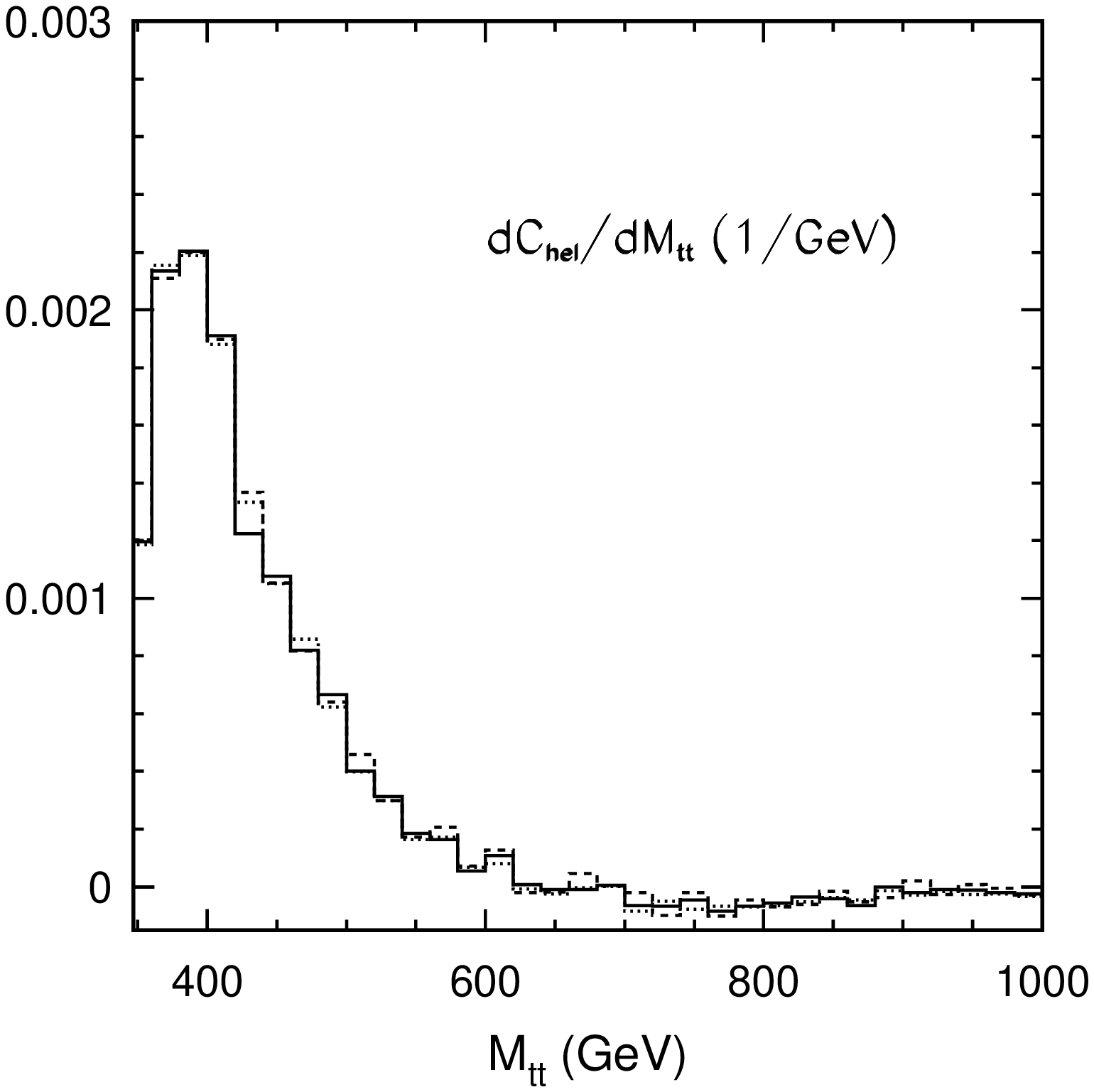}
\includegraphics[width=8cm]{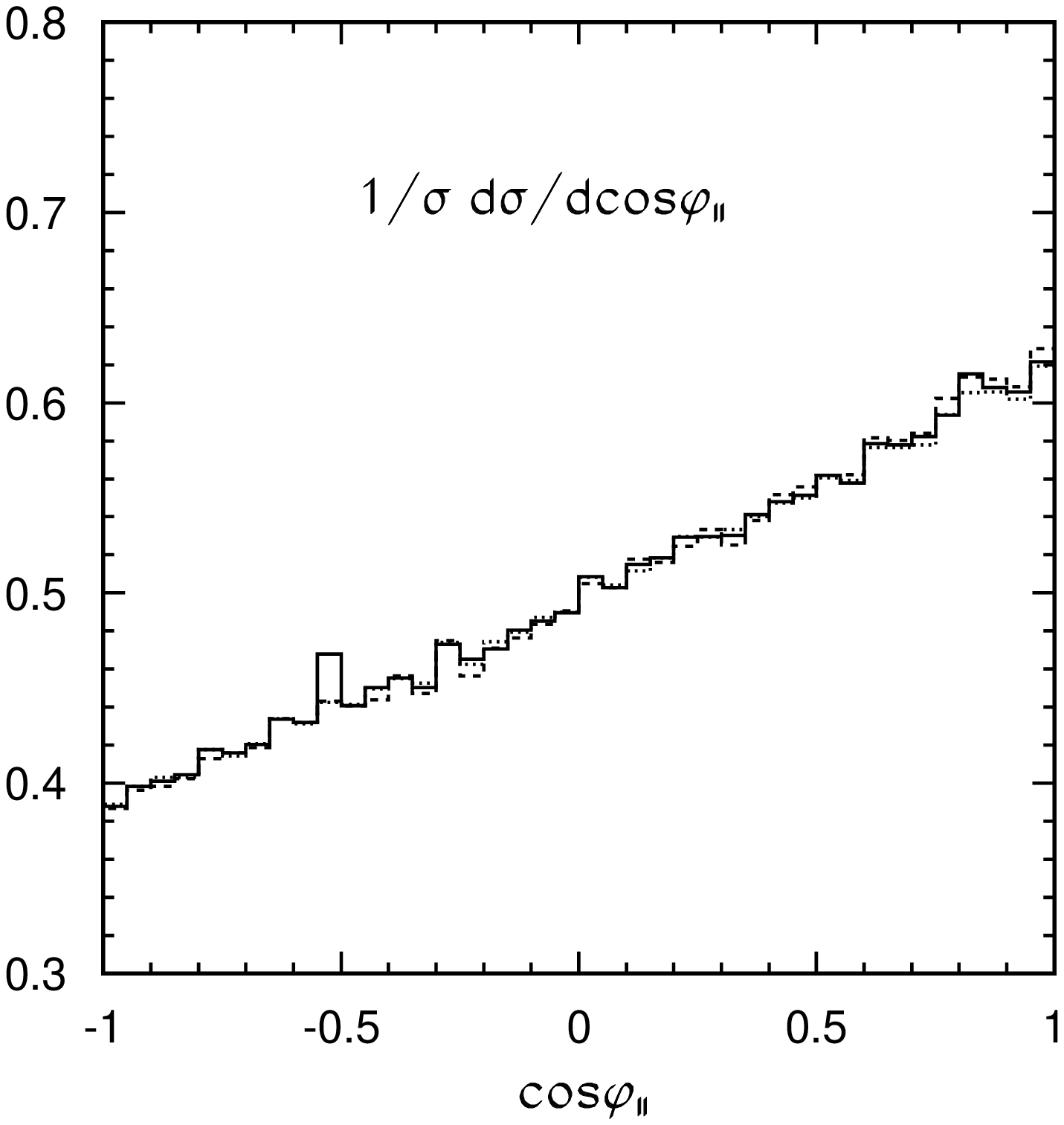}
\end{center}
\caption{Distributions  at NLOW for $\ell \ell$ final states  at the
  LHC  ($\sqrt{s}=10$ TeV):
 $dC_{\rm hel}/ d\mtt$
   (left panel) and the opening angle distribution  (\ref{eq:ddist2}) (right panel). The solid, dashed, and
  dotted lines correspond to  $\mu =m_t, \, m_t/2$, and $2m_t$, respectively.}
 \label{fig:spin1lhc10}
\end{figure}

\begin{figure}
\begin{center}
\includegraphics[width=8cm]{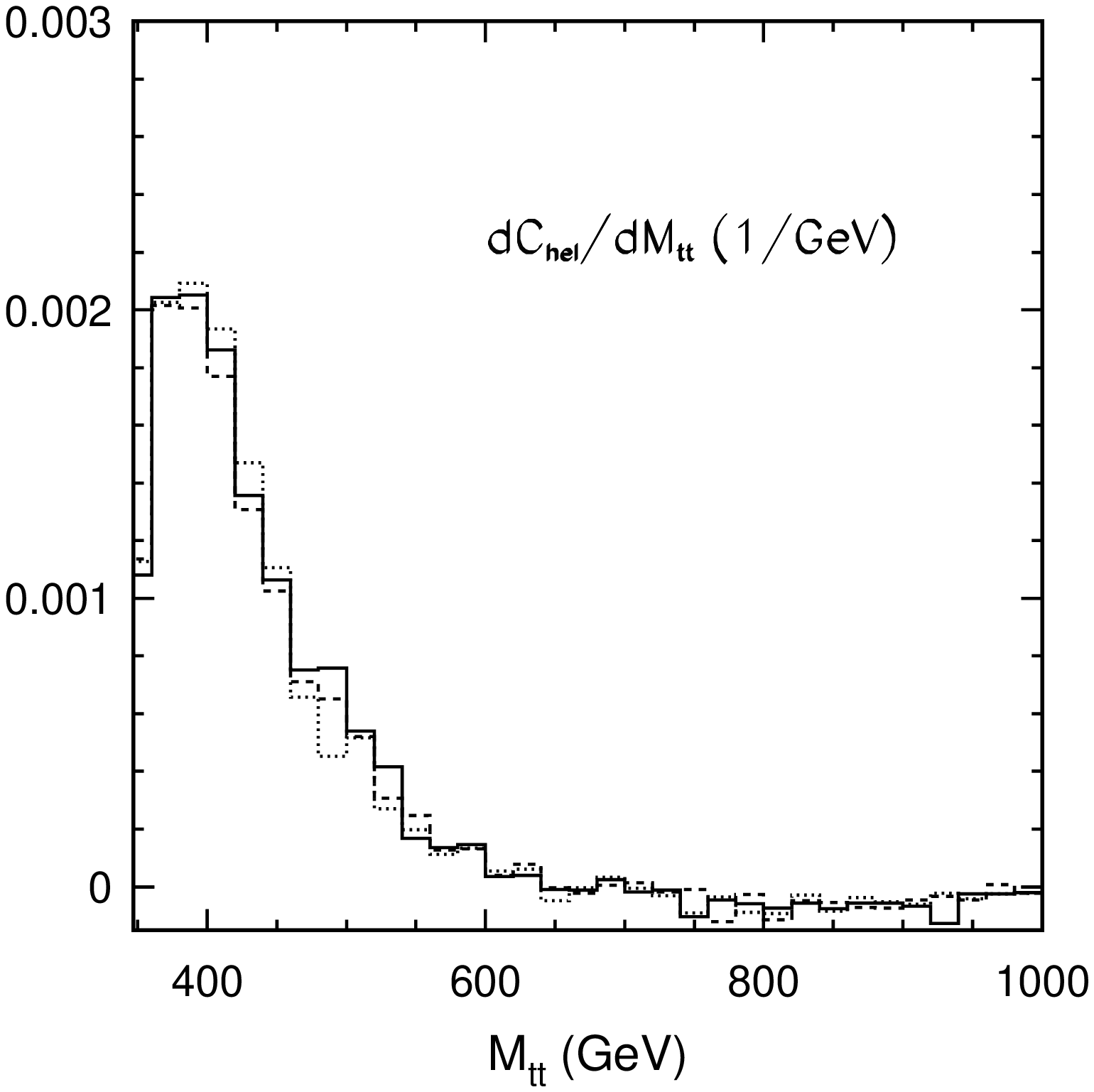}
\includegraphics[width=8cm]{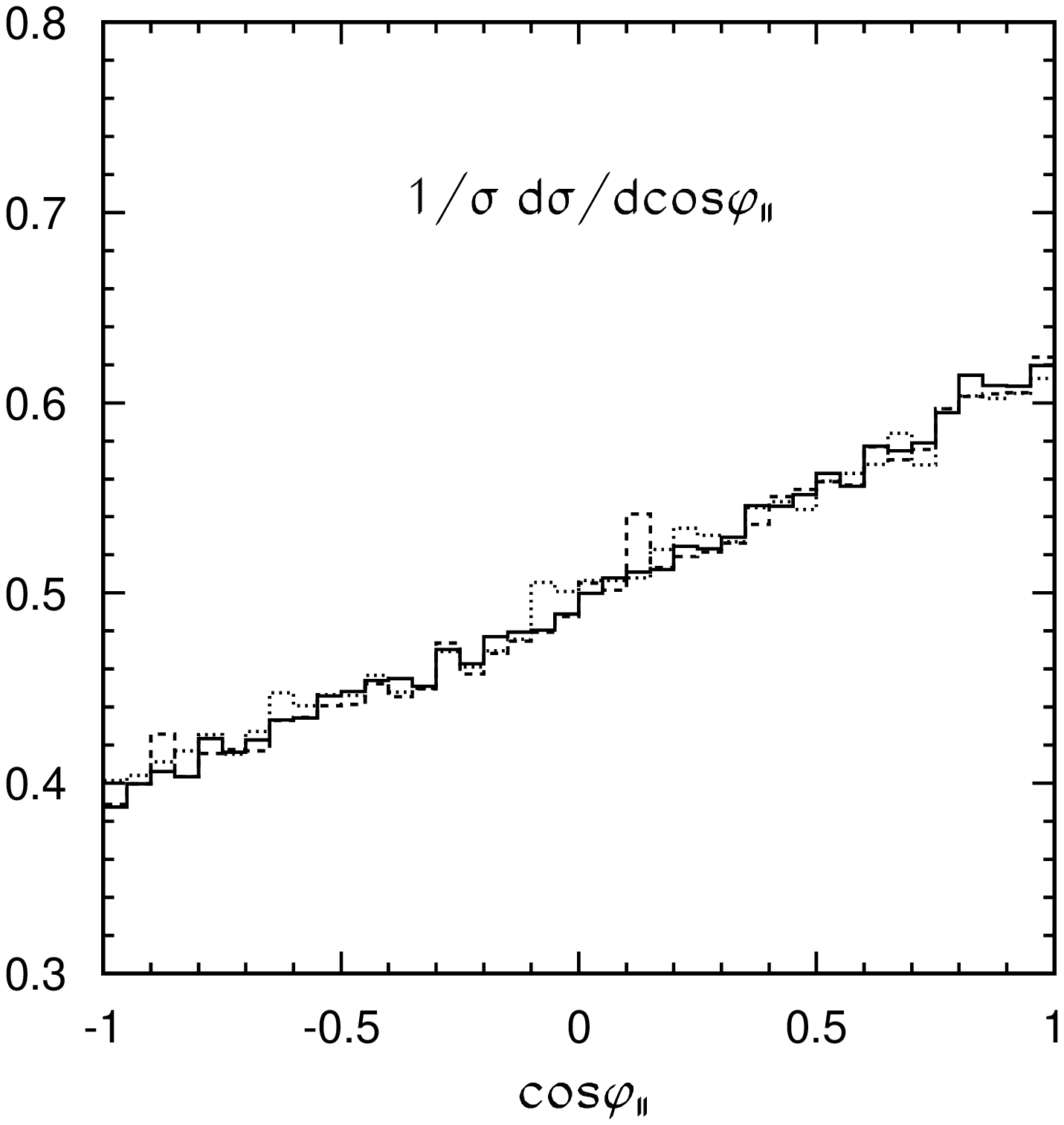}
\end{center}
\caption{Distributions  at NLOW for
 $\ell \ell$ final states  at the LHC ($\sqrt{s}=14$ TeV):
 $dC_{\rm hel}/ d\mtt$
   (left panel) and the opening angle distribution  (\ref{eq:ddist2}) (right panel). The solid, dashed, and
  dotted lines correspond to $\mu =m_t, \, m_t/2$, and $2m_t$, respectively.}
 \label{fig:spin2lhc14}
\end{figure}

\begin{figure}
\begin{center}
\includegraphics[width=8cm]{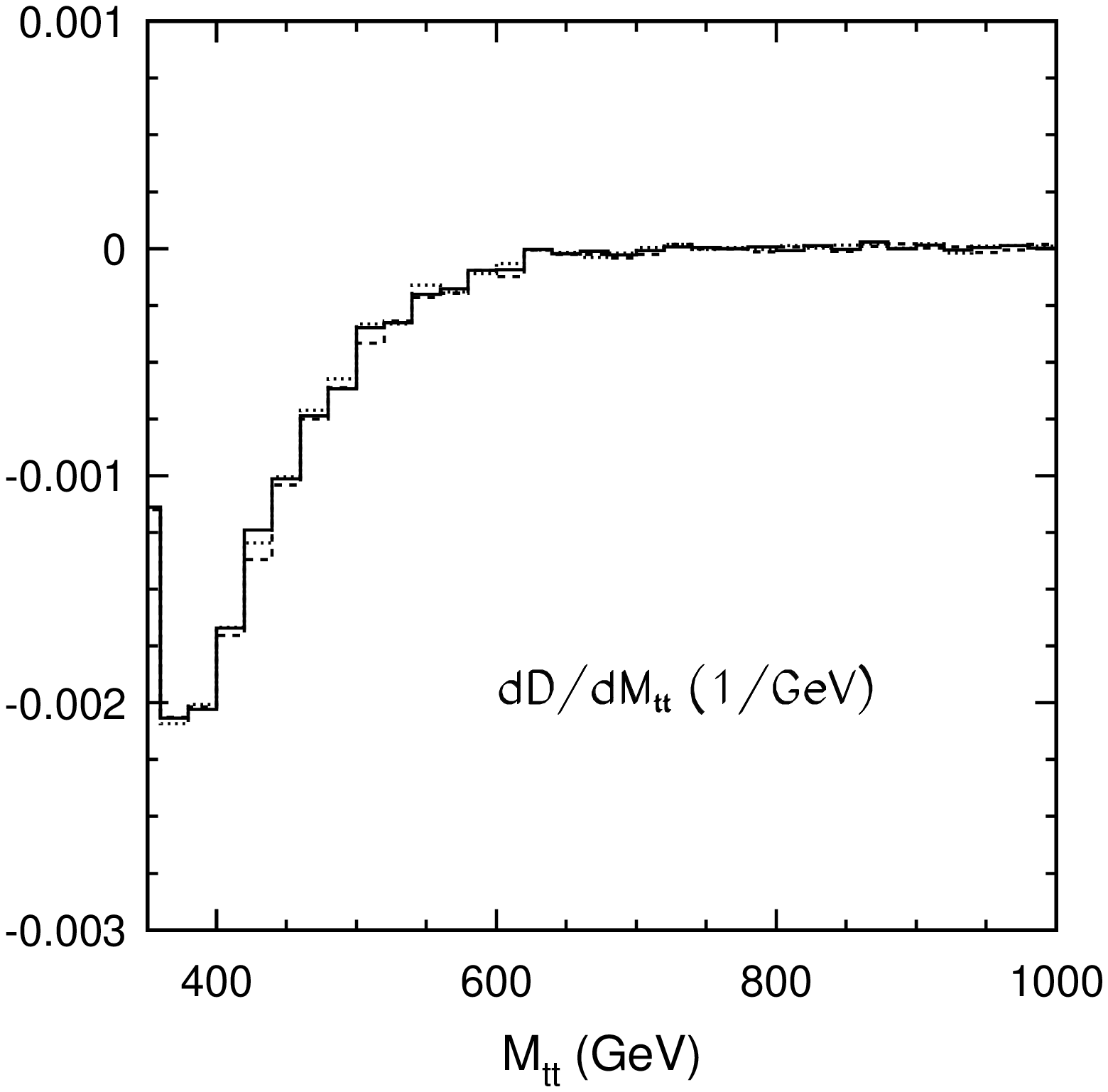}
\includegraphics[width=8cm]{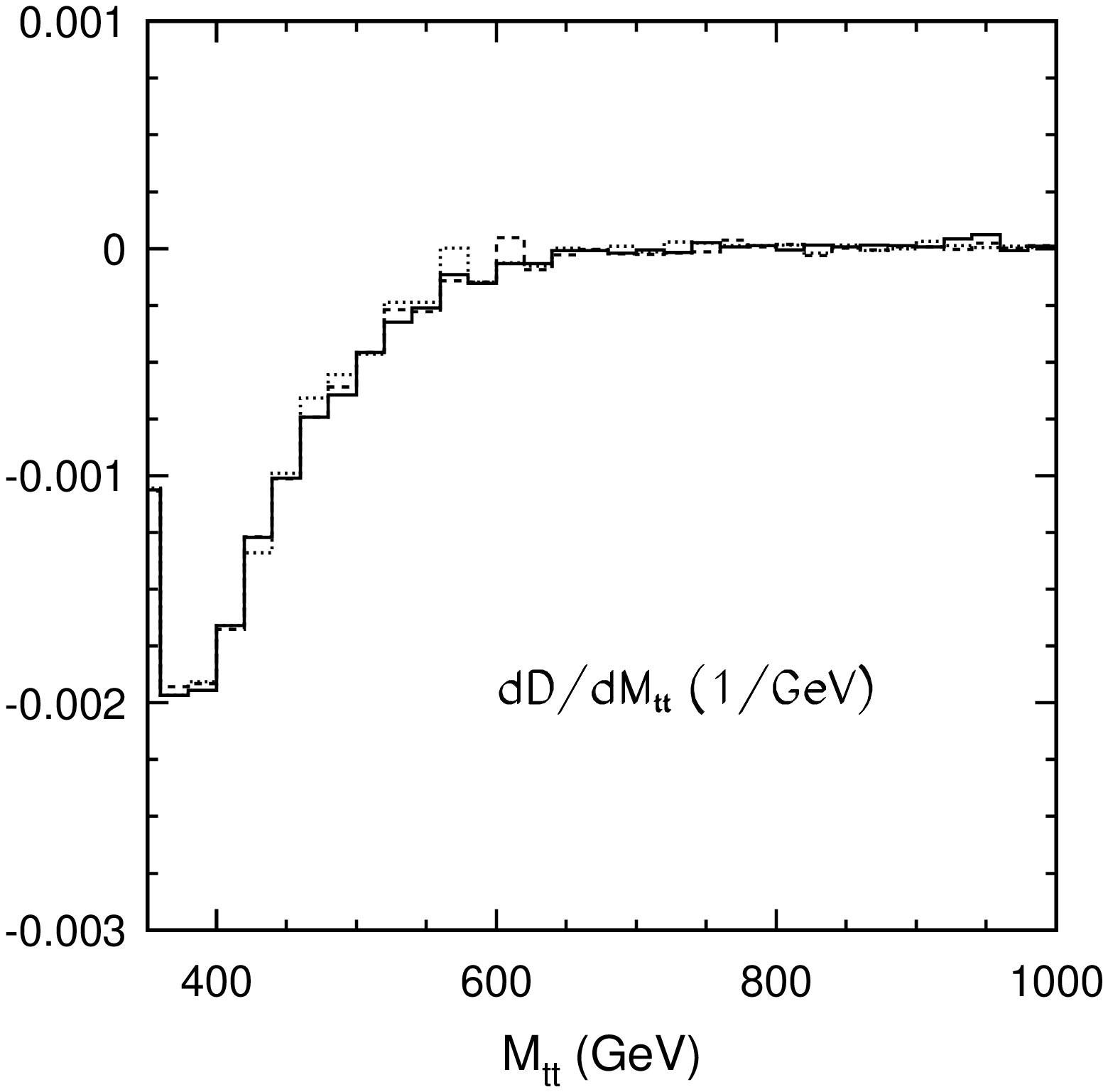}
\end{center}
\caption{The distribution $d D/d\mtt$
   at NLOW for $\ell \ell$ final states  at
  the  LHC at $\sqrt{s}=10$ TeV
   (left panel) and at 14 TeV (right panel).
  The solid, dashed, and
  dotted lines correspond to $\mu =m_t, m_t/2,$ and $2m_t$, respectively.}
 \label{fig:spin2lhc1014}
\end{figure}

Figs.~\ref{fig:spin1lhc10},~\ref{fig:spin2lhc14},
  and~\ref{fig:spin2lhc1014} show $dC_{\rm hel}/ d\mtt$, the opening angle
   distribution (\ref{eq:ddist2}), and $dD/ d\mtt$ for dileptonic final
 states at the LHC with $\sqrt{s}$ = 10 and 14 TeV. The distribution
$dC_{\rm hel}/ d\mtt$ changes sign  around $\mtt\sim 700$ GeV.
 This is because for values of
 $\mtt$ not too far above threshold, the QCD dynamics tells us
  that the number of $\tbart$ pairs
 produced with like helicities is
 larger than those with unlike helicities, 
 while at large $\mtt$ helicity conservation of 
the strong and weak interactions 
  implies that it is the other way around.
 As the event numbers become
   small very rapidly and because we have normalized to $\sigma_{\ell \ell}$,
  the distribution
    $dC_{\rm hel}/ d\mtt$ displayed
   in Figs.~\ref{fig:spin1lhc10} and~\ref{fig:spin2lhc14} is essentially
 zero above $\mtt\simeq 700$ GeV.   
 
 As to the qualitative understanding 
  of the  opening angle
  distribution,  Figs.~\ref{fig:spin1lhc10} and~\ref{fig:spin2lhc14}: 
  Because the  production of $\tbart$ pairs at the LHC 
  is dominated by  gluon fusion, the pair is, to LO QCD, 
   in a $^1S_0$ state  at threshold. 
 Hence $\langle \Sp \cdot \Sm \rangle_{gg} = -3/4$ and $D_{gg}= - 1$.
 Away from threshold, the pair is actually in a 
 mixed state already at Born level. This fact,  the 
  component from
    $q {\bar q}$ annihilation, and the radiative
 corrections deplete this correlation to $D=-0.252(5)$ (10 TeV),
 respectively $D=-0.240(10)$ (14 TeV),  cf. 
  Table~\ref{TabTEVLHC}. The effect of cuts on $D$ is less drastic than in the case of the
  Tevatron, as seen by comparing with 
the  ``no cut'' values $-0.233(4)$ (10 TeV)  and   $-0.236(3)$ (14
TeV).  On the other hand, the helicity correlation is depleted by the
cuts, as shown by comparing the respective numbers  in
Table~\ref{TabTEVLHC} with $C_{\rm hel}({\rm no \; cut}) = 0.326(2)$
(10 TeV) and $0.328(3)$ (14 TeV). \\
The helicity and opening angle correlations at the LHC can be enhanced
significantly by taking into account only events with a $\tbart$
invariant mass below a certain value ${M}_{\rm max}$. 
Table~\ref{TabTEVLHC} shows that for ${M}_{\rm max} = 550$ GeV
 the correlation coefficients
  $D$ and  $C_{\rm hel}$ can be enhanced by about $43\%$ and $50\%$,
  respectively. We have not optimized this cut.

The beam, off-diagonal, and helicity correlation and the opening angle
 distribution were computed before to NLO QCD for dileptonic, semileptonic,
 and non-leptonic final states  in
 \cite{Bernreuther:2004jv}, without cuts, and with a different set of
 PDF (CTEQ6.1M and MRST2003) and a slightly different value of the 
  top-quark mass.  The numbers 
   given in that paper can be compared with the results given here,
    using  Table~\ref{TabTevLHCsptt},  the formulae  (\ref{allofoab})
   and  (\ref{allofoD}), and the spin-analyzing powers $\kappa$ given in
 \cite{Bernreuther:2004jv}.  There are two effects:
   the PDF set CTEQ6.6M and the smaller value of the
    top-quark mass that we use  in the present paper enhances
   the correlation coefficients slightly  as compared to those
  of \cite{Bernreuther:2004jv}. On the other hand the weak-interaction 
   corrections lead to a slight reduction of the correlations at the
   LHC (cf. Sect.~\ref{sec.rett}), while their influence on the Tevatron 
  correlations is smaller. In effect this leads to slightly enhanced 
     numbers for the Tevatron, about
     $2\%$, while the numbers for the LHC given in
     Table~\ref{TabTevLHCsptt}
       are essentially the same as those in \cite{Bernreuther:2004jv}.
  
  The authors of \cite{Frixione:2007zp} computed the  opening 
  angle distribution as defined above 
   in  the MC@NLO framework -- whith parton showering, but no cuts were
   applied --  for dileptonic final states at the
   Tevatron and the LHC (14 TeV) and found agreement with the results
   given in \cite{Bernreuther:2004jv}. This shows that this leptonic
    correlation is essentially unaffected by the parton shower.
    This distribution 
   was recently also evaluated  in \cite{Melnikov:2009dn}
   for dileptonic final states at the Tevatron
   and the LHC (10 TeV), to NLO QCD with cuts.  We have computed these
   distributions also with the selection criteria of
 \cite{Melnikov:2009dn} and agree with their results.

The dileptonic angular correlations considered above require
  the reconstruction of  the $t$ and $\bar t$ rest frames, which is
  a highly non-trivial experimental task. There are
  enough kinematic constraints but, as is well-known, for  dileptonic events a
  four-fold ambiguity arises, which may increase to an eightfold one if
   the $b$ and $\bar b$ jets are incorrectly assigned.  
  Monte-Carlo based methods have been devised 
 (see, e.g., \cite{Sonnenschein:2005ed}) in order to
   reduce or  partly resolve these ambiguities. 

The beam-basis  and off-diagonal
  correlations  were  recently measured for
dileptonic events at the Tevatron by the D0 and CDF collaborations,
 respectively, with the results
\begin{equation} \label{res:tevspind0cdf}
C_{\rm beam} =   -0.17_{-0.53}^{+0.64}  \quad \mbox{\cite{D0public1}} ,   \qquad 
  \kappa_{\rm off} = 0.32^{+0.55}_{-0.78}  \quad \mbox{\cite{CDFpublic2}} .
\end{equation}
In our sign convention  $\kappa_{\rm off}= C_{\rm off}$.
  These results agree, within the still large experimental uncertainties, with
  our  predictions for the dilepton channels at the Tevatron 
 given Table~\ref{TabTEVLHC}.
 The CDF collaboration measured also the helicity correlation for
 lepton + jets events. They obtained for the respective spin
 correlation parameter at the $\tbart$ level the value
 $\kappa_{\rm hel} =   0.60 \pm 0.50 \pm 0.16$ \cite{CDFpublic3}.
  In our sign convention for the  double spin asymmetries \cite{Bernreuther:2004jv}
    this is equivalent to $-{\cal O}_{\rm hel}$ and should be compared
   with the value  $-{\cal O}_{\rm hel} = +0.368(10)$ for the Tevatron given in
   Table~\ref{TabTevLHCsptt}. The above acceptance cuts deplete this
   value to   $-C_{\rm hel}= +0.30(6)$,  cf. Table~\ref{TabTEVLHC}.

%

%

%

%


\begin{table}[]
\begin{center}
\caption{Results for several observables at NLOW for dileptonic final
   states  at the Tevatron and LHC, with cuts as specified in
   (\ref{cut:dilep}).  
  The  correlation coefficients are computed also with a cut
  $\mtt<M_{\rm max}$. Here we have chosen $M_{\rm max}$ = 550 GeV.
 }
\vspace*{0.5cm}

\label{TabTEVLHC}

\begin{tabular}{||c|c|c|c||c|c|c||c|c|c||}\hline \hline
  &\multicolumn{3}{c|}{Tevatron} 
&\multicolumn{3}{c|}{LHC (10 TeV)} 
&\multicolumn{3}{c|}{LHC (14 TeV)} \\ \hline \hline
$\mu$ & $m_t/2$ & $m_t$ & $2m_t$ & $m_t/2$ & $m_t$ & $2m_t$ &$m_t/2$ 
& $m$ & $2m$ \\ \hline \hline
$\sigma_{\ell\ell}$ \, (pb) & 0.043 & 0.042 & 0.038 &2.31 &2.03 &1.76
&5.00&4.38 & 3.82 \\ \hline \hline
$D$ &0.139&0.145  &0.151 & -0.257 &-0.252 &-0.257 &-0.240&-0.247&-0.230\\ \hline
$D({M}_{\rm max})$ &0.125 &0.132 &0.138 & -0.344 & -0.340 & -0.347 & -0.340 &
-0.353 & -0.338 \\ \hline \hline
$C_{\rm hel}$  &-0.294 &-0.299 &-0.306 & 0.249  & 0.247 & 0.252 &0.225&0.237 &0.229\\ \hline
$C_{\rm hel}({M}_{\rm max})$ &-0.256 &-0.262&-0.269 &0.350 & 0.351 & 0.362 & 0.336 & 0.360 & 0.345 
\\ \hline \hline
$B_1$       & & & &0.174 &0.176 &0.174 &0.162 &0.162 & 0.178  \\ \hline \hline
$C_{\rm beam}$& 0.605 &0.614 &0.624 &  & & & & & \\ \hline
$C_{\rm beam}(M_{\rm max})$&0.577  &0.586 &0.596 &  & & & & & \\ \hline
$C_{\rm off}$ & 0.612 &0.621 &0.631&   & & & & & \\ \hline  \hline
$C_{\rm off}(M_{\rm max})$ &0.582 &0.591 &0.601&   & & & & & \\ \hline  \hline
\end{tabular}
\end{center}
\end{table}


A simulation study \cite{Hubaut:2005er} within the ATLAS collaboration
concluded that $D$ and $C_{\rm hel}$ can be measured at the LHC 
 with a relative uncertainty of $\delta D \simeq 5\% $ and $\delta C_{\rm hel} \simeq
 7\%$, respectively, using
  dileptonic and semileptonic $\tbart$ events. \\ 

\begin{figure}
\begin{center}
\includegraphics[width=8cm]{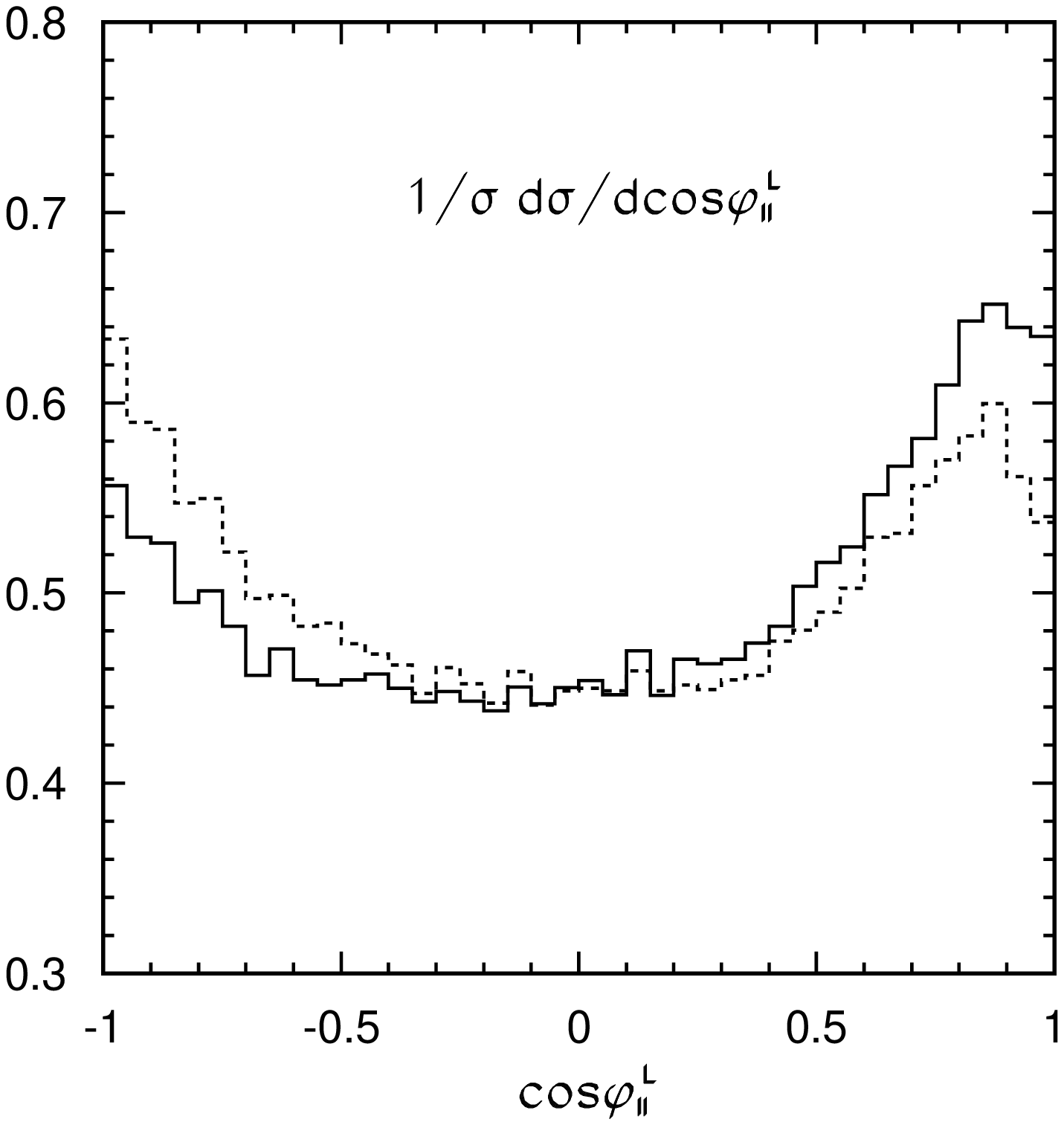}
\includegraphics[width=8cm]{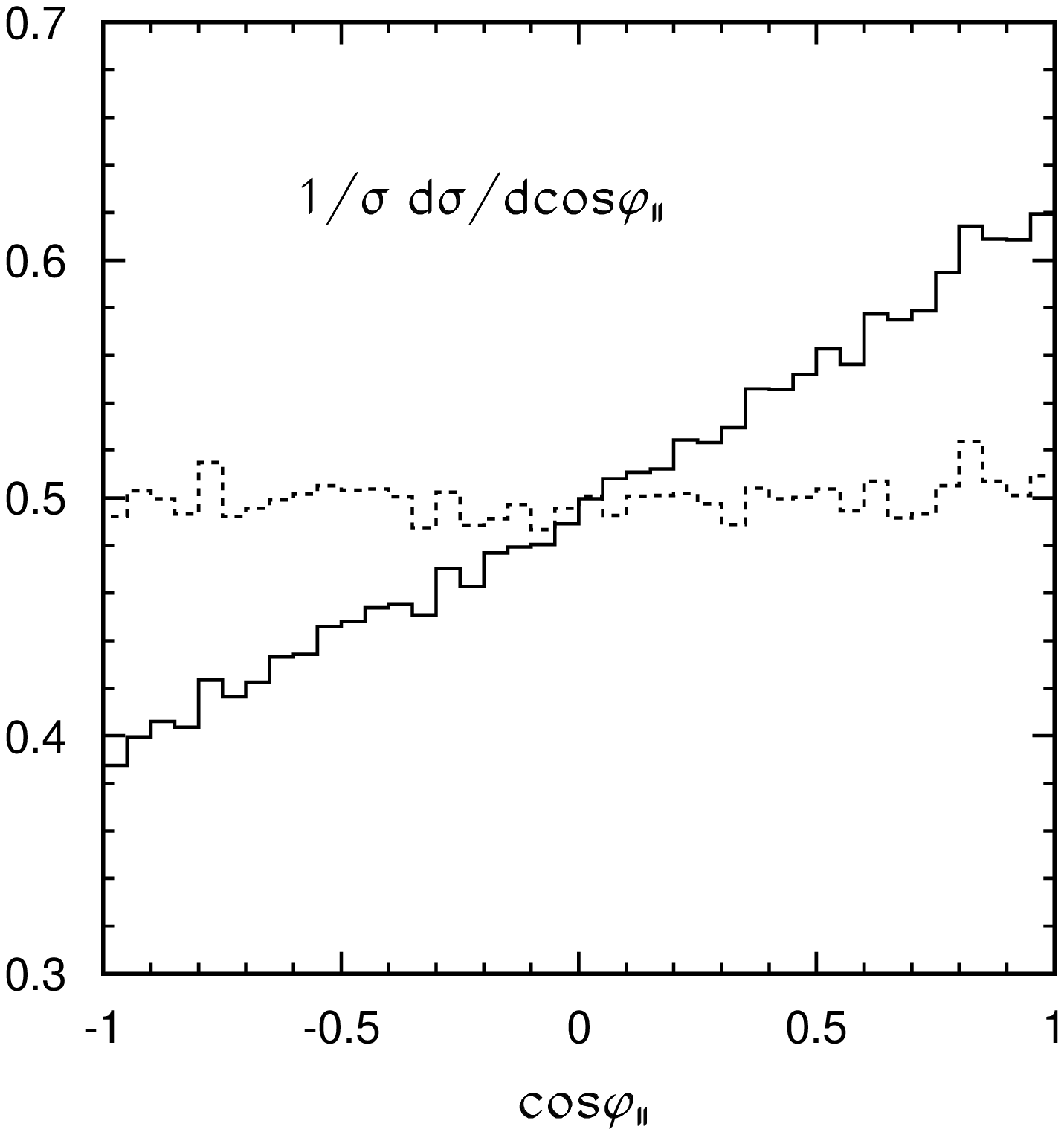}
\end{center}
\caption{Left: The opening angle distribution  $\sigma^{-1}d\sigma/d\cos\varphi^L$ in the
  laboratory frame. Right: The opening angle  (\ref{eq:ddist2}) defined 
   with respect to the $\tbart$ ZMF.
   Both are computed at  NLOW with $\mu =m_t$ 
  for $\ell \ell$ final states at the LHC ($\sqrt{s}=14$ TeV).
   The solid lines are the SM predictions, the dashed lines
    are the distributions which result when the $\tbart$
   spin correlations are switched off.}
 \label{fig:spin3lhc14}
\end{figure}

Next we analyze two leptonic angular correlations defined in the laboratory frame. 
The distribution of the opening angle between the lepton momenta in the
 laboratory frame, 
${\sigma}^{-1}{d\sigma / d\cos\varphi^L}$, which was recently considered 
   at NLO QCD in \cite{Melnikov:2009dn}, is almost 
 insensitive to $\tbart$ spin correlations \cite{Bernreuther:1997gs}.
  This is demonstrated in
   Fig.~\ref{fig:spin3lhc14} (left panel) where this distribution is shown 
   for the LHC with\footnote{For comparison we have computed this distribution
  also with the cuts used in \cite{Melnikov:2009dn} and find agreement
   with Fig.~4 of this work.}
     and without $\tbart$ spin correlations. The  correlation of the boosted leptons 
in the laboratory frame traces the 
    correlation of the $t$ and $\bar t$ spins  inefficiently.
    On the other hand, the distribution (\ref{eq:ddist2}) discriminates by
    construction\footnote{This is of course also the case for the
      beam, off-diagonal, and helicity correlations.} between spin 
    correlations ``switched-on'' and ``switched-off'',
     cf.  Fig.~\ref{fig:spin3lhc14} (right panel). 
      
\begin{figure}
\begin{center}
\includegraphics[width=8cm]{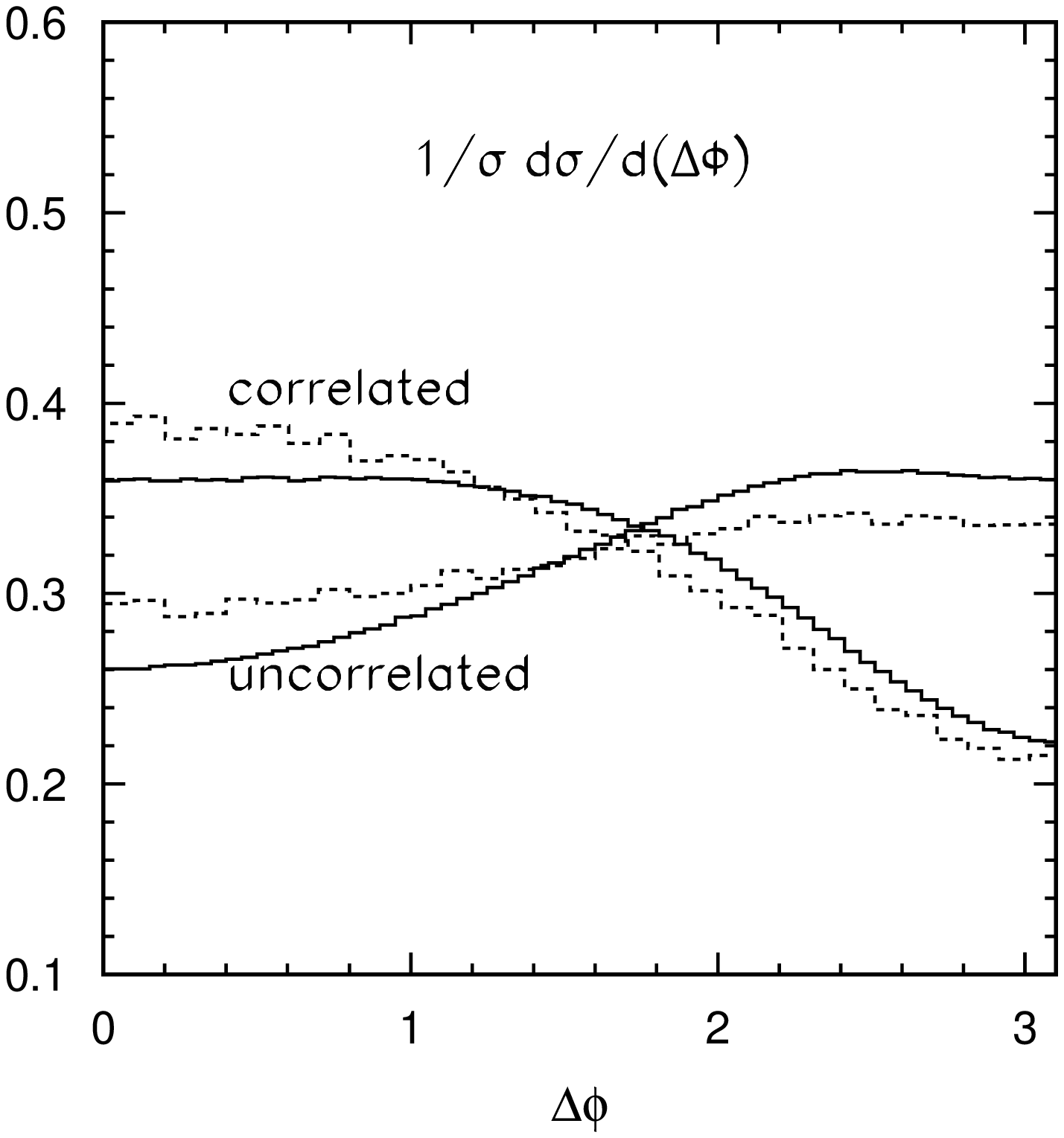}
\includegraphics[width=8cm]{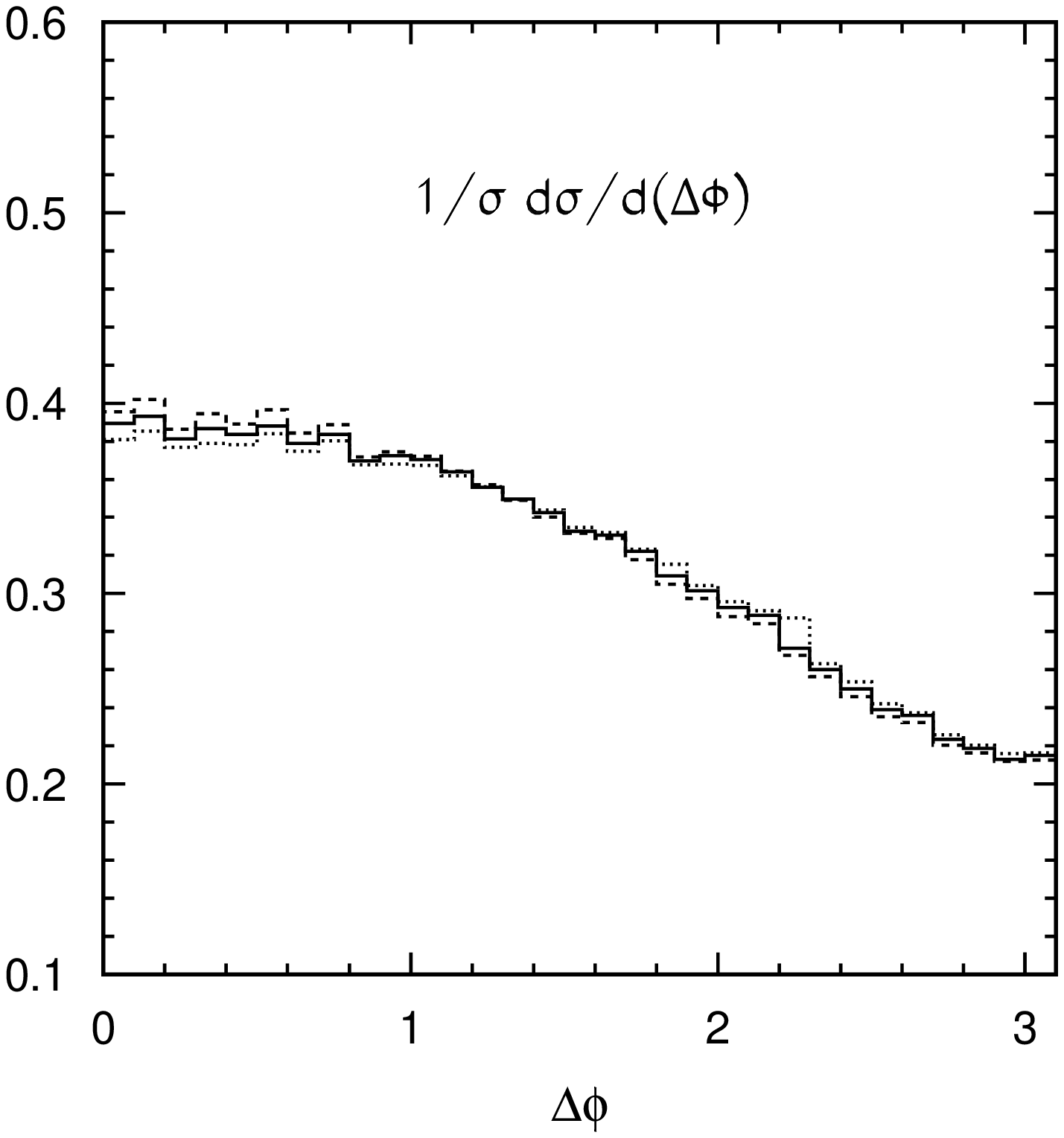}
\end{center}
\caption{The azimuthal angle distribution
     $\sigma^{-1}(M_{\rm max}=400 \,{\rm GeV}) d\sigma/d\Delta\phi$   in the
  laboratory frame for  $\ell\ell$ final states at the LHC (14 TeV)
 with and without $\tbart$ spin correlations,
        with the additional cut $\mtt<M_{\rm max}$. Solid
  and dashed lines are LO and NLOW, respectively, for $\mu=m_t$.
   Right: The distribution at NLOW for  $\mu =m_t$ (solid), $m_t/2$ (dashed), and
  $2m_t$ (dotted).}
 \label{fig:spin4lhc14}
\end{figure}

 As shown recently  \cite{Mahlon:2010gw} by a tree-level analysis, 
 the dileptonic azimuthal angle correlation at the  LHC, 
   $\sigma^{-1}{d\sigma/ d\Delta\phi}$,
     does discriminate between correlated and uncorrelated $\tbart$ events
  if dileptonic final states with $\mtt < 400$ GeV are taken into account only.
  Here
  $\Delta\phi$ is the difference of the 
      azimuthal angles of  $\ell^+$ and $\ell'^-$ in the laboratory
      frame. 
 This distribution is plotted in Fig.~\ref{fig:spin4lhc14} at LO and NLOW
 for correlated and uncorrelated dileptonic $\tbart$ events. 
  In  \cite{Mahlon:2010gw} only the LO distribution is given with which we agree.
  For the  cross section ratio 
$r=\sigma_{\ell \ell}(\mtt < 400\, {\rm GeV})/\sigma_{\ell \ell}$ we get with
   the above cuts $r \simeq 18.6\%$ at NLO QCD with $\mu= m_t$. Assuming 
   an integrated luminosity of 10 ${\rm fb}^{-1}$ at the LHC (14 TeV),
    this corresponds to $\sim 32000$ dilepton events $(\ell=
    e,\mu)$.

\begin{figure}
\begin{center}
\includegraphics[width=8cm]{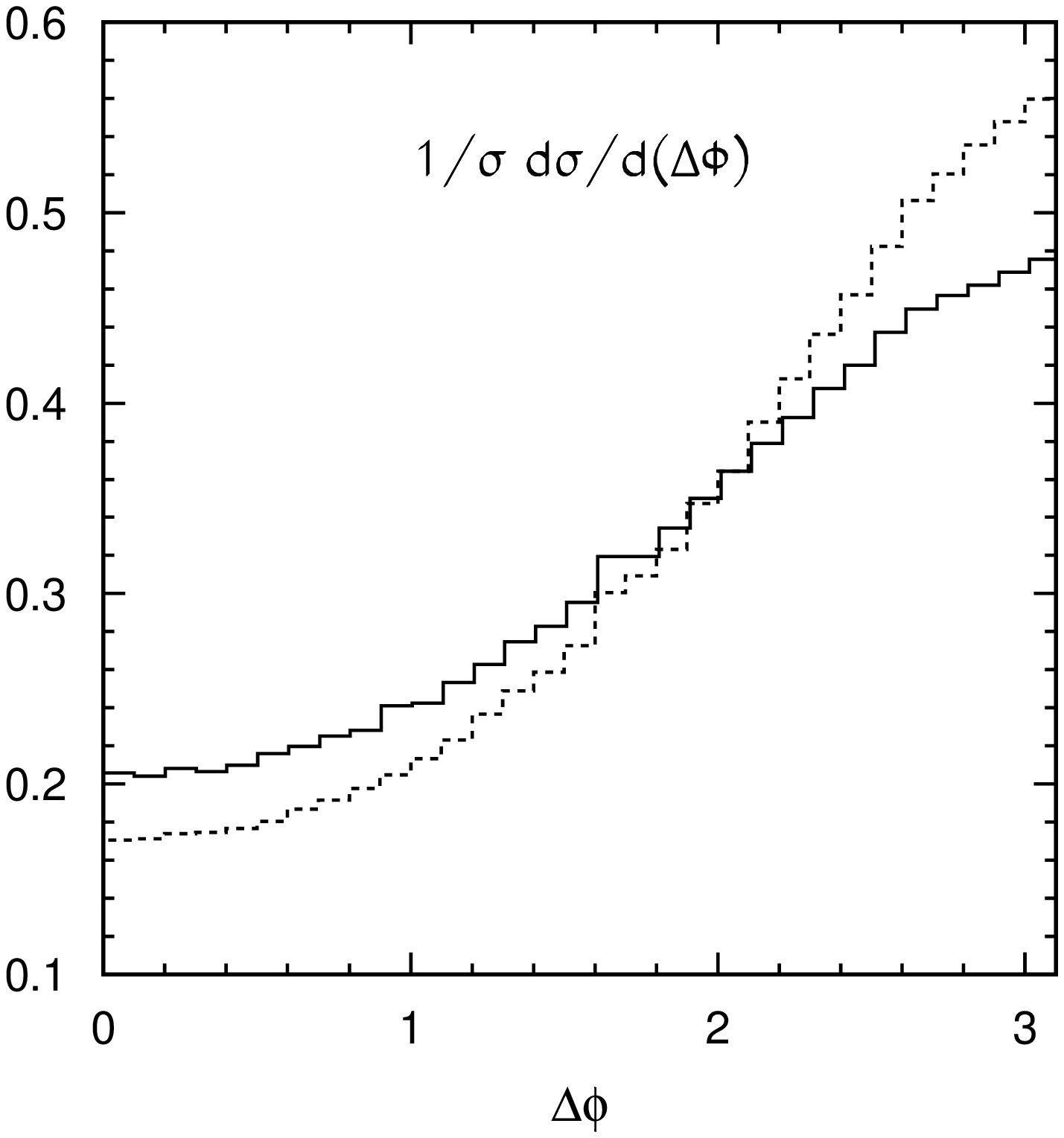}
\end{center}
\caption{Left: The azimuthal angle distribution
     $\sigma^{-1} d\sigma/d\Delta\phi$   in the
  laboratory frame for  $\ell\ell$ final states at the LHC (14 TeV)
  at NLOW for $\mu =m_t$ and no cut on $\mtt$. 
 The solid line is with and the dashed line 
     without $\tbart$ spin correlations.}
 \label{fig:spin4lhc14no}
\end{figure}

   The shapes of the correlated and uncorrelated distribution depend
   very sensitively on the cut on $\mtt$. Thus, a critical
   (experimental) issue will be how well the theory-cut on $\mtt$ can
   be realized in the selection of dileptonic events at the LHC, given
   the  ambiguities that arise from the presence of the two neutrinos
   in the final state. This complication was analyzed in
   \cite{Mahlon:2010gw}.
    In addition, the azimuthal angle correlation looses its discriminating power the more
   the cut on $\mtt$  is relaxed to  values above $M_{max} = 400$
   GeV. Fig.~\ref{fig:spin4lhc14no} shows this distribution at NLOW
   for correlated and uncorrelated  dileptonic $\tbart$ events (with
   cuts as given in (\ref{cut:dilep})) if no cut on $\mtt$ is applied.
    The plot compares well with Fig.~1 of \cite{Frixione:2007zp}
    computed with the MC@NLO code. (The above comments on  the
     differences between \cite{Frixione:2007zp} and our
     approach in the computation of the openening angle distribution
     apply also here.)

The azimuthal angle correlation is clearly easier and more precisely
  measurable  (assuming that the problem of finding the best option 
   for selecting dileptonic events will be solved)  than the 
helicity correlation or the opening angle distribution (\ref{eq:ddist2}).
 It can be used to check the SM prediction on $\tbart$ spin
 correlations or may be employed in the search for new physics
  effects in the low energy tail of the $\mtt$ spectrum.
 If one
  one wants to keep  spin correlation observables in the tool kit,  especially  for 
       new physics searches in the upper part of the 
    $\mtt$ spectrum, where new resonances or continuum effects may show up,
     one should use observables like the helicity correlation or the
   opening angle distribution. \\
%

%

%

The beam, off-diagonal, and helicity correlation and the opening
  angle distribution  (\ref{eq:ddist2}) can also be measured for
 semileptonic $\tbart$ events at the Tevatron
and at the LHC. In this case the $t$ and $\bar t$ rest frames are 
 easier and more efficiently reconstructible than for dileptonic
 final states. Moreover the number of $\ell+$ jets events is about six times
 larger than the number of $\ell\ell'$ events. 
 In the case of a hadronically decaying top quark  the least energetic
  non-$b$ jet $j_<$ among the top-decay products may serve as analyzer of the
  top spin. It was shown \cite{Brandenburg:2002xr} 
  that $j_<$  has the  best spin-analyzing power
  for this decay mode (barring the extremely difficult task of tagging
   the flavor of the $d$-type jet from $W$ decay); $\kappa_{j_<}\simeq 0.47$ to order $\alpha_s$
  depending on the jet algorithm. Thus the  angular
  correlations for the  $\ell+$ jets events will be reduced roughly
  by this factor as compared with  the respective correlations for
  $\ell\ell'$ final states. Detailed predictions with acceptance cuts will be
  given elsewhere \cite{BeSi2010}.\\

%
\begin{figure}
\begin{center}
\includegraphics[width=8cm]{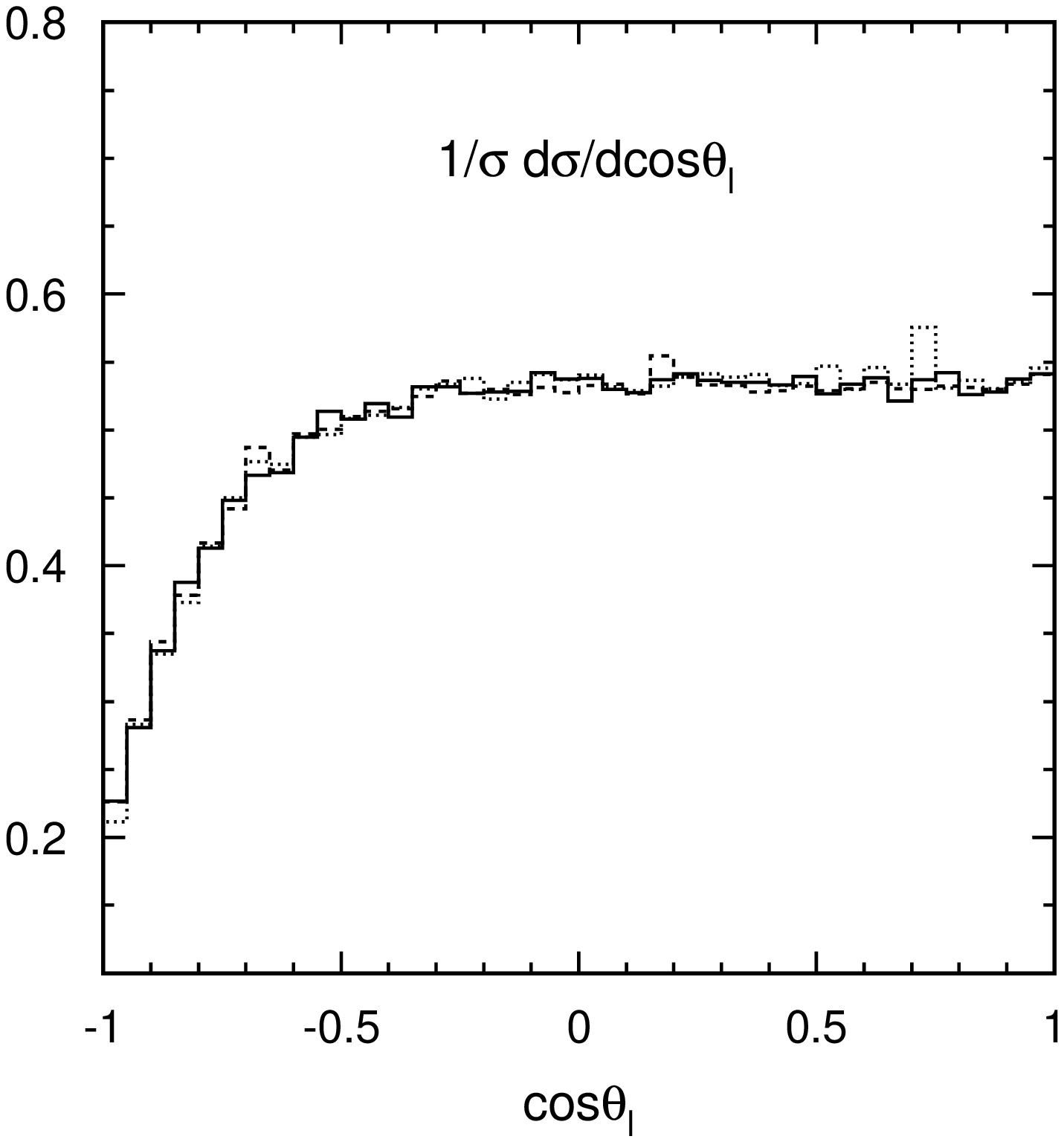}
\includegraphics[width=8cm]{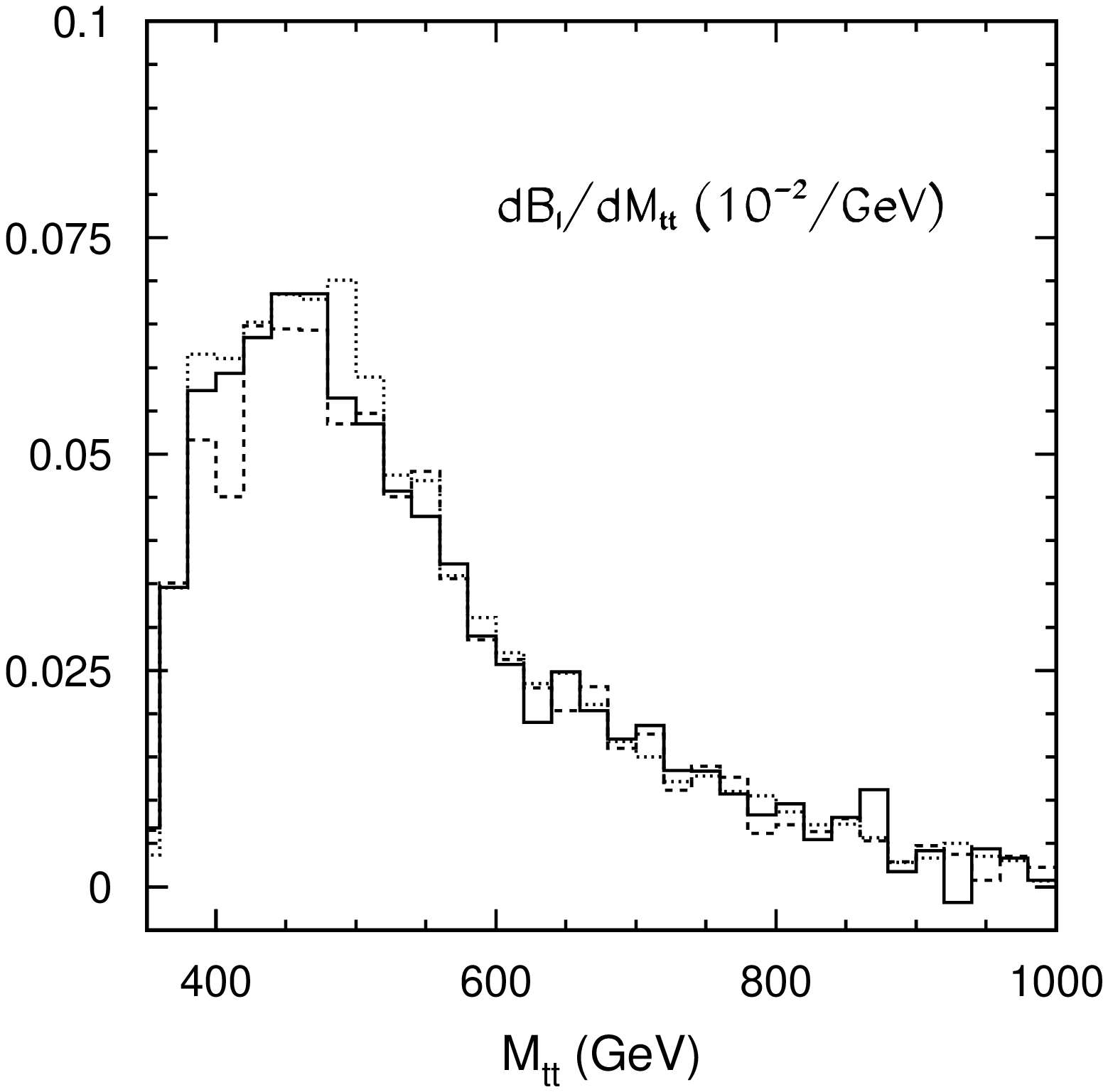}
\end{center}
\caption{The distribution
 $\sigma^{-1}d\sigma/ d\cos\theta_{\ell^+}$ (left panel) and the
       corresponding differential 
distribution $dB_1/d\mtt$ (right panel) for 
    $\ell^+ +$ jets at the LHC (14 TeV). Here $\theta_{\ell^+}$ is the polar angle
   of $\ell^+$ in the $t$ rest frame with respect to the $t$ direction in the
   $\tbart$ ZMF. The NLOW results are for $\mu = m_t$ (solid), $m_t/2$
   (dashed), and  $2m_t$ (dotted).}
 \label{fig:sinspinlhc14}
\end{figure}

Finally we consider, for  the  the $\ell^\pm + $ jets 
 modes, the one-particle distributions
 $G_{\ell^\pm}(z_\pm)$ defined in (\ref{eq:ddistcosl2})
 where $z_\pm =\cos\theta_{\ell^\pm}$ denote the cosines of the
   lepton helicity angles, 
 i.e.,  $\theta_{\ell^+}$ ($\theta_{\ell^-}$)
  is the angle between the $\ell^+$   ($\ell^-$)   direction of
   flight in the $t$ ($\bar t$)  rest frame
              and the $t$ ($\bar t$) direction in the $\tbart$ ZMF. 
    
If no cuts are applied, non-trivial, i.e. non-flat  distributions would 
  arise if the ensemble of (anti)top quarks were longitudinally
  polarized -- a  small effect in the SM
  \cite{Bernreuther:2006vg,Bernreuther:2008md}. However, the presence  of cuts severely
  distort these distributions in the vicinity of
  $\cos\theta_{\ell^\pm}=-1$. This region corresponds to leptons that
  are emitted into the hemisphere opposite to the (anti)top direction of flight in the
  $\tbart$ ZMF. These leptons are therefore less energetic on average and thus
  stronger affected by cuts than those in the remaining region. 
    This is shown  in Fig.~\ref{fig:sinspinlhc14} for $G_{\ell^+}$
   and the corresponding differential 
distribution $dB_1/d\mtt$ at the LHC. The SM value of $B_1$ is
  given, for the above cuts, in  Table~\ref{TabTEVLHC}.    
 To NLO in the gauge couplings, the distributions  $G_{\ell^-}$ and $dB_2/d\mtt$
  are identical to  $G_{\ell^+}$ and $dB_1/d\mtt$, respectively, which
  we checked with our code.
This is because 
  \begin{equation}\label{eq:ddistcosl3}
    G_{\ell^+}(z_+) =   G_{\ell^-}(z_-)
\end{equation}
 holds in the SM to a high degree of accuracy if the selection cuts/criteria
  are CP images of each other. Although the
   $pp$ state is not a CP
  eigenstate, one can show that (\ref{eq:ddistcosl3}) holds
 as long as  $gg$ and
   $q\bar q$ initiated $t \bar t$ production is governed by
   CP-invariant interactions  \cite{Bernreuther:1998qv}. \\
 One may use these distributions for checks of whether or not the
 collected ensemble
 of (anti)top
 quarks has a  longitudinal polarization, which would be due
 to some new parity-violating interaction that affects $\tbart$
 production\footnote{Related parity-violating polarization observables and
   their size in some SM extensions were investigated in
   \cite{Kao:1999kj}.}. That is, one may investigate whether or not
 the measured distributions differ from the SM results, Fig.~\ref{fig:sinspinlhc14}. \\
If one defines $\cos\theta_{\ell^+}={\bf{\hat
     a}}\cdot{\velp}$ with respect to an axial vector ${\bf{\hat
     a}}$, e.g. the vector normal to
 the $2 \to 2$ scattering plane (and likewise for ${\ell^-}$), one may check
     for a normal polarization of top
 quarks generated by absorptive parts in the scattering amplitude. At
 NLO QCD this polarization amounts to a few percent
  \cite{Bernreuther:1995cx,Dharmaratna:xd}. \\

We close this section with a few  remarks on some
distributions/correlations  suitable
   for CP symmetry tests
respectively searches for non-standard CP violation (cf., e.g.,
\cite{Atwood:2000tu} for a review).  
  This can be made
  with dileptonic and semileptonic $\tbart$ events at the Tevatron and
  especially  at the LHC, once sufficiently large data samples will have
  been collected.  Although the (expectation values of the) observables below
 cannot be classified a priori with respect to CP  in the case of $pp$
 collisions, they may nevertheless be used at the LHC.
 As shown in
  \cite{Bernreuther:1993df,Bernreuther:1993hq} by a
  general kinematic analysis,
       P- and CP-violating interactions that
  affect hadronic $\tbart$ production induce, in general, at the level of the
  $\tbart$ intermediate states two types of spin
  correlations/asymmetries:
 \begin{enumerate}
\item A difference in the longitudinal polarization of $t$ and $\bar t$
  quarks. The most useful asymmetry  corresponds to 
  $\kh \cdot(\Sp - \Sm)$ 
   where $\kh$ is the direction of the $t$ quark  in the $\tbart$ ZMF. 
  In the dileptonic or semileptonic $\tbart$ events these correlations
   lead to  a difference in the distributions  $G_{\ell^\pm}$.
  One may consider  the ratio
\begin{equation} \label{eq:ddistcosl4}
   \Delta_{CP}^\ell (z) = \frac{G_{\ell^+}(z)-
     G_{\ell^-}(z)}{G_{\ell^+}(z)+ 
    G_{\ell^-}(z)} \, 
\end{equation}
As shown above,  $\Delta_{CP}^\ell =0$ to NLOW in the SM. \\
 A non-zero  $\Delta_{CP}^\ell$ requires, apart from P and CP
 violation, also an absorptive part in the scattering matrix. 
 A non-zero polarization asymmetry  $\kh \cdot(\Sp - \Sm)$ amounts to a
   difference $N(t_L{\bar t}_L)- N(t_R{\bar t}_R) \neq 0$  of the
   number of $\tbart$ with negative  and positive helicity \cite{Schmidt:1992et}.
   In the case of a CP-violating Higgs sector, in particular
     if heavy spin-zero
   states with undefined CP parity 
   exist that are resonantly produced in the $\tbart$ channel,
    then suitable asymmetries evaluated in suitable bins around the
   position of the resonance \cite{Bernreuther:1997gs} may be as large as a few percent
   \cite{Bernreuther:1998qv}. \\
 In the context of leptonic
  charge asymmetries discussed in Sect.~\ref{sub:chaasy} it is natural to  consider
\begin{equation} \label{rapdischl22}
 A_{CP}^\ell(y) =  \frac{N_{\ell^+}(y) - N_{\ell^-}(-y)}{N_{\ell^+}(y) + N_{\ell^-}(-y)} \, ,
\end{equation}
 or its integrated version,
\begin{equation} \label{rapdischl2i}
 A_{CP}^\ell =  \frac{\int\limits_{y>0} dy \, N_{\ell^+}(y) - \int\limits_{y<0} dy \,
 N_{\ell^-}(y)}
{\int\limits_{y>0} dy \, N_{\ell^+}(y) + \int\limits_{y<0} dy \,
 N_{\ell^-}(y)} \, .
\end{equation}
 The asymmetries (\ref{rapdischl22}) and (\ref{rapdischl2i})
 correspond to a top-spin asymmetry with respect to the beam direction.
  They are  rather insensitive to P- and
  CP-violating Higgs-boson effects. Larger effects would result from
  $s$-channel exchanges of  new heavy $J\neq 0$ resonances with
   CP-violating couplings to top quarks.  A possible difference in the
  transverse energies of $\ell^+$ and $\ell^-$, which derives also from
 these single spin asymmetries, is below the percent level for
 non-resonant  CP-violating exchanges of Higgs bosons
  \cite{Schmidt:1992et} or of  squarks/gluinos \cite{Schmidt:1992kt}.

  \item  P- and CP-odd/T-odd $\tbart$ spin correlations. The most
   useful one is  proportional to
   $\kh \cdot(\Sp \times \Sm)$. 
  It can be induced already at tree level.  In dileptonic events
  it would show up in correlations proportional to
     ${\cal O}_{CP}=\kh \cdot ({\velp} \times
  {\velm})$, and
  likewise for semileptonic events. When the unit momenta
   ${\velp}$ $({\velm})$ and
   $\kh$    are 
  taken in the $t$ $({\bar t})$ rest frame and
 in the $\tbart$ ZMF, respectively, then  ${\cal O}_{CP}$ corresponds,
 in the absence of cuts, to a term proportional to $\sin\Delta\phi^*$,
 where $\Delta\phi^*$ denotes the difference of the  $\ell^+$,
 $\ell^-$   azimuthal angles in the plane orthogonal to
  $\kh$.  As the construction of these frames with
 reasonably small experimental error will be  difficult, one may use
 pseudo rest-frames in analogy to what was  shown 
in  \cite{Berge:2008dr}. \\
In the case of resonant $s$-channel exchange of a Higgs particle
  with scalar and pseudoscalar Yukawa couplings,  suitable
 asymmetries $A_{CP}\propto N({\cal O}_{CP}>0) - N({\cal O}_{CP}<0)$ 
   may be  as large as a few
 percent when evaluated  in appropriate bins around the
   position of the resonance
   \cite{Bernreuther:1997gs,Bernreuther:1998qv}, and  do not
  receive SM contributions at  the few per mille level.

The  above dileptonic azimuthal angle correlation in the laboratory frame,
  $\sigma^{-1}{d\sigma/ d\Delta\phi}$, is symmetric with respect
    to $\Delta\phi \to 2\pi - \Delta\phi$ in the case of CP
    invariance. CP-violating interactions would, in general, modulate it
     by a term proportional to $\sin\Delta\phi$ (in the absence of
     cuts); i.e. one should measure it in the whole range 
 $0\leq \Delta\phi < 2\pi$. In the terminology of
 \cite{Bernreuther:1993df}
    this term corresponds
 to a correlation of $\Sp \times \Sm$ with the beam direction in the
  laboratory frame, i.e., to a correlation observable ${\bf\hat p}_L \cdot ({\velpL} \times
  {\velmL})$.  As far as possible
   Higgs sector CP violation is concerned, this correlation is
   rather insensitive to resonant terms if the mass of
   the resonance is significantly above the $\tbart$ production
   threshold.  The  non-resonant Higgs-boson exchange terms lead to
    rather small CP-violating effects \cite{Bernreuther:1993hq}.

\end{enumerate}

\section{Conclusions}
\label{sec.conc}
We have investigated a number of  observables  
that are and will be instrumental in the exploration of 
$\tbart$ production and decay at the Tevatron and the LHC. 
For this analysis we made a computer program that incorporates 
 besides the NLO QCD corrections to $\tbart$ production and decay also
   mixed weak-QCD corrections to the production amplitudes, and that allows for
 studies of correlated versus uncorrelated $\tbart$ events. \\

 We assessed, for a few 
 distributions and correlations
  at the LHC, the relative size of 
 the weak-interaction contributions as compared to
the NLO QCD results. In the  $p_T$ or $\mtt$ range where most of the
 $\tbart$ events at the LHC are located,  these contributions deplete the NLO
 QCD results somewhat, for instance, the $p_T$ and $\mtt$ distribution
   by about $- 2\%$ for
 $p_T \sim 400$ GeV  and  ${\mtt} \sim 1.2$ TeV. For larger $p_T$,
 $\mtt$ the depletion  grows to several percent. As to the 
    charge asymmetries $A$ and $A^{\tbart}$ computed at the level 
  of intermediate $\tbart$: our results agree 
  with \cite{Antunano:2007da}; yet we find 
  that the  weak-interaction contributions are slightly 
     smaller than those 
  determined in \cite{Kuhn:1998kw}, due to additional 
 terms that were previously not  taken into account. \\

Our main physics results are the investigations at NLOW in the
  gauge couplings of the asymmetries and
  angular correlations for dileptonic events which were
  presented in Sect.~\ref{sec:diljet}.
  First, we considered charge asymmetries at the
   Tevatron.  The pair-asymmetry 
     $A^{\tbart}$ decreases by about $10\%$ if acceptance cuts are
   applied. Our result, $A^{\tbart}=0.071(7)$ (where the given uncertainty
   due to scale variations certainly underestimates the true
   theoretical error), may be compared with the respective D0 measurement
    \cite{:2007qb}. Both results agree within the rather large
     uncertainties.
         We computed also
     a leptonic charge asymmetry $A^{\ell}$  and a  pair-asymmetry
      $A^{\ell\ell}$ for the Tevatron, 
    which are related to $A$ and $A^{\tbart}$,
      respectively. As in the case of
     $A^{\tbart}$, we determined these asymmetries
    for correlated and uncorrelated $\tbart$ events, in order to
     study the effect of spin correlations on these 
  distributions.  We showed that $\tbart$ spin correlations affect
    the pair asymmetry $A^{\ell\ell}$ by about $7\%$ for the
    acceptance cuts used in   Sect.~\ref{sec:diljet}.  These leptonic
   asymmetries have, to our knowledge, not yet been measured, probably
   because of  limited numbers of (di)leptonic events.
    As these leptonic asymmetries should eventually be measurable  more precisely
    than $A$ or $A^{\tbart}$, their experimental determination might
    provide a more conclusive comparison with SM results.
     Our NLOW predictions  are given  in Table~\ref{chasTEVdilep}.
    One may speculate that top-spin effects are more pronounced
     in more exclusive charge or forward-backward asymmetries
     \cite{Bowen:2005ap}, but
     this remains to be investigated. \\

Furthermore, we  determined several dileptonic angular correlations,
 which reflect $\tbart$ spin correlations with respect to different
 spin bases, namely the beam, off-diagonal and helicity correlation,
  and an opening angle distribution defined in a specific 
 way  \cite{Bernreuther:2004jv}, when selection cuts are applied.
Our NLOW predictions for the beam, off-diagonal, and helicity
correlation for the Tevatron agree with recent measurements 
  by the CDF and D0 experiments; albeit the experimental uncertainties are still large,
  mainly because of limited statistics.

We have also made predictions for estimators (which have
 1-dimensional distributions)  of these correlations  as functions
of $\mtt$. These estimators
    may prove useful for the Tevatron and 
 also  in the early rounds of LHC data-analyses,  where the  event numbers will likely not be abundant.

We have also considered two angular 
 correlations  defined in the laboratory frame  for dileptonic
  final states at the LHC, and we have computed
 them to NLOW for correlated and uncorrelated $\tbart$ events. Our
 results confirm the findings of \cite{Mahlon:2010gw}, namely that  the  azimuthal
 angle correlation is a sensitive variable for exploring $\tbart$ spin
 correlations at the LHC in the low-energy tail of the $\mtt$
 spectrum (${\mtt} <  400$ GeV). It should be recalled that the helicity correlation and the opening angle distribution
 as defined in \cite{Bernreuther:2004jv} are sensitive to spin
 correlations also when applied to  event samples with 
  ${\mtt}^{\rm cut} > 400$ GeV.
 An crucial experimental issue in the precise 
  measurement of  both the azimuthal
 angle correlation and the latter observables is the kinematic
 reconstruction  of the dileptonic events. \\

Moreover, we have briefly discussed a few P- and
CP-odd observables that should be useful in  the search 
 for P- and/or CP-violating interactions in dileptonic and
 semileptonic $\tbart$ events. Our NLOW results on
 distributions and correlations indicate that the above
 observables do not receive SM contributions  at the level of a few per mille
 or below. This issue deserves further, more detailed studies. \\

It is  also worth recalling here that, in case the measurements of the
$\tbart$ spin correlations at the LHC should eventually match 
 the SM predictions -- for instance predictions that involve the low-energy
 tail of the $\mtt$ spectrum -- and would reach a reasonable level
   of precision, one may use these correlations also for the further
   exploration of 
   the  parton content of the proton. They are, in fact,
   quite sensitive to the proton's relative quark and gluon content, 
   because the contributions from $gg$ and $q {\bar q}$ initiated
   $\tbart$ production have opposite 
   signs \cite{Bernreuther:2001rq,Bernreuther:2004jv}. \\

The above investigations and comparisons
   with existing experimental results  show that the precision, with which
   some distributions, especially  angular
 correlations can be predicted,  will
  likely not be matched quickly by experiments in the near future. 
 Nevertheless, in view of the potential importance of observables such
 as the leptonic charge asymmetries and spin correlations in exploring
  the interactions of top quarks, further detailed investigations on
   expected measurement uncertainties at the LHC
     would certainly not be futile exercises. \\

 Future extensions of our work may include the implementation
  of  threshold resummations into our code and, as
to more phenomenological issues,  an analysis of charge asymmetries
  and angular correlations at NLOW also for lepton + jets final states, and a more
 systematic error analysis by taking  into account also 
 other available PDF sets \cite{Martin:2009iq,Alekhin:2009ni} including
  the PDF uncertainties.

\subsubsection*{Acknowledgements}
We thank  Jorgen D'Hondt, Eric Laenen, Christian
Schwanenberger, Peter Uwer and Wolfgang Wagner for discussions and
information about their work. 
Z.G. Si wishes to thank RWTH Aachen for its hospitality
 during a stay where this work was completed, and 
   the Deutsche Forschungsgemeinschaft (DFG) and the Ministry of Education (China)  for an exchange
   grant. His work was supported in part by NSFC and by Natural Science Foundation of
Shandong Province and that of W.B. by DFG, SFB TR9.

\end{document}